\documentclass[a4paper,11pt]{article}
\pdfoutput=1 

\usepackage{jheppub} 

\usepackage[T1]{fontenc} 

\usepackage[utf8]{inputenc}

\usepackage{titlesec}
\usepackage{dcolumn}
\usepackage{bm}
\usepackage{amsmath,amssymb,amsfonts}

\usepackage{graphicx}
\usepackage{subfigure}
\usepackage{color}
\usepackage{bbold}

\usepackage{slashed}
\usepackage[svgnames]{xcolor}
\usepackage[english]{babel}
\usepackage{blindtext}
\usepackage{microtype}
\usepackage{tikz}

\usepackage[colorlinks=true,urlcolor=blue,citecolor=blue]{hyperref}
\usepackage{ifpdf}
\usepackage{float}

\graphicspath{{.}{figures/}}

\title{\boldmath Kaon GTMDs in the Dyson-Schwinger equations using contact interaction}

\author[a,1]{Jin-Li Zhang\note{Corresponding author.}}

\affiliation[a]{Department of Mathematics and Physics, Nanjing Institute of Technology, Nanjing 211167, China}

\emailAdd{jlzhang@njit.edu.cn}

\abstract{An array of the kaon twist-two, three, four generalised transverse momentum dependent parton distributions (GTMDs) has been investigated within the framework of covariant and confining Dyson-Schwinger equations using contact interaction. GTMDs are of great significance as they encompass information regarding both the generalised parton distributions (GPDs) and the transverse momentum dependent parton distributions (TMDs), thus being considered as the parent distribution. From GTMDs, we can derive the twist-two, three, four GPDs and TMDs. GPDs are obtained through the integration of $\bm{k}_{\perp}$ from GTMDs, with a focus on the twist-two GPDs. The first Mellin moments of GPDs yield the form factors of local currents. The second Mellin moments of vector GPDs are related to gravitational form factors. The Wigner distribution can be obtained from a Fourier transform in the transverse space of the GTMDs at skewedness parameter $\xi=0$. The Wigner distributions of an unpolarized, longitudinally polarized, and transversely polarized quark inside the kaon have been calculated. Through the three twist-two Wigner distributions, we study the dynamical spin effects of the quarks inside kaon to reveal its multidimensional structure. The spin-orbit correlations between a hadron and a quark can be explained based on the phase-space average of Wigner distributions. We investigate the correlation between the longitudinal spin and orbital angular momentum of valence quarks within the pion and kaon. Our findings reveal that $C_z^{u,K}=-0.336$, $C_z^{s,K}=0.242$, and $C_z^{u,\pi}=-0.374$. The parton distribution function in impact parameter space can be derived from the Wigner distribution. The study focuses on the light-front transverse-spin distributions $\rho_u^1\left(\bm{b}_{\bot },\bm{s}_{\perp}\right)$ and $\rho_u^2\left(\bm{b}_{\bot },\bm{s}_{\perp}\right)$, which exhibit distortions, and we calculate their average shift. This comprehensive analysis will enhance our understanding of the parton distribution picture of kaons. While there have been a few theoretical studies investigating the GTMDs in experiments, no experimental data is currently available. The results of our model calculation offer qualitative insights into these distributions.}

\begin{document}
\maketitle
\flushbottom

\section{Introduction}
The internal structure of hadrons and the fundamental rules of interactions have always been at the forefront of research in high-energy hadronic physics. Quantum Chromodynamics (QCD) is the underlying theory describes the strong interactions between quarks and gluons. Over the past two decades, significant attention has been devoted to GPDs~\cite{Mueller:1998fv,Ji:1996nm,Radyushkin:1997ki,Ji:1998pc,Diehl:2003ny,Boffi:2007yc,Zhang:2021mtn,Zhang:2021shm,Zhang:2021tnr,Zhang:2021uak,Zhang:2022zim,Deja:2023tuc,Bertone:2023jeh,Riberdy:2023awf}  and TMDs~\cite{Collins:2007ph,Meissner:2007rx,Hautmann:2007uw,Barone:2010zz,Boglione:2024dal,Deganutti:2023qct,Puhan:2023hio,Kou:2023ady}, as well as their interrelation~\cite{Courtoy:2013oaa}. GPDs are actual amplitudes for quarks or gluons that are probed in a hard process and then recombined to form a scattered hadron. They are obtained through exclusive electroproduction of vector bosons with the nucleon, such as deeply virtual Compton scattering (DVCS)~\cite{Goeke:2001tz,Radyushkin:1996nd,Hobart:2023knc,Xie:2023xkz}, deep virtual meson production (DVMP)~\cite{Muller:2013jur,Favart:2015umi,Cuic:2023mki} and timelike Compton scattering (TCS)~\cite{Berger:2001xd,Boer:2015fwa,Xie:2022vvl,CLAS:2021lky,Peccini:2021rbt}. TMDs refer to the distributions of various spin configurations of quarks and gluons within the nucleon. Semi-Inclusive Deep Inelastic Scattering (SIDIS)~\cite{Anselmino:2007fs,Anselmino:2008sga,Anselmino:2008jk,Anderle:2021wcy,Anselmino:2013vqa,Leader:2010rb} provides access to the longitudinal and transverse momenta of TMDs. GPDs characterize the distribution of a parton's longitudinal momentum and transverse position, while TMDs describe the distribution of the parton's longitudinal momentum and transverse momentum. The significance of TMDs in current hot topics, such as the precise determination of the mass of the W boson, underscores their relevance, leading to substantial efforts being devoted to their determination~\cite{Delcarro:2018lbr,Scimemi:2019cmh,Bacchetta:2022awv}. In addition, the recent establishment of electron-ion colliders in China (EicC)~\cite{Anderle:2021wcy} and in the US (EIC)~\cite{AbdulKhalek:2021gbh}, has sparked increased research interest in GPDs, leading to a rapid expansion of this field. Furthermore, advancements in lattice technologies~\cite{Ji:2013dva,Radyushkin:2017cyf} have made it more feasible to conduct first principle calculations of GPDs, thereby attracting additional attention to their study. Significant quantities such as the transverse position of partons and their angular momentum can be derived from these distributions.

GTMDs~\cite{Kanazawa:2014nha,Hagiwara:2016kam,Zhou:2016rnt,Zhang:2020ecj,Luo:2020yqj,Kumano:2020ijt,Boer:2021upt,Bertone:2022awq,Sharma:2023tre,Boer:2023mip,Jana:2023btd,Setyadi:2023wks,Bhattacharya:2023yvo,Sharma:2023qgb,Agrawal:2023mzm,Tan:2024dmz}  contain the maximum message with respect to the parton structure of strong interacting system. They are often referred to as ``mother distribution'' since they can be reduced to GPDs and to TMDs in specific kinematical limits. Therefore, GTMDs inherit properties of both TMDs and GPDs and any subtle problems relevant. Furthermore, GTMDs provide information on the orbital angular momentum (OAM) of quarks and gluons, OAM could help enhance our comprehension of the partonic structure of hadron and the related QCD dynamics. However, up to now, very little is known about the OAM of partons. From the viewpoint of both theoretical and experimental, it is still very challenging to extract parton OAM from high energy scattering experiment.

The Wigner distributions~\cite{Wigner:1932eb,Boussarie:2018zwg} are defined as the Fourier transform of GTMDs with respect to the impact parameter $\bm{b}_{\perp}$. As such, they represent the phase space distributions in momenta and impact parameters. TMDs and GPDs can also be viewed as different limits of Wigner distributions. These distributions contain valuable information about dynamical spin effects, allowing for a comprehensive visualization of the partonic structure of hadrons in five dimensions. This multidimensional representation is crucial for revealing the intricate nature of hadronic systems~\cite{Belitsky:2003nz,Lorce:2011ni,Mukherjee:2015aja,Liu:2015eqa,Chakrabarti:2017teq,Pasechnik:2023mdd}. Wigner distributions also allow us to study the correlation between quark spin and OAM. The spin-orbit correlations (SOCs) between a hadron and a quark can be explained by considering the phase-space average of Wigner distributions. This provides a professional and formal explanation for the relationship between the spin and orbital motion of these particles. Understanding the spin composition of hadrons has recently become a fascinating subject of research~\cite{Jaffe:1989jz,Ji:1996ek,Leader:2013jra,Kaur:2020vkq}. By comprehending the multidimensional structure, it becomes possible to analyze characteristics such as SOCs, spin-spin correlations, quark-gluon correlations, and other interactions of this nature. It would be of great interest to investigate the spin-$0$ mesons and their correlations with spin-orbital angular momentum, and then proceed to calculate analytical results. In particular, examining the correlations between hadron spin and the orbital motion of partons within the hadron can provide valuable insights into the spin structure of hadrons~\cite{Tan:2021osk}.

There is limited quantitative knowledge regarding GTMDs, and the majority of existing studies are based on theoretical models. Despite recent proposals for experimental access to GTMDs, there is currently a scarcity of data sensitive to these distributions, which makes phenomenological studies challenging and laborious. So far, the studies of GTMDs primarily focus on the formal theory side, including model calculations~\cite{Meissner:2009ww,Mukherjee:2014nya,Courtoy:2013oaa,Kanazawa:2014nha,Liu:2015eqa,Chakrabarti:2016yuw}, the analysis of the multipole structure associated with GTMDs, as well as the investigation of their QCD evolution properties.

We will assess the kaon GTMDs using the contact interaction (CI), which preserves many characteristics of the leading-order truncation of QCD's Dyson-Schwinger equations (DSEs), thereby offering a robust approach to numerous hadron observables. This builds upon our previous work on pion GTMDs~\cite{Zhang:2020ecj}. The key distinction from the pion lies in the fact that the isospin symmetry of the kaon is broken, allowing us to investigate the impact of this symmetry breaking. Furthermore, it can be compared with the results for pion and kaon GPDs and light-front wave functions (LFWFs) in the Nambu--Jona-Lasinio (NJL) model~\cite{Mineo:2005vs,Volkov:2005kw,Courtoy:2019cxq,Broniowski:2020jwk,Zhang:2021shm,Zhang:2021tnr,Zhang:2024dhs}.

Functions at various twists have been defined based on the helicity of hadrons in both the initial and final states of the process. Recently, higher twist distributions have been receiving a significant amount of attention. In this work, we discuss the twist-two, three, and four GTMDs for kaon mesons in the DSE. From the kaon GTMDs, we are able to obtain the twist-two, three, and four GPDs as well as TMDs. Furthermore, we also evaluate the parton distribution functions (PDFs) and form factors (FFs) of the kaon. The three twist-two Wigner distributions of $u$ and $s$ quarks in kaons are studied and compared with those of $u$ quarks in pions.

This paper is structured as follows: In Sec.~\ref{good}, we provide a concise introduction to the generalised parton correlation function and DSE, followed by an explanation of the calculation process for twist-two $u$ quark GTMDs of kaon. In Sec.~\ref{good1}, we derive the twist-two $u$ quark GPDs of kaon from GTMDs. The diagrams of the vector and tensor GPDs, as well as an analysis of the Mellin moments of the vector GPD, are provided. Additionally, the tensor form factors from the tensor GPD are obtained. In Sec.~\ref{good2}, we investigate the $u$ quark twist-two, three, four TMDs of kaon derived from GTMDs. In Sec.~\ref{good3}, we analyze the quark Wigner distributions for unpolarized (U), longitudinally polarized (L) and transversely polarized (T) quarks inside an unpolarized kaon, in both the impact parameter space (IPS) and the transverse space. Finally, a brief summary and outlook are provided in Sec.~\ref{excellent}.

\section{Kaon GTMDs}\label{good}

\subsection{Generalised parton correlation function}
\begin{figure}
\centering
\includegraphics[width=0.40\textwidth]{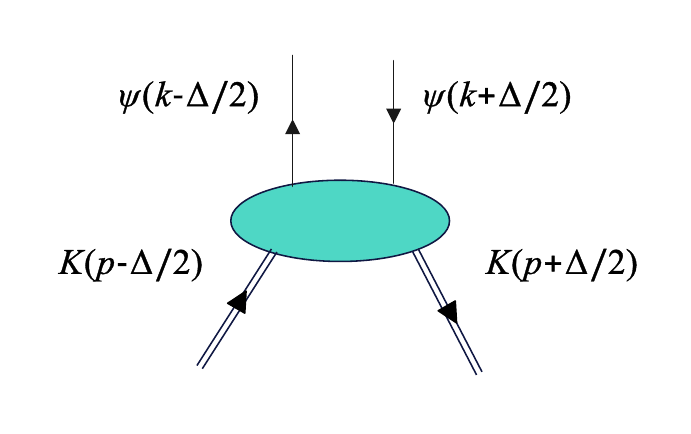}
\caption{Momentum-space conventions used in defining the in-kaon quark-quark correlator in Eq. (\ref{1}).}\label{gpcf}
\end{figure}

The kaon quark-quark correlator is defined as~\cite{Meissner:2008ay}
\begin{align}\label{1}
W_{ij}(P,k,\Delta,\bar{N};\eta)&=\int \frac{d^4z}{(2\pi)^4}e^{i k \cdot z} \langle K( p^{'})|\bar{\psi}_j(-\frac{1}{2}z) \mathcal{W} (-\frac{1}{2}z,\frac{1}{2}z;\bar{n})\psi_i(\frac{1}{2}z)|K( p)\rangle,
\end{align}
where $p$ is the incoming and $p^{'}$ the outgoing kaon momentum, and
\begin{align}\label{2}
P=(p+p^{'})/2, \quad \Delta=p^{'}-p,\quad P\cdot \Delta=0,
\end{align}
and $k$ is the relative quark-antiquark momentum. These conventions are depicted in Fig. \ref{gpcf}.
The Wilson line is denoted as $\mathcal{W} (-\frac{1}{2}z,\frac{1}{2}z;\bar{n})$ in Eq. (\ref{1}). Here, $\bar{n}$ represents a light-like four vector with the property of $\bar{n}^2=0$, being antiparallel to $P$, and satisfying the condition $\bar{n}\cdot P=P^-$. It is necessary to choose the path:
\begin{align}\label{3}
-\frac{z}{2}\rightarrow-\frac{z}{2}+\frac{1}{\epsilon}\bar{n}\rightarrow\frac{z}{2}+\frac{1}{\epsilon}\bar{n}\rightarrow\frac{z}{2},\epsilon\rightarrow0^+.
\end{align}
by rescaling $\bar{n}\rightarrow \lambda \bar{n} $, $\lambda\in \mathbb{R}$, $\lambda>0$, the same path is achieved; with $\hat{P}^2=1$, Eq. (\ref{1}) relies on
\begin{align}\label{4}
\bar{N}=\bar{n}/\bar{n}\cdot \hat{P}.
\end{align}
The quantity $\eta$ in Eq. (\ref{1}) represents the sole remaining degree of freedom, namely $\eta=\text{sign}(\bar{n}_0)$. The values of $\eta=\pm 1$ correspond to the distinct future and past trajectories of the Wilson line.

One can obtain the GTMDs by considering the following partially integrated quantity:
\begin{align}\label{5}
W_{ij}(P,x,\bm{k}_{\perp},\Delta,N;\eta)=\int \frac{d^4z}{(2\pi)^4}e^{i k \cdot z}\delta(n\cdot z) \langle K( p^{'})|\bar{\psi}_j(-\frac{1}{2}z)\mathcal{W} (-\frac{1}{2}z,\frac{1}{2}z;\bar{n})\psi_i(\frac{1}{2}z)|K( p)\rangle,
\end{align}
where $n$ is a light-like four-vector, $n\cdot P=P^+$.

We use $\mathcal{H}$ represent the combination of Dirac matrices, then GTMDs are obtained through
\begin{align}\label{6}
&W^{\mathcal{H}}(P,x,\bm{k}_{\perp},\Delta,N;\eta)=\frac{1}{2}W_{ij}(P,x,\bm{k}_{\perp},\Delta,N;\eta)\mathcal{H}_{ji}\nonumber\\
=&\int \frac{d^4z}{2(2\pi)^4}e^{i k \cdot z}\delta(n\cdot z) \langle K( p^{'})|\bar{\psi}_j(-\frac{1}{2}z)\mathcal{H}_{ji}\mathcal{W} (-\frac{1}{2}z,\frac{1}{2}z;\bar{n})\psi_i(\frac{1}{2}z)|K( p)\rangle,
\end{align}
from Fig. \ref{1}, this operation corresponds to inserting $\mathcal{H}$ as a connection between the open quark and antiquark lines: $\psi(k\mp \Delta/2)$, respectively.

\subsection{Contact interaction}
The quark-quark scattering kernel, introduced in the symmetry-preserving truncation scheme, is at leading-order and incorporates the rainbow-ladder (RL) kernel as referenced in Refs.~\cite{Munczek:1994zz,Bender:1996bb}
\begin{subequations}\label{dqpv1}
\begin{align}\label{7}
\mathfrak{K}_{\alpha_1\alpha_1^{'},\alpha_2\alpha_2^{'}}&=\mathcal{G}_{\mu\nu}(k) [i\gamma_{\mu}]_{\alpha_1\alpha_1^{'}}[i\gamma_{\mu}]_{\alpha_2\alpha_2^{'}}\,,\\
\mathcal{G}_{\mu\nu}(k)&=\tilde{\mathcal{G}}(k^2)T_{\mu\nu}(k),
\end{align}
\end{subequations}
where $k=p_1-p_1^{'}=p_2^{'}-p_2$, $k^2T_{\mu\nu}(k)=k^2\delta_{\mu\nu}-k_{\mu}k_{\nu}$. $\tilde{\mathcal{G}}$ saturates at infrared momenta:
\begin{align}\label{8}
\tilde{\mathcal{G}}(k^2) \overset{k^2\simeq0}{=}\frac{4\pi\alpha_0}{m_G^2}.
\end{align}
In QCD~\cite{Cui:2019dwv}: $\alpha_0\approx 0.93*\pi$ and $m_G\approx 0.8$ GeV  $\approx m_N$, $m_N$ is the nucleon mass.

The CI RL kernel are defined as
\begin{align}\label{ci}
\mathcal{K}_{\alpha_1\alpha_1^{'},\alpha_2\alpha_2^{'}}^{\text{CI}}&= \frac{4\pi \alpha_0}{m_G^2}[i\gamma_{\mu}]_{\alpha_1\alpha_1^{'}}[i\gamma_{\mu}]_{\alpha_2\alpha_2^{'}}.
\end{align}
In the DSE, an ultraviolet regularization scheme should be implemented for the above equation. The associated regularization mass-scale, $\Lambda_{\text{uv}}$, is an additional physical parameter. It is momentum-independent and can be interpreted as an upper bound on the momentum domain. We also include an infrared regularization scale, $\Lambda_{\text{ir}}$, to fully define the CI when calculating the bound-state problems. By introducing the $\Lambda_{\text{ir}}$, one obtains a rudimentary expression of confinement via elimination of quark production thresholds. A natural choice for this scale is $\Lambda_{\text{ir}} \sim \Lambda_{\text{QCD}}$. In this work, we choose $\Lambda_{\text{ir}}=0.24$ GeV.

We will implement the proper time regularization (PTR) scheme
\begin{align}\label{9}
\frac{1}{X^n}&=\frac{1}{(n-1)!}\int_0^{\infty}d\tau \tau^{n-1}e^{-\tau X} \rightarrow \frac{1}{(n-1)!} \int_{1/\Lambda_{\text{UV}}^2}^{1/\Lambda_{\text{IR}}^2}d\tau \tau^{n-1}e^{-\tau X},
\end{align}
where $X$ donates a product of propagators that have been combined using Feynman parametrization.

For the kaon, unlike the pion, the isospin symmetry is broken. The current mass of the $u$ quark is denoted as $m_u$, and the current mass of the $\bar{s}$ quark is denoted as $m_s$. In the following discussion, we will use $s$ to represent the $\bar{s}$ quark. The current quark masses can be determined by solving the kaon bound state problem specified by the kernel in Eq. (\ref{ci}). The gap equation for the dressed $u$ or $s$ quark propagator is as follows:
\begin{align}\label{ps}
S_q^{-1}(p)=i\gamma\cdot p +m_q+\frac{16\pi}{3}\frac{\alpha_0}{m_G^2}\int \frac{d^4l}{(2\pi)^4}\gamma_{\mu}S_q(l)\gamma_{\mu},
\end{align}
where $q=u,s$ represents either the $u$ quark or the $s$ quark. The integral exhibits quadratic divergence, which can be regularised in a Poincar$\acute{e}$ invariant manner. The resulting solution is
\begin{align}\label{pro}
S_q^{-1}(p)=i\gamma\cdot p+M_q,
\end{align}
where $S_q^{-1}(p)$ represents the propagator of the $u$ quark or the $s$ quark, and $M_q$ denotes the dressed mass of the $u$ quark or $s$ quark. In the context of CI, $M_q$ is independent of momentum and is determined by
\begin{align}\label{dqm}
M_q=m_q+M_q\frac{4\alpha_0}{3\pi m_G^2}\left[\int_0^{\infty} ds s \frac{1}{s+M_q^2}\right]_{\text{reg}}.
\end{align}
Applying the PTR to the above equation
\begin{align}\label{dqm1}
M_q=m_q+M_q\frac{4\alpha_0}{3\pi m_G^2}\int_0^{\infty} ds \int_{\tau_{\text{uv}}^2}^{\tau_{\text{ir}}^2} d\tau s e^{-\tau (s+M_q^2)},
\end{align}
where $\tau_{\text{ir},\text{uv}}=1/\Lambda_{\text{ir},\text{uv}}$ represent the infrared and ultraviolet regulators as described previously. As a result, the gap equation is then given by:
\begin{align}\label{dqm2}
M_q=m_q+M_q\frac{4\alpha_0}{3\pi m_G^2}\mathcal{C}_0(M_q^2),
\end{align}
where
\begin{align}\label{dqm3}
\mathcal{C}_0(\sigma)&=\int_0^{\infty}ds s \int_{\tau_{\text{uv}}^2}^{\tau_{\text{ir}}^2} d\tau  e^{-\tau (s+\sigma)}=\sigma[\Gamma(-1,\sigma\tau_{\text{uv}}^2)-\Gamma(-1,\sigma\tau_{\text{ir}}^2)].
\end{align}
In the CI, the $K$-meson Bethe-Salpeter amplitude can be expressed in a consistent manner using vector $\times$ vector treatment, and it takes the following form:
\begin{align}\label{dqm4}
\Gamma_K(Q)&=\gamma_5\left[iE_K(Q)+\frac{1}{2M_{us}}\gamma \cdot Q F_K(Q)\right],
\end{align}
where $M_{us}=M_uM_s/(M_u+M_s)$. $Q$ represents the total momentum of the kaon with $Q^2=-m_K^2$, and $m_K$ denotes the kaon mass. Additionally, $M_q$ refers to the dressed quark mass of either the $u$ or $s$ quark as defined in Eq. (\ref{dqm2}). It is important to note that both $E_K$ and $F_K$ are independent of the relative quark-antiquark momentum.

$\Gamma_K$ is derived from the homogeneous Bethe-Salpeter equation:
\begin{align}\label{dqm5}
\Gamma_K(Q)&=-\frac{16\pi}{3} \frac{\alpha_0}{m_G^2}\int  \frac{d^4l}{(2\pi)^4}\gamma_{\mu}S_u(l+Q)\Gamma_K(Q)S_s(l)\gamma_{\mu}.
\end{align}
using the symmetry-preserving regularization scheme as described in Refs.~\cite{Gutierrez-Guerrero:2010waf,Chen:2012txa},
\begin{align}\label{dqm6}
0=\int_0^1d\alpha [\mathcal{C}_0(\omega(\alpha,Q^2))+\mathcal{C}_1(\omega(\alpha,Q^2))],
\end{align}
where $\mathcal{C}_1$ is defined in Eqs. (\ref{app1}), (\ref{app2}) and
\begin{align}\label{dqm7}
\omega(\alpha,Q^2)=\bar{\alpha}M_u^2+\alpha M_s^2+\alpha\bar{\alpha}Q^2,
\end{align}
of which $\bar{\alpha}=1-\alpha$, one obtain the pair of coupled equations:
\begin{align}\label{dqm8}
\left[
\begin{array}{cc}
E_K (Q) \\
E_K (Q) \\
\end{array}
\right]=\frac{4\alpha_0}{3\pi m_G^2}\left[
\begin{array}{cc}
\mathcal{K}_{EE}^K &  \mathcal{K}_{EF}^K  \\
\mathcal{K}_{FE}^K  &\mathcal{K}_{FF}^K \\
\end{array}
\right]\left[
\begin{array}{cc}
E_K (Q) \\
E_K (Q) \\
\end{array}
\right],
\end{align}
the matrix elements ${\mathcal{K}_{EE}^K, \mathcal{K}_{EF}^K, \mathcal{K}_{FE}^K, \mathcal{K}_{FF}^K}$ are defined in Eq. (\ref{app4}).

Inserting of Eq. (\ref{dqm8}) and Eq. (\ref{app4}) shows that a nonzero value for $E_K$ results in $F_K\neq 0$. Eq. (\ref{dqm8}) represents an eigenvalue problem, with a solution existing when $Q^2=-m_K^2$, at which point the eigenvector corresponds to the meson's Bethe-Salpeter amplitude. This leads us to obtain the kaon's decay constant
\begin{align}\label{dqm9}
f_K=\frac{N_c}{4\pi^2}\frac{1}{M_{us}}[E_K\mathcal{K}_{FE}^K+F_K\mathcal{K}_{FF}^K]_{Q^2=-m_K^2},
\end{align}
our results yield $f_K=0.111$ GeV.

The parameters utilized in this study are outlined in Table \ref{tb1}.
\begin{center}
\begin{table}
\centering
\caption{Parameter set used in our work. The dressed quark mass and regularization parameters are in units of GeV.}\label{tb1}
\begin{tabular}{p{1.0cm} p{1.0cm} p{1.0cm} p{1.0cm}p{1.0cm}p{1.0cm}p{1.0cm}}
\hline\hline
$\Lambda_{\text{IR}}$&$\Lambda _{\text{UV}}$&$M_u$&$M_s$&$E_{K}$&$F_{K}$&$m_{K}$\\
\hline
0.240&0.905&0.368&0.533&3.83&0.6&0.5\\
\hline\hline
\end{tabular}
\end{table}
\end{center}

\subsection{Dressed quark photon vertex}
Using Eq. (\ref{ci}), one can obtain the dressed $q$ quark-photon vertex, $\Gamma_{\mu}^{q,\gamma}$,
\begin{align}\label{qfv}
\Gamma_{\mu}^{q,\gamma}(\Delta)=\gamma_{\mu}-\frac{16\pi \alpha_0}{3m_G^2}\int \frac{d^4l}{(2\pi)^4}\gamma_{\alpha}S_q(l_{+\Delta})\Gamma_{\mu}^{\gamma}(\Delta)S_q(l_{-\Delta})\gamma_{\alpha},
\end{align}
where $l_{\pm\Delta}=l\pm\frac{\Delta}{2}$. According to the Ward-Green-Takahashi identity (WGTI) preserved in our regularization of the CI, we obtain
\begin{align}\label{qfvs}
\Gamma_{\mu}^{\gamma}(\Delta)=\Gamma_{\mu}^TP_T(\Delta^2)+\gamma_{\mu}^L,
\end{align}
where $\Delta\cdot \Gamma_{\mu}^T=0 $, $\Gamma_{\mu}^T+\gamma_{\mu}^L=\gamma_{\mu} $
\begin{subequations}\label{dff}
\begin{align}
P_T^q(Q^2)&=\frac{1}{1+K_V^q(Q^2)}\,, \\
K_V^{q}(Q^2)&=\frac{2\alpha_0}{3\pi m_G^2}\int_0^1 d\alpha\alpha\bar{\alpha}Q^2  \mathcal{C}_1(M_q^2+x(1-x)Q^2).
\end{align}
\end{subequations}
Consistent with RL truncation studies of the photon-$u$-quark vertex, the dressing function $P_T^{u}(Q^2)$ displays a simple pole at $Q^2=-m_{\rho}^2$, where $m_{\rho}=0.928$ GeV represents the mass of the $\rho$ meson generated by the interaction. Additionally, $P_T^{s}(Q^2)$ also exhibits a pole at 
\begin{align}\label{pole}
1+K_V^{us}(-m_{ss}^2)=0.
\end{align}

\subsection{Twist-two kaon GTMDs}\label{qq}
In this paper, we will employ the symmetry notation as presented in the papers~\cite{Ji:1998pc,Diehl:2003ny}. Here, $n$ denotes a light-like four-vector with $n^2=0$. The kinematics of this process and its associated quantities are defined as follows:
\begin{subequations}\label{notations}
\begin{align}\label{4}
p^{'2}&=p^2=-m_K^2, \quad t=(p^{'}-p)^2=-\Delta^2,\quad P^2=t/4-m_K^2\,, \\
\xi&=-\Delta \cdot n/2P\cdot n,\quad p\cdot \Delta =-\Delta^2/2=p^{'}\cdot \Delta, \quad \Delta^2=Q^2\,,
\end{align}
\end{subequations}
$\xi$ represents the parameter of skewness.
%
%

The twist-two kaon GTMDs are obtained with the following choices:
\begin{align}\label{gcpf1}
\mathcal{H}\rightarrow\{\mathcal{H}_1= i\gamma\cdot n,\mathcal{H}_2= i\gamma\cdot n\gamma_5,\mathcal{H}_3=i\sigma_{j \mu}\cdot n_{\mu} \},
\end{align}
$\mathcal{H}_1$ is the most basic, it involved with the kaon electromagnetic form factor and valence-quark distribution function. We will use $\mathcal{H}_1$ to illustrate the computational process.

In Euclidean space
\begin{align}\label{gcpf1}
W^{[\mathcal{H}_1]}(P,x,\Delta,N,\eta)\rightarrow F_1(x,\bm{k}_{\perp}^2,\xi,t),
\end{align}
the vector GTMD of the $u$ quark in kaons are defined as follows:
\begin{align}
F_1^u(x,\bm{k}_{\perp}^2,\xi,t)=2 N_c\text{tr}_D \int \frac{dk_3dk_4}{(2 \pi )^4}\delta_n^x (k) \Gamma_{K}(-p^{'}) S_u \left(k_{+\Delta}\right) n\cdot \Gamma_{\gamma} (\Delta)S_u\left(k_{-\Delta}\right)\Gamma_{K}(p) S_s\left(k_P\right),
\end{align}
where $\text{tr}_D$ indicates a trace over spinor indices, $\delta_n^x (k)=\delta(n\cdot k-xn\cdot P)$, $k_{\pm \Delta}=k \pm \frac{\Delta }{2}$, $k_P=k-P$ and $S_q$ are defined in Eq. (\ref{pro}).

By utilizing a WGTI-preserving evaluation of the above equation, we derived the expression ($k^2=D_k^q-M_q^2$)
\begin{subequations}\label{dqpv1}
\begin{align}
2k\cdot \Delta&= D_{k_{+\Delta}}^u-D_{k_{-\Delta}}^u\,, \\
2k\cdot p&=D_{k_{-\Delta}}^u-D_{k_P}^s-M_u^2+M_s^2+p^2-\Delta^2/2\,, \\
2k\cdot p^{'}&=D_{k_{+\Delta}}^u- D_{k_P}^s-M_u^2+M_s^2+p^{'2}-\Delta^2/2\,, \\
2k^2&=D_{k_{+\Delta}}^u+D_{k_{-\Delta}}^u-2M_u^2-\Delta^2/2,
\end{align}
\end{subequations}
using these equations we arrived at ($r=\bm{k}_{\perp}^2$)
\begin{align}\label{a9f}
&\quad  F_1^u\left(x,r,\xi,t\right)\nonumber\\
&=\frac{\bar{N}^{FF}N_c}{4\pi^3} \frac{(\xi+x)(1-x)}{(1+\xi)^2 }\frac{\theta_{\bar{\xi} 1}}{\sigma_1^{r,1}}\bar{\mathcal{C}}_2(\sigma_1^{r,1})+ \frac{\bar{N}^{FF}N_c }{4 \pi ^3} \frac{(1-x)(x-\xi)}{(1-\xi)^2} \frac{\theta_{\xi 1}}{\sigma_1^{r,-1}}\bar{\mathcal{C}}_2(\sigma_1^{r,-1})\nonumber\\
&+\frac{N^{EE}N_c}{4\pi^3} \left(\frac{\theta_{\bar{\xi} 1}}{\sigma_1^{r,1}}\bar{\mathcal{C}}_2(\sigma_1^{r,1})+\frac{\theta_{\xi 1}}{\sigma_1^{r,-1}}\bar{\mathcal{C}}_2(\sigma_1^{r,-1})\right)\nonumber\\
&+\frac{\bar{N}^{EF}N_c}{4\pi^3} \frac{2x+\xi-1}{1+\xi} \frac{\theta_{\bar{\xi} 1}}{\sigma_1^{r,1}}\bar{\mathcal{C}}_2(\sigma_1^{r,1})+ \frac{\bar{N}^{EF}N_c }{4 \pi^3} \frac{2x-\xi-1}{1-\xi} \frac{\theta_{\xi 1}}{\sigma_1^{r,-1}}\bar{\mathcal{C}}_2(\sigma_1^{r,-1})\nonumber\\
&+\frac{\bar{N}^{EE}N_c}{4\pi^3} \theta_{\bar{\xi} \xi} \frac{x}{\xi }\frac{\bar{\mathcal{C}}_2(\sigma_2^r)}{\sigma_2^r}-  \frac{N^{FF}N_c}{8\pi^3}\frac{\theta_{\bar{\xi} \xi}}{\xi} (1-\frac{x^2}{\xi^2})t \frac{\bar{\mathcal{C}}_2(\sigma_2^r)}{\sigma_2^r}\nonumber\\
&+ \frac{N^{EF}N_c}{16\pi^3}\int_0^1  d\alpha \frac{\theta_{\alpha \xi}}{\xi} \left(2x (m_K^2-\left(M_u-M_s\right)^2)+(1-x)t\right)\frac{6\bar{\mathcal{C}}_3(\sigma_3^r)}{[\sigma_3^r]^2},
\end{align}
where $\theta$ represents the step function, and $x$ is only valid within these specified regions.
\begin{subequations}\label{region1}
\begin{align}
\theta_{\bar{\xi} 1}&=x\in[-\xi, 1]\,, \\
\theta_{\xi 1}&=x\in[\xi, 1]\,, \\
\theta_{\bar{\xi} \xi}&=x\in[-\xi, \xi]\,, \\
\theta_{\alpha \xi}&=x\in[\alpha (\xi +1)-\xi , \alpha  (1-\xi)+\xi ]\cap x\in[-1,1],
\end{align}
\end{subequations}
and
\begin{subequations}\label{region1}
\begin{align}
N^{FF} &=F_K^2 \frac{(M_s+M_u)^2}{4M_u^2M_s^2}\,, \\
\bar{N}^{FF}&=F_K^2\frac{(M_s+M_u)^2\left(M_u^2-M_s^2\right)}{4M_u^2M_s^2}\,, \\
\bar{N}^{EF}&=( \frac{F_KE_K}{2M_uM_s} -\frac{F_K^2(M_s+M_u)^2}{4M_u^2M_s^2})(M_s^2-M_u^2)\,, \\
N^{EF}&=E_K^2-\frac{F_KE_K(M_s+M_u)^2}{M_uM_s} +\frac{F_K^2(M_s+M_u)^4}{4M_u^2M_s^2}\,, \\
N^{EE}&=E_K^2-E_KF_K\frac{(M_s+M_u)^2}{2M_uM_s}\,, \\
\bar{N}^{EE}&=E_K^2-E_KF_K\frac{\left( M_s^2-M_u^2\right)}{M_uM_s}\,,
\end{align}
\end{subequations}
here we express $\theta_{\bar{\xi} \xi}/\xi=\Theta(1-x^2/\xi^2) $, where $\Theta(x)$ represents the Heaviside function. Additionally, we have $\theta_{\alpha\xi}/\xi=\Theta((1-\alpha)^2-(x-\alpha)^2/\xi^2)\Theta(1-x^2)$. These findings are valid in the region where $\xi > 0$, and under the transformation $\xi \rightarrow -\xi $: $\theta_{\bar{\xi} 1} \leftrightarrow \theta_{\xi 1}$; and both $\theta_{\bar{\xi} \xi}/\xi$ and $\theta_{\alpha \xi}/\xi$ remain invariant. GTMD is expected to satisfy certain fundamental properties, such as yielding the Wigner distribution when Fourier transformed, and reducing to GPD and TMD in specific limits. We will verify these properties individually.

The dressed $u$ quark vector GTMD of kaon is $\bar{P}_T^u F_1^u\left(x,r,\xi,t\right)$, $\bar{P}_T^u$ is defined as
\begin{align}\label{gcpf1}
\bar{P}_T^u=\left[\theta_{\bar{\xi}{\xi}}+ P_T^u(-t)(1-\theta_{\bar{\xi}{\xi}})\right],
\end{align}
where $P_T^u(-t)$ is defined in Eq. (\ref{dff}) for $q=u$, for the $s$ quark, the dressed GTMD should multiply $\bar{P}_T^s(t)$.

For $\mathcal{H}_2= i n\cdot\gamma \gamma_5$, where the definitions $\varepsilon_{ij}^{\perp} =\varepsilon_{\alpha\beta ij}\bar{n}_{\alpha}n_{\beta}$,
\begin{align}\label{gcpf1}
W^{[\mathcal{H}_2]}(P,x,\bm{k}_{\perp},\Delta,N;\eta)\rightarrow \frac{i\varepsilon_{ij}^{\perp}k_i\Delta_j}{m_K^2}\tilde{G_1},
\end{align}
then we obtain ($r=\bm{k}_{\perp}^2$)
\begin{align}\label{gcpf1}
\tilde{G }_1^u\left(x,r,\xi,t\right)=&m_K^2(M_s+M_u)^2\frac{N^{FF}N_c }{4\pi^3}\frac{\theta_{\bar{\xi} \xi}}{\xi} \frac{\bar{\mathcal{C}}_2(\sigma_2^r)}{\sigma_2^r}-\frac{N^{EF}N_c}{8\pi ^3}\int_0^1 d\alpha m_K^2 \frac{\theta_{\alpha \xi}  }{\xi}\frac{6\bar{\mathcal{C}}_3(\sigma_3^r)}{[\sigma_3^r]^2}.
\end{align}
For $\mathcal{H}_3= i\sigma_{j \mu}n_{\mu}$ correspond to two terms
\begin{align}\label{gcpf1}
W^{[\mathcal{H}_3]}(P,x,\bm{k}_{\perp},\Delta,N;\eta)\rightarrow  \frac{k_j}{m_K}H_1^k+\frac{ \left(P\cdot n \Delta_j- P_j \Delta \cdot n \right)}{P\cdot n m_K} H_1^{\Delta},
\end{align}
we obtain the two terms ($r=\bm{k}_{\perp}^2$)
\begin{align}\label{tGPD1}
H_1^{u,\Delta}(x,r,\xi,t)=&-  \frac{N^{FF}N_c}{4\pi^3} m_K\frac{\theta_{\bar{\xi} \xi}}{ \xi}  \frac{\bar{\mathcal{C}}_2(\sigma_2^r)}{\sigma_2^r}\nonumber\\
+&\frac{N^{EF}N_c}{8\pi^3} \int_0^1  d\alpha m_K \frac{\theta_{\alpha \xi}}{\xi} (M_u+(M_s-M_u)\alpha ) \frac{6\bar{\mathcal{C}}_3(\sigma_3^r)}{[\sigma_3^r]^2},
\end{align}
\begin{align}\label{gcpf1}
H_1^{u,k}(x,r,\xi,t)=& - \frac{\tilde{N}^{EF}N_c}{2\pi^3}m_K\left( \frac{\theta_{\bar{\xi} 1}}{\sigma_1^{r,1}} \bar{\mathcal{C}}_2(\sigma_1^{r,1})-\frac{\theta_{\xi 1} }{\sigma_1^{r,-1}}\bar{\mathcal{C}}_2(\sigma_1^{r,-1})\right)\nonumber\\
+&  \frac{\hat{N}^{EF}N_c }{\pi ^3}m_K\frac{\theta_{\bar{\xi} \xi} }{\sigma_2^r}\bar{\mathcal{C}}_2(\sigma_2^r)+ \frac{N^{EF}N_c}{8\pi^3} \int_0^1  d\alpha \theta_{\alpha \xi}(M_s-M_u)m_K \frac{6\bar{\mathcal{C}}_3(\sigma_3^r)}{[\sigma_3^r]^2},
\end{align}
where
\begin{subequations}\label{region2}
\begin{align}
\tilde{N}^{EF}&=F_KE_K \frac{(M_s+M_u)}{2M_uM_s}-F_K^2\frac{(M_s+M_u)^3}{4M_u^2M_s^2}\,, \\
\hat{N}^{EF}&=F_KE_K \frac{(M_s+M_u)}{2M_uM_s}-F_K^2\frac{(M_s+M_u)^3}{8M_u^2M_s^2}\,,
\end{align}
\end{subequations}
from the above equations, it is evident that $H_1^k$ will be zero when $F_K=0$. This distinguishes it from the NJL model, where this term is zero. Among the twist-two GTMDs, $F_1$, $H_1^{\Delta}$, and $\tilde{G}_1$ are T-even, while $H_1^k$ is T-odd.

The twist-three, four GTMDs can be found in the appendices \ref{AppendixT2} and \ref{AppendixT3}.

\section{Kaon twist-two GPDs}\label{good1}
\subsection{Algebraic results}
The relationship between the twist-two kaon $u$ quark GTMDs and the twist-two GPDs are
\begin{subequations}\label{gpd2}
\begin{align}
H_K^u(x,\xi,t)&=\int d^2\bm{k}_{\perp}F_1^u(x,\bm{k}_{\perp}^2,\xi,t)\,, \\
E_K^{u,T}(x,\xi,t)&=\int d^2\bm{k}_{\perp}H_1^{u,\Delta}(x,\bm{k}_{\perp}^2,\xi,t),
\end{align}
\end{subequations}
where $H_K^u$ and $E_K^{u,T}$ represent the $u $ quark vector (no spin-flip) and tensor (spin-flip) GPDs of kaon. The Mellin moments of the vector GPD are linked to the kaon's elastic electromagnetic form factor and gravitational form factors, including mass and pressure/stress. The tensor GPD contains information about the distribution of kaon quarks with polarization perpendicular to the direction of motion of the pion (transversity).

The vector GPD
\begin{align}\label{t2gpd}
H_{K}^u\left(x,\xi,t\right)=&\frac{\bar{N}^{FF}N_c}{8\pi^2} \theta_{\bar{\xi} 1}\frac{(\xi+x)(1-x)}{(1+\xi)^2 }\bar{\mathcal{C}}_1(\sigma_1^{0,1})+ \frac{\bar{N}^{FF}N_c }{8 \pi ^2} \theta_{\xi 1}\frac{(1-x)(x-\xi)}{(1-\xi)^2}\bar{\mathcal{C}}_1(\sigma_1^{0,-1})\nonumber\\
+&\frac{N^{EE}N_c}{8\pi^2} \left(\theta_{\bar{\xi} 1}\bar{\mathcal{C}}_1(\sigma_1^{0,1})+ \theta_{\xi 1}\bar{\mathcal{C}}_1(\sigma_1^{0,-1})\right)\nonumber\\
+&\frac{\bar{N}^{EF}N_c}{8\pi^2} \theta_{\bar{\xi} 1}\frac{2x+\xi-1}{1+\xi} \bar{\mathcal{C}}_1(\sigma_1^{0,1})+ \frac{\bar{N}^{EF}N_c }{8 \pi ^2} \theta_{\xi 1}\frac{2x-\xi-1}{1-\xi} \bar{\mathcal{C}}_1(\sigma_1^{0,-1})\nonumber\\
+&\frac{\bar{N}^{EE}N_c}{8\pi^2}   \frac{\theta_{\bar{\xi} \xi}}{\xi }x\bar{\mathcal{C}}_1(\sigma_2^0)-  \frac{N^{FF}N_c}{16\pi^2}\frac{\theta_{\bar{\xi} \xi}}{\xi} (1-\frac{x^2}{\xi^2})t \bar{\mathcal{C}}_1(\sigma_2^0)\nonumber\\
+ &\frac{N^{EF}N_c}{8 \pi^2}\int_0^1  d\alpha\frac{\theta_{\alpha \xi}}{\xi}\left(2x (m_K^2-\left(M_u-M_s\right)^2)+(1-x)t\right)\frac{\bar{\mathcal{C}}_2(\sigma_3^0)}{\sigma_3^0},
\end{align}
from the above equation, we can find that vector GPD satisfy the time-reversal-invariance,
\begin{align}\label{gcpf1}
H_K^u(x,\xi,t)=H_K^u(x,-\xi,t),
\end{align}
which means $H_K^u$ is even function of $\xi$. In general, GPDs are neither even nor odd of $x$, the combinations,
\begin{subequations}\label{conjugation}
\begin{align}\label{8}
H_K^{I=0}(x,\xi,t)&=H_K^u(x,\xi,t)-H_K^u(-x,\xi,t)\,, \\
H_K^{I=1}(x,\xi,t)&=H_K^u(x,\xi,t)+H_K^u(-x,\xi,t),
\end{align}
\end{subequations}
are useful, the first equation corresponds to charge conjugation $C = +1$ in the T-channel. The second equation corresponds to $C =-1$ in the T-channel, tensor GPD $E_K^u(x,\xi,t)$ is analogy. For the kaon $s$ quark vector GPD, we can obtained through the symmetry property of the GPD,
\begin{align}\label{t2sgpd}
H_K^{u}(x,\xi,t)&=-H_K^s(-x,\xi,t;M_u\rightleftharpoons M_s)\,,
\end{align}
where $M_u\rightleftharpoons M_s$ denotes the exchange of $M_u$ with $M_s$, utilizing the aforementioned equation allows us to derive the kaon's $s$ quark vector GPD, denoted as $H_K^s(x,\xi,t)$.

The twist-two tensor GPD is
\begin{align}\label{t2tgpd}
E_K^{u,T}(x,\xi,t)=&-  N^{FF} \frac{N_c}{8\pi^2}  m_K\frac{\theta_{\bar{\xi} \xi}}{ \xi} \bar{\mathcal{C}}_1(\sigma_2^0)\nonumber\\
+& \frac{N^{EF}N_c}{4\pi^2} \int_0^1  d\alpha\frac{\theta_{\alpha \xi}}{\xi} m_K(M_u+(M_s-M_u)\alpha) \frac{\bar{\mathcal{C}}_2(\sigma_3^0)}{\sigma_3^0},
\end{align}
the tensor GPDs also satisfy the time-reversal-invariance,
\begin{align}\label{gcpf2}
E_K^{u,T}(x,\xi,t)=E_K^{u,T}(x,-\xi,t),
\end{align}
it indicates that $E_K^{T}$ is T-even. The kaon's $s$ quark tensor GPD can be obtained through
\begin{align}\label{t2stgpd}
E_K^{u,T}(x,\xi,t)&=-E_K^{s,T}(-x,\xi,t;M_u\rightleftharpoons M_s).
\end{align}

\subsection{Vector GPDs: images}
\begin{figure}
\centering
\includegraphics[width=0.47\textwidth]{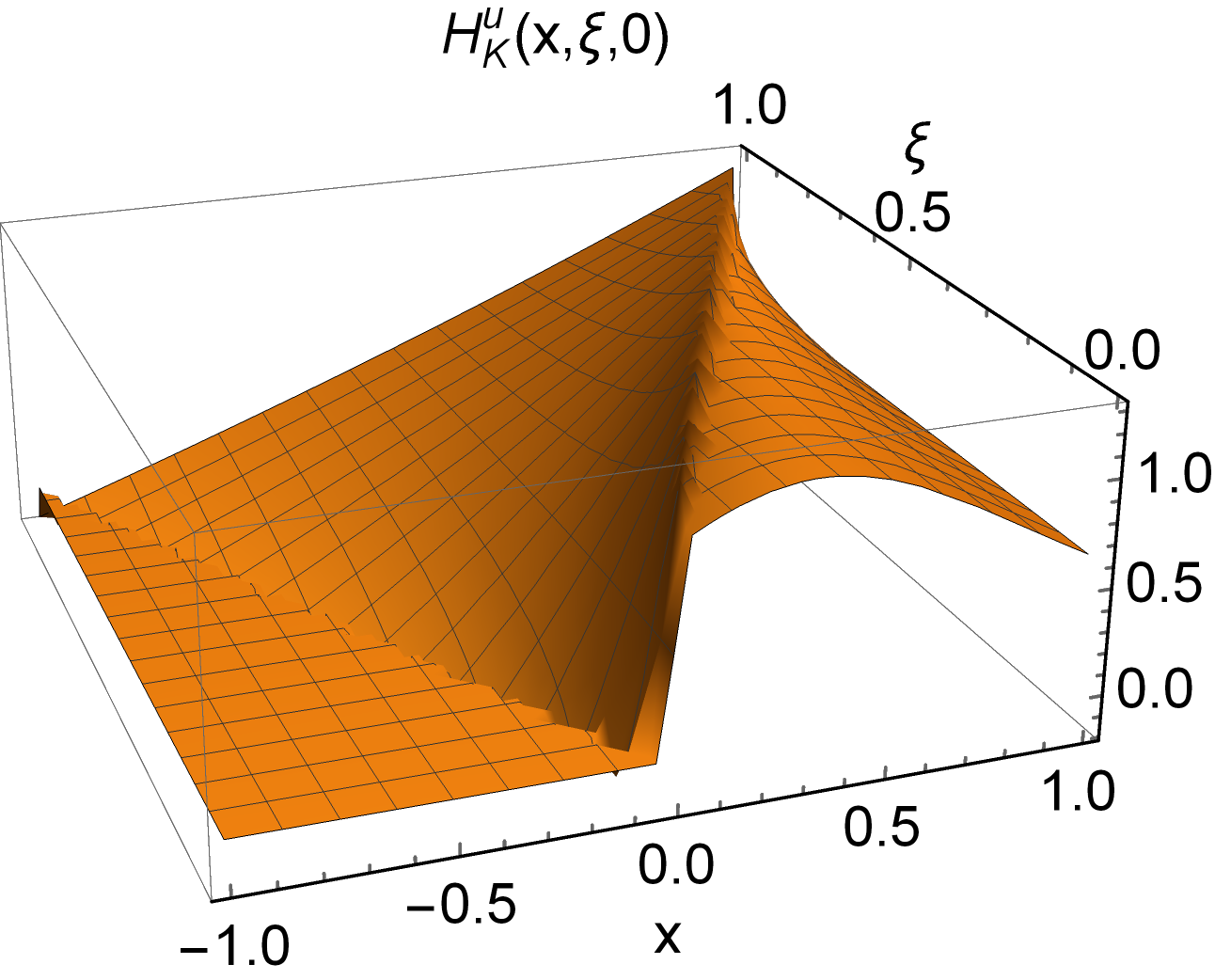}
\qquad
\includegraphics[width=0.47\textwidth]{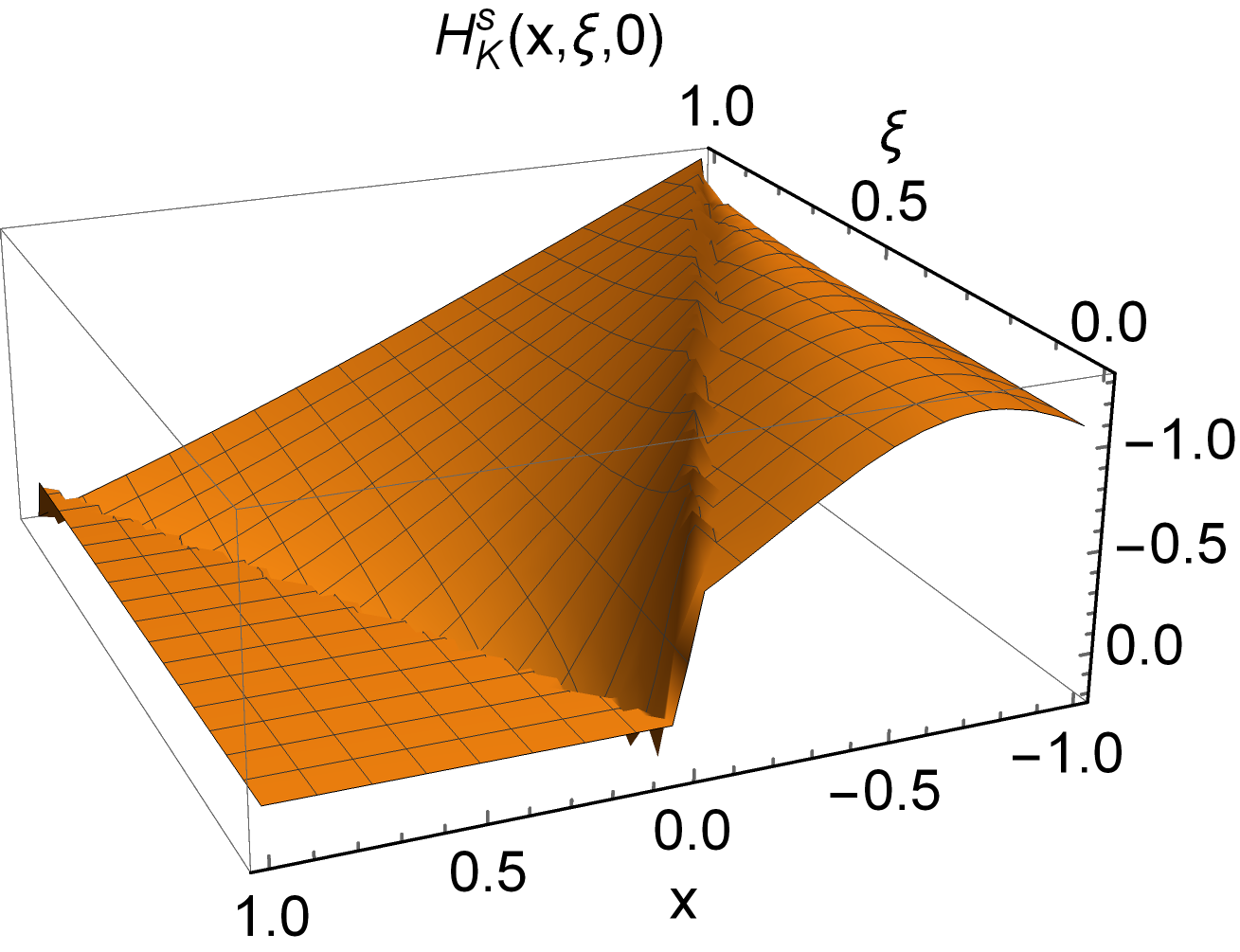}
\caption{\emph{Left panel} -- Kaon $u$ quark vector GPD $H_K^u(x,\xi,0)$ in Eq. (\ref{t2gpd}). \emph{Right panel} -- Kaon $s$ quark vector GPD $H_K^s(x,\xi,0)$ in Eq. (\ref{t2sgpd}). Owing to Eq. (\ref{gcpf1}), only $\xi>0$ is plotted. }\label{t2gpdp}
\end{figure}
The vector GPDs of the kaon, involving the $u$ quark and $s$ quark, as represented in Eq. (\ref{t2gpd}) and Eq. (\ref{t2sgpd}), are depicted in Fig. \ref{t2gpdp}. For the $u$ quark vector GPD: (a) $H_K^u(x,0,0)=u_K(x)$ represents the CI valence $u$-quark PDF. (b) $H_K^u(x,\xi,0)=0$ on $x<\xi$, similar to the pion vector GPDs. (c) Using a CI, the GPD is continuous but not differentiable at $x=\pm\xi$. For the $s$ quark vector GPD: (a) $H_K^s(x,0,0)=s_K(x)$ represents the CI valence $s$-quark PDF. (b) $H_K^s(x,\xi,0)=0$ on $x>\xi$. (c) Using a CI, the GPD is continuous but not differentiable at $x=\pm\xi$.

The electromagnetic FF is defined as
\begin{align}\label{2ff}
F_K^{u,\text{em}}(\Delta^2)=\int_{-1}^1 dx H_K^u(x,\xi,t),
\end{align}
the integral is independent of $\xi$. The expression of $F_K^{u,\text{em}}$ is
\begin{align}\label{2ffj}
F_K^{u,\text{em}}(Q^2)=&\frac{N^{EE}N_c}{4\pi^2}\int_0^1 dx\bar{\mathcal{C}}_1(\sigma_1^{0,0})+\frac{N^{FF}N_c }{2\pi^2}\int_0^1dx  x(1-x)Q^2\bar{\mathcal{C}}_1(\sigma_5)\nonumber\\
-&\frac{\bar{N}^{EF}N_c}{4\pi^2}\int_0^1dx \left(1-2x\right)\bar{\mathcal{C}}_1(\sigma_1^{0,0})+\frac{N^{FF}N_c}{2\pi^2}\int_0^1 dx x(1-x)(1-2x)m_K^2 \bar{\mathcal{C}}_1(\sigma_1^{0,0})\nonumber\\
-&\frac{N^{FF}N_c}{4\pi^2}\int_0^1 dx\bar{\mathcal{C}}_1(\sigma_1^{0,0}) ((2-3x)xM_u^2+(1-3x)(1-x)M_s^2)\nonumber\\
+&\frac{N^{EF}N_c}{4\pi^2}\int_0^1 dx \int_0^{1-x}dy \frac{\bar{\mathcal{C}}_2(\sigma_6)}{\sigma_6}  (2(1-x-y)(m_K^2-(M_u-M_s)^2)-Q^2(x+y)),
\end{align}
from which one obtains the associated radius:
\begin{align}\label{ab17}
\langle r_K^2\rangle &=-6\frac{\partial F_K^{u,\text{em}}\left(Q^2\right)}{\partial Q^2}|_{Q^2=0},
\end{align}
the electromagnetic radii for the dressed $u$ and $s$ quarks of the kaon are found to be $r_K^{u,\text{em}}=0.454$ fm and $r_K^{s,\text{em}}=0.362$ fm, respectively, while the bare values are $r_K^{u,\text{em}}=0.322$ fm and $r_K^{s,\text{em}}=0.266$ fm. These results can be compared with the electromagnetic radius of the pion, which is found to be $r_{\pi}^{u,\text{em}}=0.459$ fm for dressed quarks and  $r_{\pi}^{u,\text{em}}=0.327$ fm for bare quarks.

It is observed that the electromagnetic radius of the kaon's $u$-quark is similar to that of the pion's $u$-quark. Furthermore, it is noted that in this study, a hard CI form factor was used which approximates a non-zero constant value as $Q^2$ approaches infinity.

In comparison with NJL model results, it is found that both $u$-quark and $s$-quark electromagnetic radii are smaller due to non-zero values of $F_k$ making the electromagnetic form factors harder than those predicted by the NJL model.

The kaon's gravitational form factors correspond to the $n = 1$ Mellin moment of the twist-two vector GPD,
\begin{align}\label{2gff}
\int_{-1}^1 dx x H_K^u(x,\xi,t)&=\theta_2^{u,K}(t)-\xi^2\theta_1^{u,K}(t)=A_{20}^u(t)+\xi^2A_{22}^u(t),
\end{align}
where $\theta_2$ relevant to the $u$ quark mass distribution within the kaon and $\theta_1$ is linked to the $u$ quark pressure distribution.

\begin{align}\label{2gff1}
A_{20}^u(Q^2)=&\frac{\bar{N}^{FF}N_c}{4\pi^2}\int_0^1 dx x^2 (1 - x)\bar{\mathcal{C}}_1(\sigma_1^{0,0})+\frac{N^{EE}N_c}{4\pi^2}\int_0^1 dx   x \bar{\mathcal{C}}_1(\sigma_1^{0,0}) - \frac{\bar{N}^{EF}N_c }{4\pi^2} \frac{1}{m_K^2}\mathcal{C}_0(M_s^2)\nonumber\\
-& \frac{\bar{N}^{EF}N_c}{4\pi^2} \int_0^1dx \frac{ x\left(M_s^2-M_u^2\right)}{ m_K^2}\bar{\mathcal{C}}_1(\sigma_1^{0,0})+ \frac{\bar{N}^{EF} N_c}{4\pi^2}\int_0^1dx  \frac{1}{m_K^2} \mathcal{C}_0(\sigma_1^{0,0})\nonumber\\
+&\frac{N^{EF}N_c }{4\pi ^2}  \int_0^1 dx \int_0^{1-x} dy(1-x-y) \frac{\bar{\mathcal{C}}_2(\sigma_6)}{\sigma_6}\nonumber\\
\times & (2 (m_K^2-(M_s-M_u)^2) (1-x-y)-Q^2 (x+y)),
\end{align}
\begin{align}\label{2gff1}
A_{22}^u(Q^2)=&-\frac{\bar{N}^{FF}N_c}{4\pi^2}\int_0^1 dx x (1 - x)^2\bar{\mathcal{C}}_1(\sigma_1^{0,0})- \frac{N^{EE}N_c}{4\pi^2}\int_0^1dx (1-x)\bar{\mathcal{C}}_1(\sigma_1^{0,0})\nonumber\\
-& \frac{\bar{N}^{EF}N_c}{4\pi^2}  \frac{1}{m_K^2}\mathcal{C}_0(M_u^2)+\frac{ \bar{N}^{EF}N_c}{4\pi^2} \int_0^1 dx \frac{ \left(M_s^2-M_u^2\right)(1-x)}{ m_K^2}\bar{\mathcal{C}}_1(\sigma_1^{0,0})\nonumber\\
+& \frac{\bar{N}^{EF}N_c}{4\pi^2} \int_0^1 dx \frac{1}{m_K^2}\mathcal{C}_0(\sigma_1^{0,0})-\frac{\bar{N}^{EE}N_c}{2\pi ^2}  \int_0^1 dx x (1-2x) \bar{\mathcal{C}}_1(\sigma_5)\nonumber\\
+&\frac{N^{EF}N_c}{4\pi^2} \int_0^1 dx \frac{(1-x )}{Q^2}\left(Q^2+2 (m_K^2-\left(M_s-M_u\right)^2)\right)\bar{\mathcal{C}}_1(\sigma_1^{0,0})\nonumber\\
-& \frac{N^{EF}N_c }{4\pi^2}\int_0^1 dx \int_0^{1-x} dy \frac{1}{Q^2}\left(Q^2+2(m_K^2-\left(M_s-M_u\right)^2)\right)\bar{\mathcal{C}}_1(\sigma_6).
\end{align}
The FFs have been previously studied in our earlier papers~\cite{Zhang:2020ecj,Zhang:2021shm,Zhang:2021tnr}. The results obtained from the DSE are consistent with the findings reported in our previous research.

A light-cone energy radius can be defined in relation to the generalized FF $A_{20}(Q^2)$ defined as~\cite{Freese:2019bhb}
\begin{subequations}
\begin{align}\label{a93}
\langle r_{E,LC}^2\rangle&=-4\frac{\partial A_{20}(Q^2)}{\partial Q^2}|_{Q^2=0}\,,\\
\langle r_{c,LC}^2\rangle&=-4\frac{\partial F_K(Q^2)}{\partial Q^2}|_{Q^2=0}.
\end{align}
\end{subequations}
The second equation is analogous to the light-cone charge radius, which can be compared with the light-cone energy radius. Our results for the bare and dressed radii of $u$ and $s$ quarks are presented in Table \ref{ttb2}. The first two rows show the results for the bare radius, while the following two lines display the dressed radius. The ratios of $r_{E,LC}^q/r_{c,LC}^q$ for both bare and dressed quarks are listed in Table \ref{tb22}.

From Table \ref{ttb2}, it is evident that the light-cone radius in the NJL model is larger than in the DSE. The light-cone radius of the $u$ quark in a kaon is also larger than that of the $s$ quark in a kaon. Furthermore, the light-cone radius of the $u$ quark in a kaon is similar to that of the $u$ quark in a pion.

Additionally, as shown in Table \ref{tb22}, it is evident that the dressed ratios of the light-cone radius exceed those of the bare ratios. Specifically, for dressed light-cone radii, the ratios in the NJL model are smaller than those in DSE; however, this trend does not hold true for bare light-cone radii.

The Breit-frame pressure distributions of the pion and kaon, as well as the shear pressure distribution defined in Ref.~\cite{Polyakov:2018zvc}, have been thoroughly examined in our previous papers~\cite{Zhang:2020ecj,Zhang:2021mtn}. Therefore, we will not analyze them here.
\begin{center}
\begin{table}
\centering
\caption{The bare and dressed light-cone energy radius and light-cone charge radius of $u$ quark and $s$ quark in the kaon, as well as the results for the $u$ quark in the pion, are provided.}\label{ttb2}
\begin{tabular}{p{1.4cm}p{1.3cm} p{1.3cm} p{1.3cm} p{1.3cm}p{1.3cm}p{1.3cm}}
\hline
Bare &$r_{E,LC}^{u,\pi}$&$r_{c,LC}^{u,\pi}$&$r_{E,LC}^{u,K}$&$r_{c,LC}^{u,K}$&$r_{E,LC}^{s,K}$&$r_{c,LC}^{s,K}$\\
\hline
NJL &0.188&0.374& 0.187&0.390&0.167& 0.296\\
DSE &0.141&0.267&0.126&0.263&0.116&0.217\\
\hline
Dressed &$r_{E,LC}^{u,\pi}$&$r_{c,LC}^{u,\pi}$&$r_{E,LC}^{u,K}$&$r_{c,LC}^{u,K}$&$r_{c,LC}^{s,K}$&$r_{c,LC}^{s,K}$\\
\hline
NJL &0.312&0.513&0.297&0.525&0.238&0.372\\
DSE &0.233&0.375&0.218&0.371&0.205&0.296\\
\hline\hline
\end{tabular}
\end{table}
\end{center}
\begin{center}
\begin{table}
\centering
\caption{The ratios of the bare and dressed light-cone energy radius with the light-cone charge radius of $u$ quark and $s$ quark in the kaon and pion, as well as the $u$ quark in the NJL model and DSE.}\label{tb22}
\begin{tabular}{p{3.5cm}p{2.4cm} p{2.4cm} p{2.4cm} }
\hline
 &$r_{E,LC}^{u,\pi}/r_{c,LC}^{u,\pi}$&$r_{E,LC}^{u,K}/r_{c,LC}^{u,K}$&$r_{E,LC}^{s,K}/r_{c,LC}^{s,K}$\\
\hline
NJL (Bare) &0.503&0.479& 0.564\\
DSE (Bare) &0.528&0.479&0.534\\
\hline
NJL (Dressed) &0.608&0.566& 0.640\\
DSE (Dressed) &0.621&0.588&0.693\\
\hline\hline
\end{tabular}
\end{table}
\end{center}

\subsection{Tensor GPDs: images}
\begin{figure}
\centering
\includegraphics[width=0.47\textwidth]{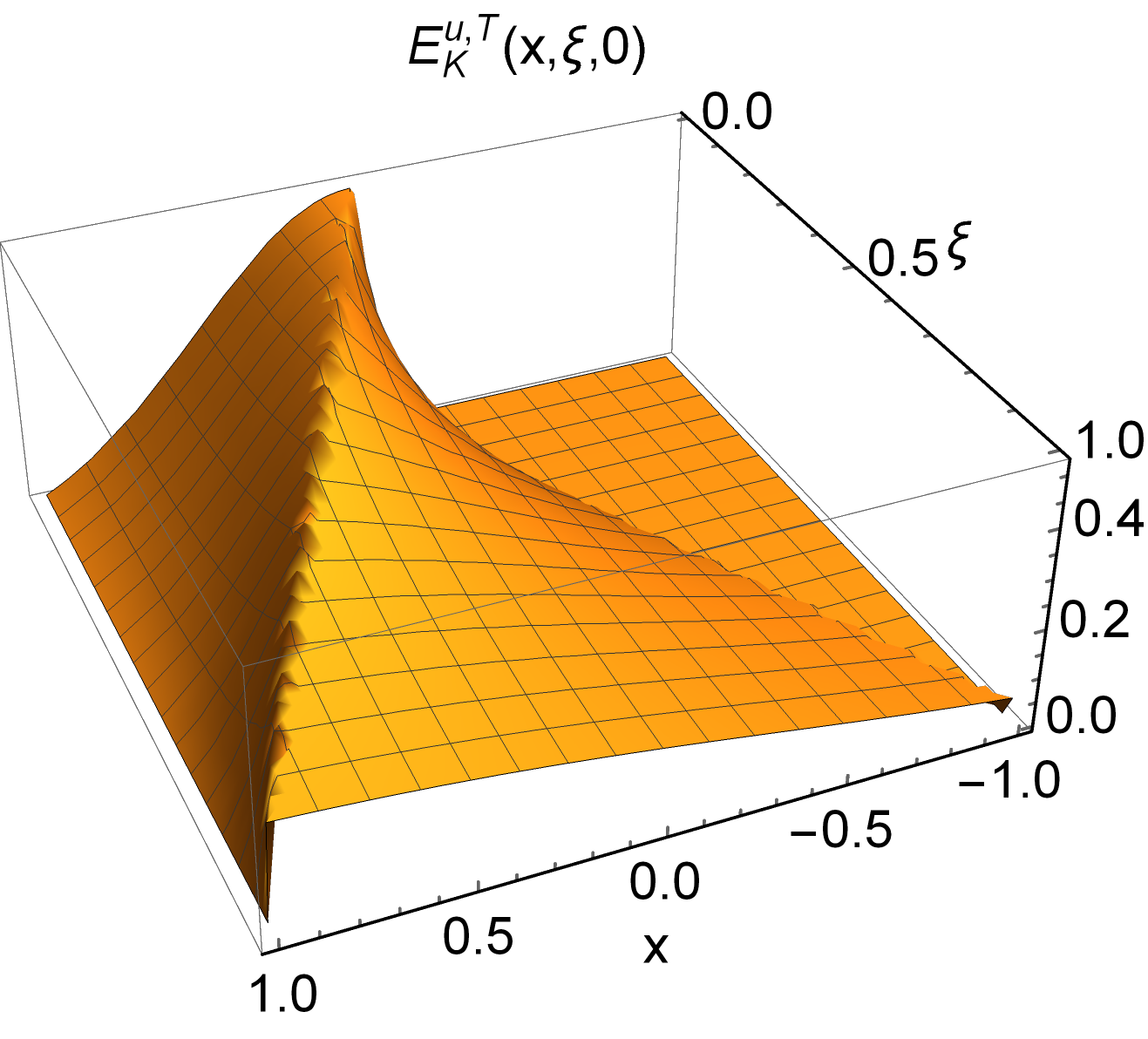}
\qquad
\includegraphics[width=0.47\textwidth]{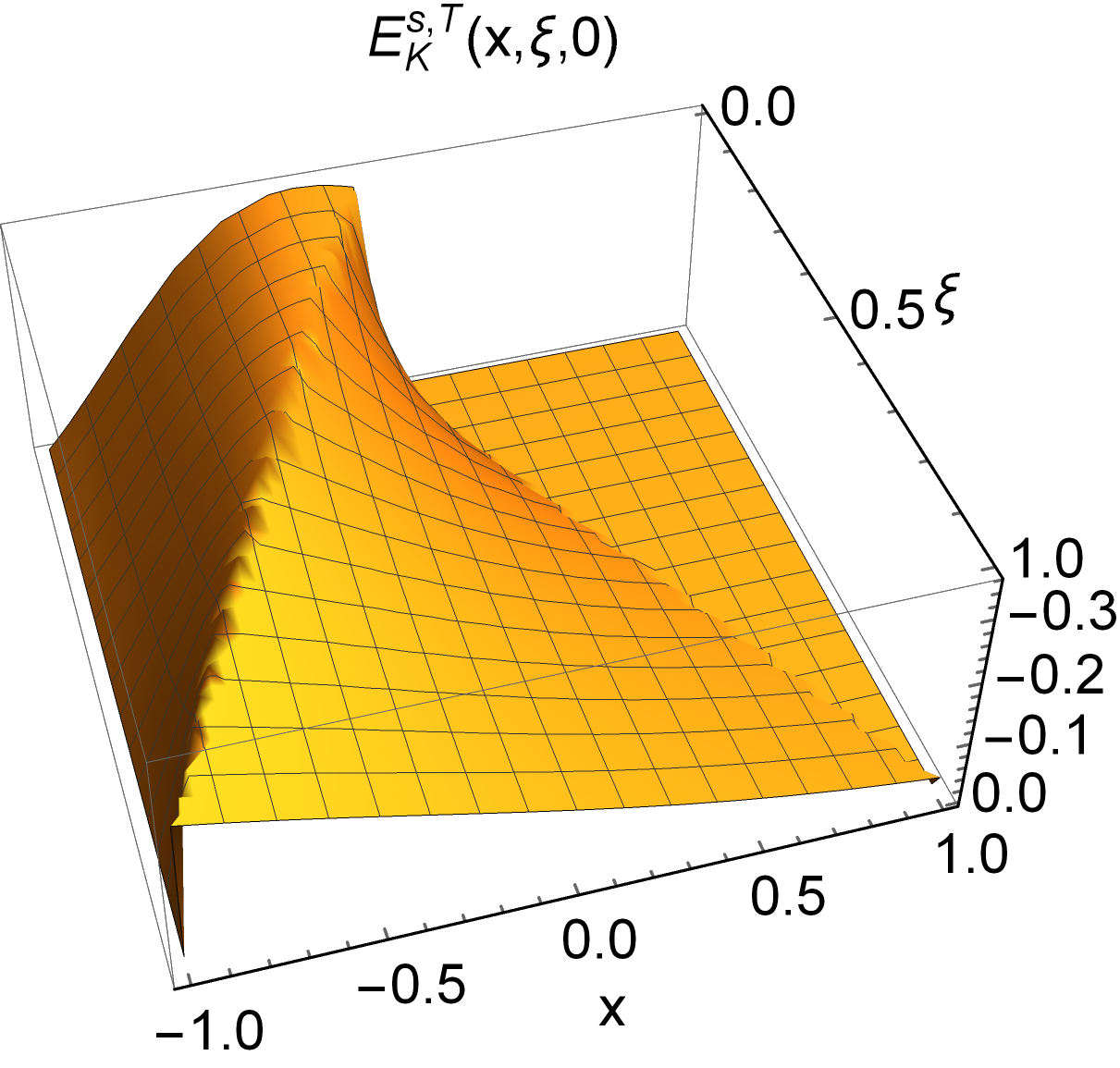}
\caption{\emph{Left panel} -- Kaon $u$ quark tensor GPD $E_K^{u,T}(x,\xi,0)$ in Eq. (\ref{t2tgpd}). \emph{Right panel} -- Kaon $s$ quark tensor GPD $E_K^{s,T}(x,\xi,0)$ in Eq. (\ref{t2stgpd}). Owing to Eq. (\ref{gcpf2}), only $\xi>0$ is plotted. }\label{t3gpdp}
\end{figure}
The tensor GPDs for the twist-two kaon, involving the $u$ quark and $s$ quark, are presented in Eq. (\ref{t2tgpd}) and Eq. (\ref{t2stgpd}), as shown in Fig. \ref{t3gpdp}. It is observed that for the kaon $u$ quark, the GPD is nonzero in the region $-\xi \leq x \leq 1$, while for the kaon $s$ quark GPD, it is nonzero in the region $-1 \leq x \leq \xi$.

\begin{figure}
\centering
\includegraphics[width=0.47\textwidth]{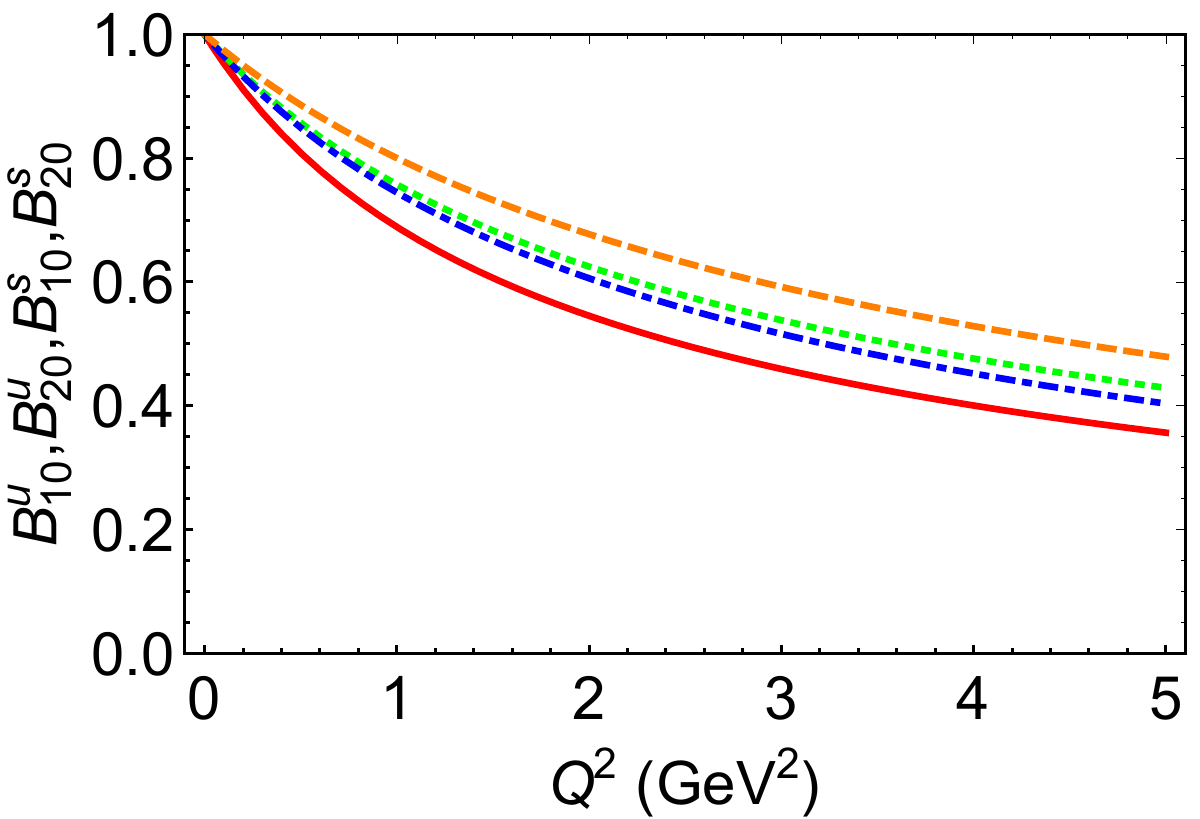}
\caption{Kaon tensor form factors normalized to their own charge: $u$ quark $B_{10}^u$--red solid curve, $u$ quark $B_{20}^u$--green dotted curve, $s$ quark $B_{10}^s$--blue dotdashed curve, $s$ quark $B_{20}^s$-- orange dashed curve.}\label{t2tff}
\end{figure}

The leading Mellin moments of the twist-two tensor GPD correspond to tensor form factors
\begin{subequations}\label{b10b20}
\begin{align}\label{2ff}
B_{10}^{u,K}(Q^2)=&\int_{-1}^1 dx E_K^{u,T}(x,\xi,-Q^2)\,, \\
B_{20}^{u,K}(Q^2)=&\int_{-1}^1 dx x E_K^{u,T}(x,\xi,-Q^2)\,,
\end{align}
\end{subequations}
we can obtain the following tensor form factors
\begin{subequations}\label{b10b20}
\begin{align}\label{2ff}
B_{10}^{u,K}(Q^2)
=&-\frac{N^{FF}N_c}{4\pi^2}\int_0^1 dx  m_K\bar{\mathcal{C}}_1(\sigma_5) \nonumber\\
+ &\frac{N^{EF}N_c}{2\pi^2} \int_0^1  dx\int_0^{1-x}dy m_K (M_u+(M_s-M_u)(1-x-y))\frac{\bar{\mathcal{C}}_2(\sigma_6)}{\sigma_6}\,, \\
B_{20}^{u,K}(Q^2)
=& \frac{N^{EF}N_c}{2\pi^2} \int_0^1  dx\int_0^{1-x} dy m_K(1-x-y)(M_u+(M_s-M_u)(1-x-y) ) \frac{\bar{\mathcal{C}}_2(\sigma_6)}{\sigma_6}.
\end{align}
\end{subequations}
These are the tensor form factors of the $u$ quark for kaon, which are shown in Fig. \ref{t2tff} and normalized by their $Q^2=0$ values. Additionally, the tensor form factors of the $s$ quark are also plotted. From the diagrams, it is evident that both the $u$ and $s$ quark $B_{10}$ are softer than $B_{20}$. Furthermore, the $u$ quark's $B_{10}$ and $B_{20}$ are softer than those of the $s$-quark respectively. This finding is in line with the outcomes achieved in our prior publication.
The radii of pion and kaon are presented in Table. \ref{tbrd}. The table indicates that the NJL results exhibit larger values compared to the DSE results.

The ratios $r_{B_{10}}/r_{B_{20}}$ of pion and kaon are presented in Table. \ref{tbrd1}. From the table, it can be observed that the results from the DSE are smaller than those from the NJL model. Lattice QCD yields $r_{B_{10}}^{\pi}/r_{B_{20}}^{\pi}=1.48(17)$, which is smaller than our results.
%
\begin{center}
\begin{table}
\centering
\caption{The comparison of pion and kaon $r_{B_{10}}$ and $r_{B_{20}}$, which are in units of fm.}\label{tbrd}
\begin{tabular}{p{1.2cm}p{1.1cm} p{1.1cm} p{1.1cm} p{1.1cm}p{1.1cm}p{1.1cm}}
\hline
 &$r_{B_{10}}^{u,\pi}$&$r_{B_{20}}^{u,\pi}$&$r_{B_{10}}^{u,K}$&$r_{B_{20}}^{u,K}$&$r_{B_{10}}^{s,K}$&$r_{B_{20}}^{s,K}$\\
\hline
NJL &0.145&0.065&0.283&0.124&0.206&0.104\\
DSE &0.079&0.045&0.148&0.087&0.127&0.076\\
\hline\hline
\end{tabular}
\end{table}
\end{center}
\begin{center}
\begin{table}
\centering
\caption{The ratios $r_{B_{10}}/r_{B_{20}}$ of pion and kaon in the NJL model and DSE.}\label{tbrd1}
\begin{tabular}{p{2.2cm}p{2.0cm} p{2.0cm} p{2.0cm} }
\hline
 &$r_{B_{10}}^{u,\pi}/r_{B_{20}}^{u,\pi}$&$r_{B_{10}}^{u,K}/r_{B_{20}}^{u,K}$&$r_{B_{10}}^{s,K}/r_{B_{20}}^{s,K}$\\
\hline
NJL  &2.230&2.282&1.981\\
DSE  &1.756&1.701&1.671\\
\hline\hline
\end{tabular}
\end{table}
\end{center}
%

\section{Kaon TMDs}\label{good2}
\subsection{Twist-two TMDs}
\begin{figure}
\centering
\includegraphics[width=0.47\textwidth]{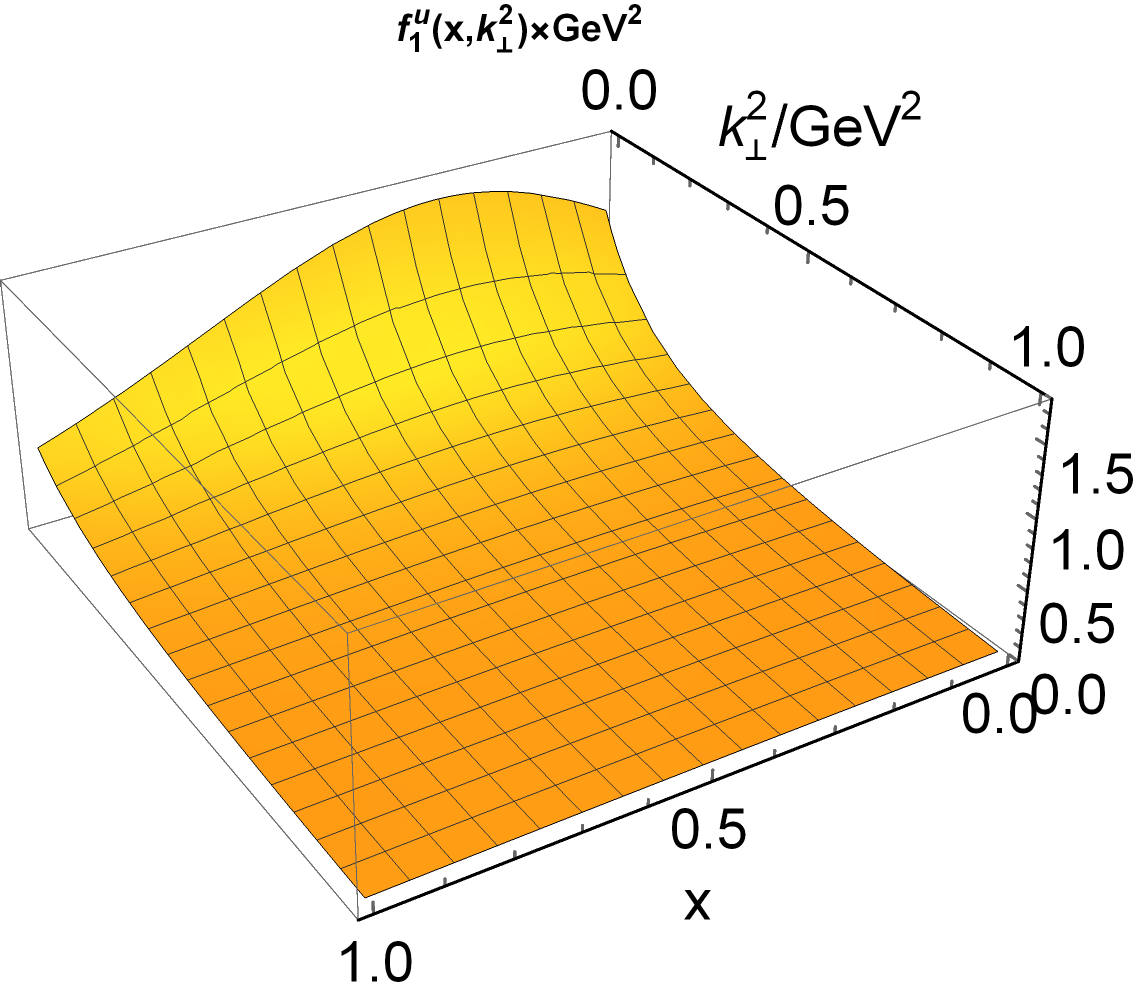}
\qquad
\includegraphics[width=0.47\textwidth]{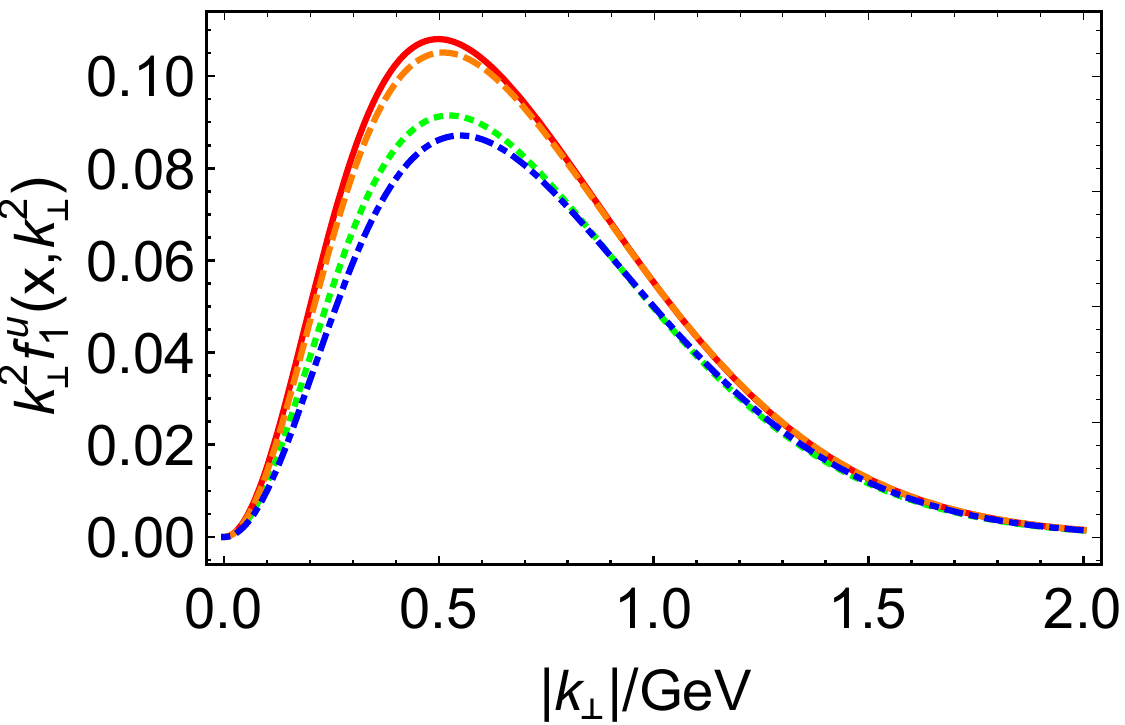}
\caption{\emph{Left panel} -- $f_1^u\left(x,\bm{k}_{\bot }^2\right)$ kaon $u$ quark twist-two TMD defined in Eq. (\ref{t2tmd}); \emph{Right panel} -- $\bm{k}_{\bot}^2 f_1^u\left(x,\bm{k}_{\bot }^2\right)$ at: $x = 0$ - dotted green curve; $x=1/4$ - solid red curve; and $x=1/2$ - dashed orange curve; and $x=3/4$ - dot-dashed blue curve.}\label{tmdp1}
\end{figure}

At twist-two, by setting $\xi=0$ and $t=0$, we can obtain the twist-two TMDs from the GTMDs ($r=\bm{k}_{\perp}^2$),
\begin{align}\label{t2tmd}
f_1^u(x,\bm{k}_{\perp}^2)=&F_1(x,\bm{k}_{\perp}^2,0,0)\nonumber\\
=&\frac{\bar{N}^{FF}N_c}{2\pi^3}x(1-x)\frac{\bar{\mathcal{C}}_2(\sigma_1^{r,0})}{\sigma_1^{r,0}}+\frac{N^{EE}N_c}{2\pi^3} \frac{\bar{\mathcal{C}}_2(\sigma_1^{r,0})}{\sigma_1^{r,0}}+\frac{\bar{N}^{EF}N_c}{2\pi^3} (2x-1) \frac{\bar{\mathcal{C}}_2(\sigma_1^{r,0})}{\sigma_1^{r,0}}\nonumber\\
+& \frac{N^{EF}N_c}{4\pi^3}x(1-x)(m_K^2+(M_u-M_s)^2) \frac{6\bar{\mathcal{C}}_3(\sigma_1^{r,0})}{[\sigma_1^{r,0}]^2},
\end{align}
this is the valence $u$-quark TMD in the $K^+$, which is depicted in Fig. \ref{tmdp1}. From the diagram, it is evident that unlike the pion $u$ quark TMDs, $f_1^u(x,\bm{k}_{\perp}^2)$ does not exhibit symmetry at $x=1/2$. The peaks of the kaon $u$ quark TMD shift to $x<1/2$.

The root-mean-square (r.m.s.) value of $\bm{k}_{\perp}^2$ is defined as
\begin{align}\label{rms1}
\langle k_{\perp}^2 \rangle^{1/2}&=\int_0^1 dx \int d^2\bm{k}_{\perp}\bm{k}_{\perp}^2f_1(x,\bm{k}_{\perp}^2),
\end{align}
we have obtained the r.m.s. values of transverse momentum for the $u$ quark, $\langle k_{\perp}^2 \rangle_u^{1/2}=0.611$ GeV, and for the $s$ quark, $\langle k_{\perp}^2 \rangle_s^{1/2}=0.614$ GeV. For the pion, the root-mean-square value of transverse momentum for the $u$ quark is $\langle k_{\perp}^2 \rangle_u^{1/2}=0.617$ GeV, which is almost identical to that of both $u$ and $s$ quarks in a kaon.

The r.m.s. transverse momentum gives a transverse size scale $R_{\bot }$
\begin{align}\label{ab20}
R_{\bot }\equiv \frac{1}{\left\langle k_{\bot }^2\right\rangle^{1/2} },
\end{align}
we obtained $R_{\bot }^u=0.322$ fm and $R_{\bot }^u=0.321$ fm, which are nearly identical to the bare $u$ quark charge radius $r_K^u$, but larger than the bare $s$ quark radius $r_K^s$. The dimensionless quantity $C=f_KR_{\bot }$~\cite{Weise:1984tc} connects the "core radius" $R_{\bot }$ with the decay constant $f_K=0.111$ GeV. Our results indicate that for kaon, $C=0.181$, and for pion, $C=0.167$ with $f_{\pi}=0.103$ GeV, which is consistent with the result $C\simeq 0.2$ in Ref.~\cite{Bernard:1985cs}. 

From the TMD we can obtain the $u$ quark distribution function
\begin{align}\label{t2pdf1}
q_K(x)=\int d^2\bm{k}_{\perp} f_1(x,\bm{k}_{\perp}^2),
\end{align}
we obtain
\begin{align}\label{kapdf}
u_1^K(x)&=\frac{\bar{N}^{FF}N_c}{4\pi^2}(1-x)x\left(M_u^2-M_s^2\right)\bar{\mathcal{C}}_1(\sigma_1^{0,0})+\frac{N^{EE}N_c}{4\pi^2} \bar{\mathcal{C}}_1(\sigma_1^{0,0})\nonumber\\
&+\frac{\bar{N}^{EF}N_c}{4\pi^2} (2x-1) \bar{\mathcal{C}}_1(\sigma_1^{0,0})+\frac{N^{EF}N_c}{2\pi^2} x(1-x)(m_K^2-(M_u-M_s)^2) \frac{\bar{\mathcal{C}}_2(\sigma_1^{0,0})}{\sigma_1^{0,0}},
\end{align}
for the mean value of the momentum fraction $x$ carried by the $u$ quark
\begin{align}\label{ab24}
\langle x\rangle^u&=\int_0^1x\, u_1^K(x)\, dx=0.556,
\end{align}
therefore, we have $\langle x\rangle^s=0.444$, indicating that the quarks do not share an equal amount of longitudinal momentum. This is different from the results of the NJL model, where $\langle x\rangle^u=0.431$ and $\langle x\rangle^s=0.569$ as reported in Ref.~\cite{Zhang:2021tnr}. In DSE, the mean value of the momentum fraction $x$ carried by the $u$ quark is larger than that carried by the $s$ quark.

\begin{figure}
\centering
\includegraphics[width=0.47\textwidth]{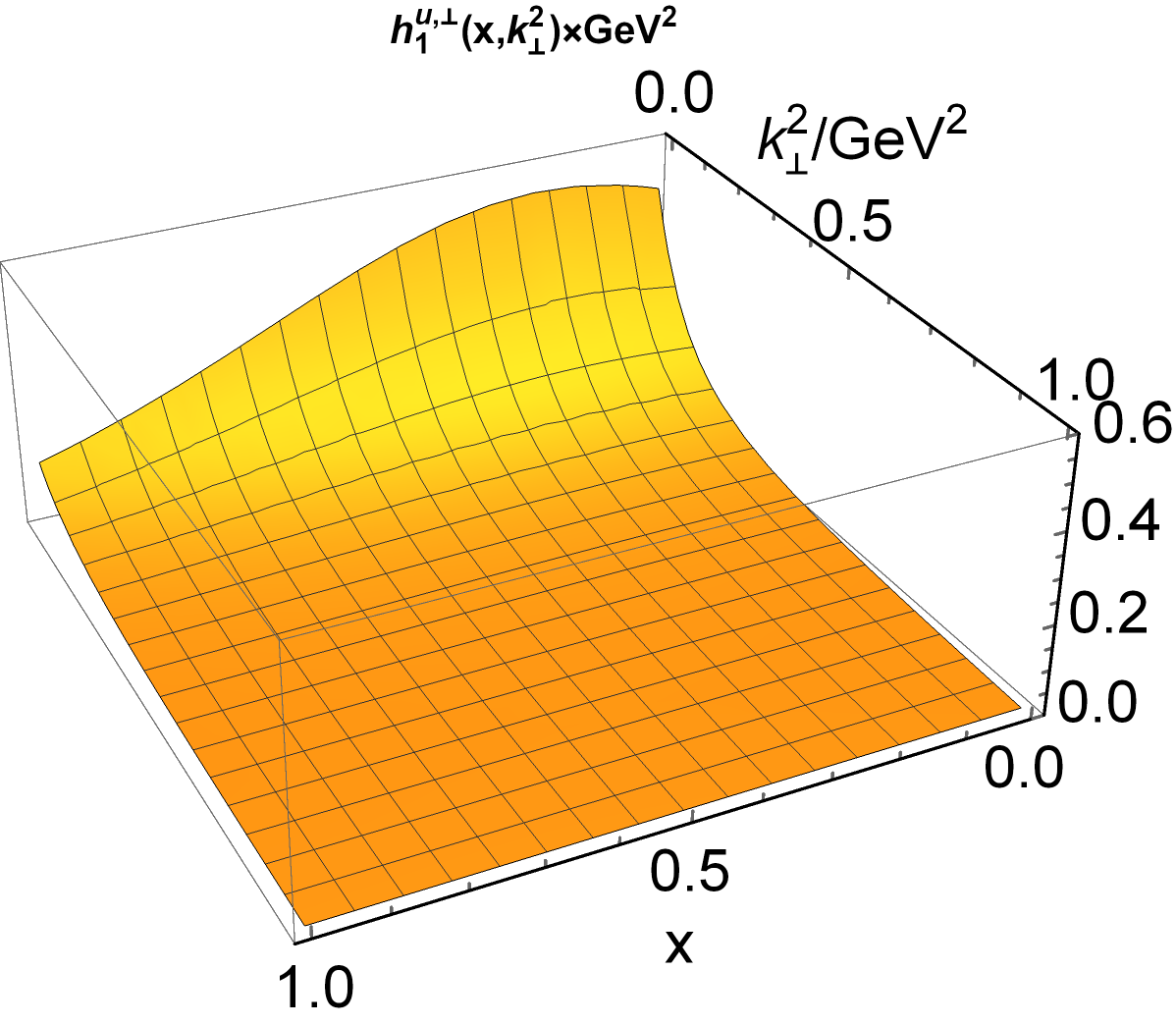}
\qquad
\includegraphics[width=0.47\textwidth]{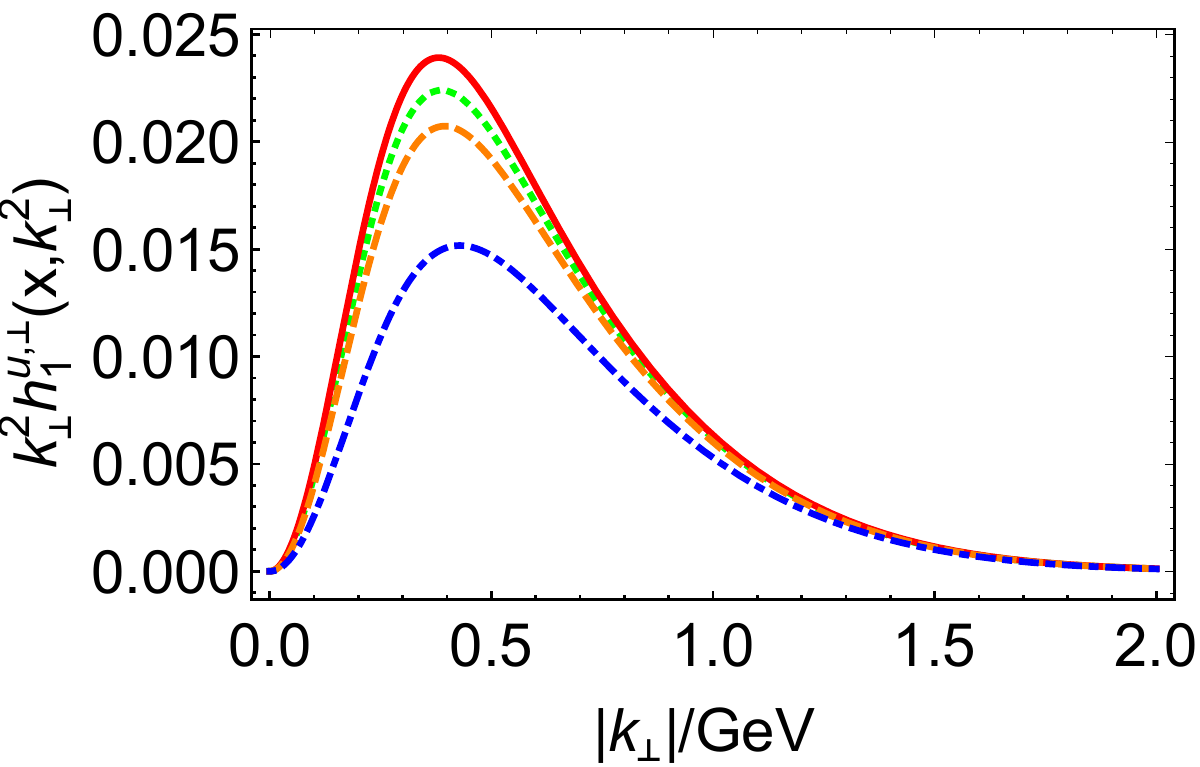}
\caption{\emph{Left panel} -- $h_1^{u,\perp}\left(x,\bm{k}_{\bot }^2\right)$ kaon $u$ quark twist-two TMD defined in Eq. (\ref{t2pdf2}); \emph{Right panel} -- $\bm{k}_{\bot}^2 h_1^{u,\perp}\left(x,\bm{k}_{\bot }^2\right)$ at: $x = 0$ - dotted green curve; $x=1/4$ - solid red curve; and $x=1/2$ - dashed orange curve; and $x=3/4$ - dot-dashed blue curve.}\label{tmdp11}
\end{figure}

Since we do not consider the Wilson line, our previous paper~\cite{Zhang:2020ecj} indicates that the Boer-Mulders function of the pion is zero. However, it is different for the kaon
\begin{align}\label{t2pdf2}
h_1^{u,\perp}(x,r)=H_1^{u,k}(x,\bm{k}_{\perp}^2,0,0)= \frac{N^{EF}N_c}{4\pi^3} (M_s-M_u)m_K \frac{6\bar{\mathcal{C}}_3(\sigma_1^{r,0})}{[\sigma_1^{r,0}]^2},
\end{align}
it can be observed that when $M_u=M_s$, this term will become zero. Therefore, for the pion, this term is also zero. $h_1^{u,\perp}(x,r)$ of the valence $u$ quark of the kaon is illustrated in Fig. \ref{tmdp11}.

\subsection{Twist-three TMDs}
\begin{figure}
\centering
\includegraphics[width=0.47\textwidth]{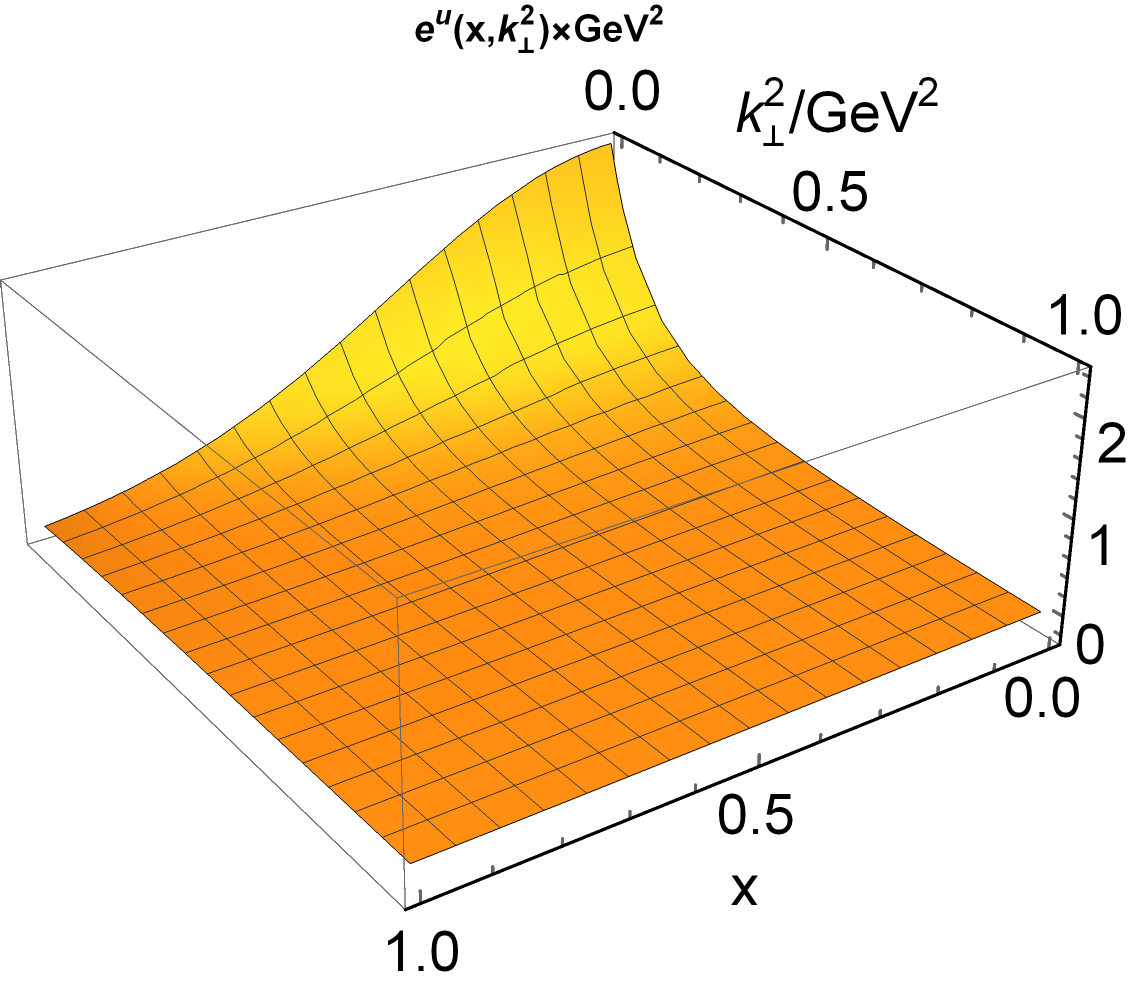}
\qquad
\includegraphics[width=0.47\textwidth]{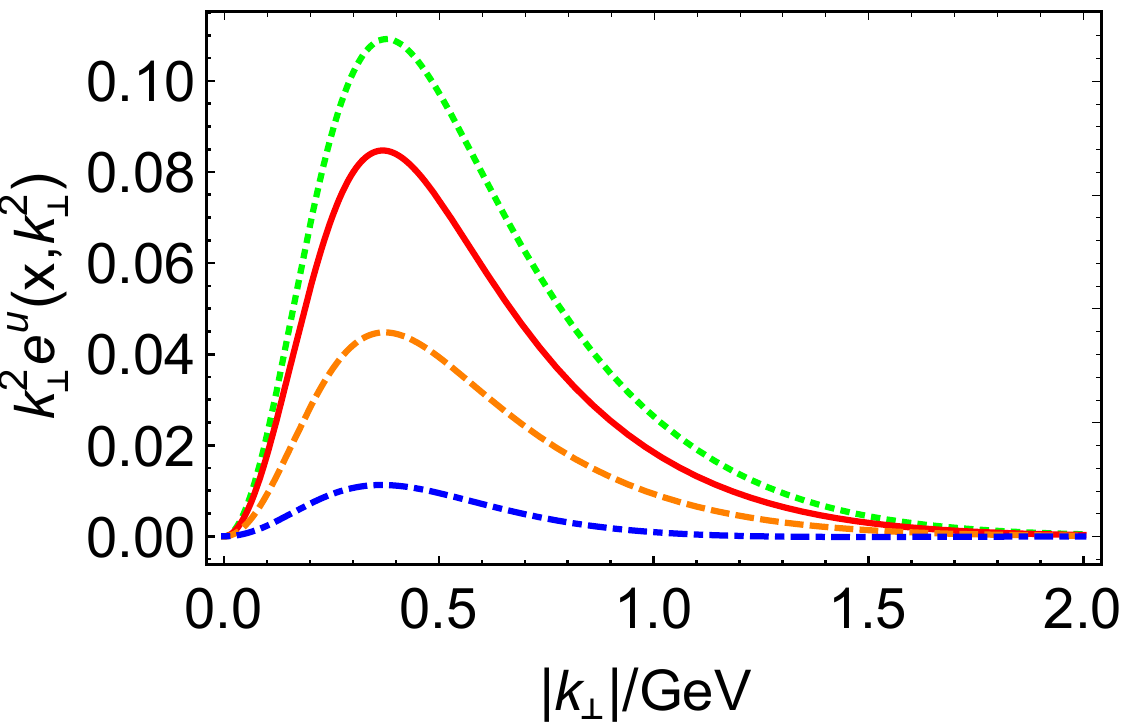}
\caption{\emph{Left panel} -- $e^u\left(x,\bm{k}_{\bot }^2\right)$ kaon $u$ quark twist-two TMD defined in Eq. (\ref{t3tmd1}); \emph{Right panel} -- $\bm{k}_{\bot}^2 e^u\left(x,\bm{k}_{\bot }^2\right)$ at: $x = 0$ - dotted green curve; $x=1/4$ - solid red curve; and $x=1/2$ - dashed orange curve; and $x=3/4$ - dot-dashed blue curve. }\label{tmdp2}
\end{figure}

The first twist-three TMD is obtained from the GTMD $E_2(x,\bm{k}_{\perp}^2,\xi,t)$ ($r=\bm{k}_{\perp}^2$)
\begin{align}\label{t3tmd1}
&e^u(x,r)=E_2^u(x,\bm{k}_{\perp}^2,0,0)= \frac{\check{N}^{EF}N_c }{2\pi^3} \frac{(M_u-M_s)}{m_K}\frac{\bar{\mathcal{C}}_2(\sigma_1^{r,0})}{\sigma_1^{r,0}}\nonumber\\
+& \frac{\tilde{N}^{EF}N_c }{2\pi^3} m_K \frac{\bar{\mathcal{C}}_2(\sigma_1^{r,0})}{\sigma_1^{r,0}}+ \frac{N^{EF}N_c}{4\pi^3}\frac{(1-x)}{m_K}(m_K^2-(M_s-M_u)^2) \frac{6\bar{\mathcal{C}}_3(\sigma_1^{r,0})}{[\sigma_1^{r,0}]^2},
\end{align}
where
\begin{align}\label{t3tmd3}
\check{N}^{EF}&=E_K^2-F_KE_K \frac{(M_s+3M_u)(M_s+M_u)}{2M_uM_s}+F_K^2\frac{(M_s+M_u)^3}{2M_uM_s^2} ,
\end{align}
this TMD is chiral even in Fig. \ref{tmdp2}, the r.m.s. value of the kaon $\bm{k}_{\perp}^2$ is $\langle k_{\perp}^2 \rangle_u^{1/2}=0.319$ GeV and $\langle k_{\perp}^2 \rangle_s^{1/2}=0.529$ GeV, for the pion $\langle k_{\perp}^2 \rangle_u^{1/2}=0.267$ GeV, which indicates that it is larger than the result for the pion. 
The twist-three quark PDF is
\begin{align}\label{t2pdf1}
e(x)=\int d^2k_{\perp} e(x,k_{\perp}^2),
\end{align}
we obtain
\begin{align}\label{agtmd}
e^u(x)&= \frac{\check{N}^{EF}N_c }{4\pi^2} \frac{(M_u-M_s)}{m_K}\bar{\mathcal{C}}_1(\sigma_1^{0,0})+ \frac{\tilde{N}^{EF}N_c }{4\pi^2} m_K \bar{\mathcal{C}}_1(\sigma_1^{0,0})\nonumber\\
&+ \frac{N^{EF}N_c}{2\pi^2}\frac{(1-x)}{m_K}(m_K^2-(M_s-M_u)^2) \frac{\bar{\mathcal{C}}_2(\sigma_1^{0,0})}{\sigma_1^{0,0}},
\end{align}

\begin{figure}
\centering
\includegraphics[width=0.47\textwidth]{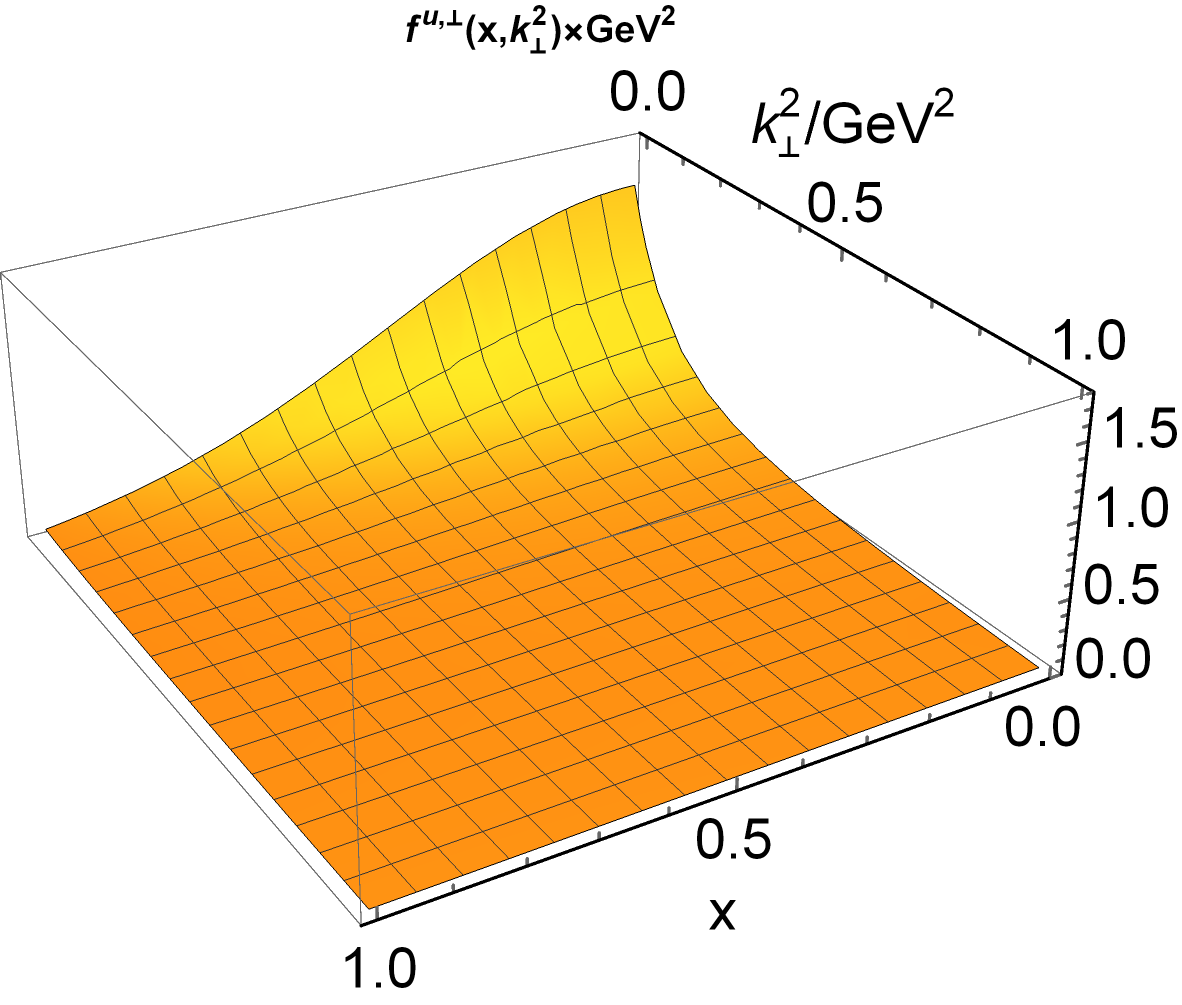}
\qquad
\includegraphics[width=0.47\textwidth]{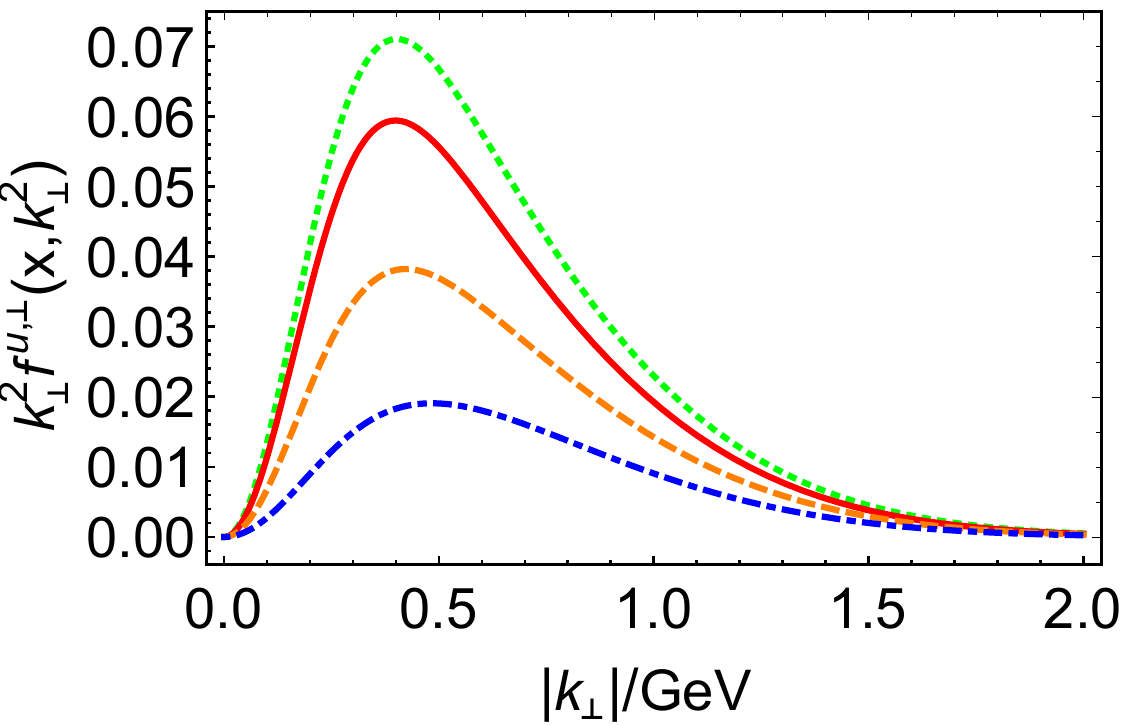}
\caption{\emph{Left panel} -- $f^{u,\perp}\left(x,\bm{k}_{\bot }^2\right)$ kaon $u$ quark twist-two TMD defined in Eq. (\ref{t3tmd2}); \emph{Right panel} -- $\bm{k}_{\bot}^2 f^{u,\perp}\left(x,\bm{k}_{\bot }^2\right)$ at: $x = 0$ - dotted green curve; $x=1/4$ - solid red curve; and $x=1/2$ - dashed orange curve; and $x=3/4$ - dot-dashed blue curve. }\label{tmdp3}
\end{figure}

\begin{figure}
\centering
\includegraphics[width=0.47\textwidth]{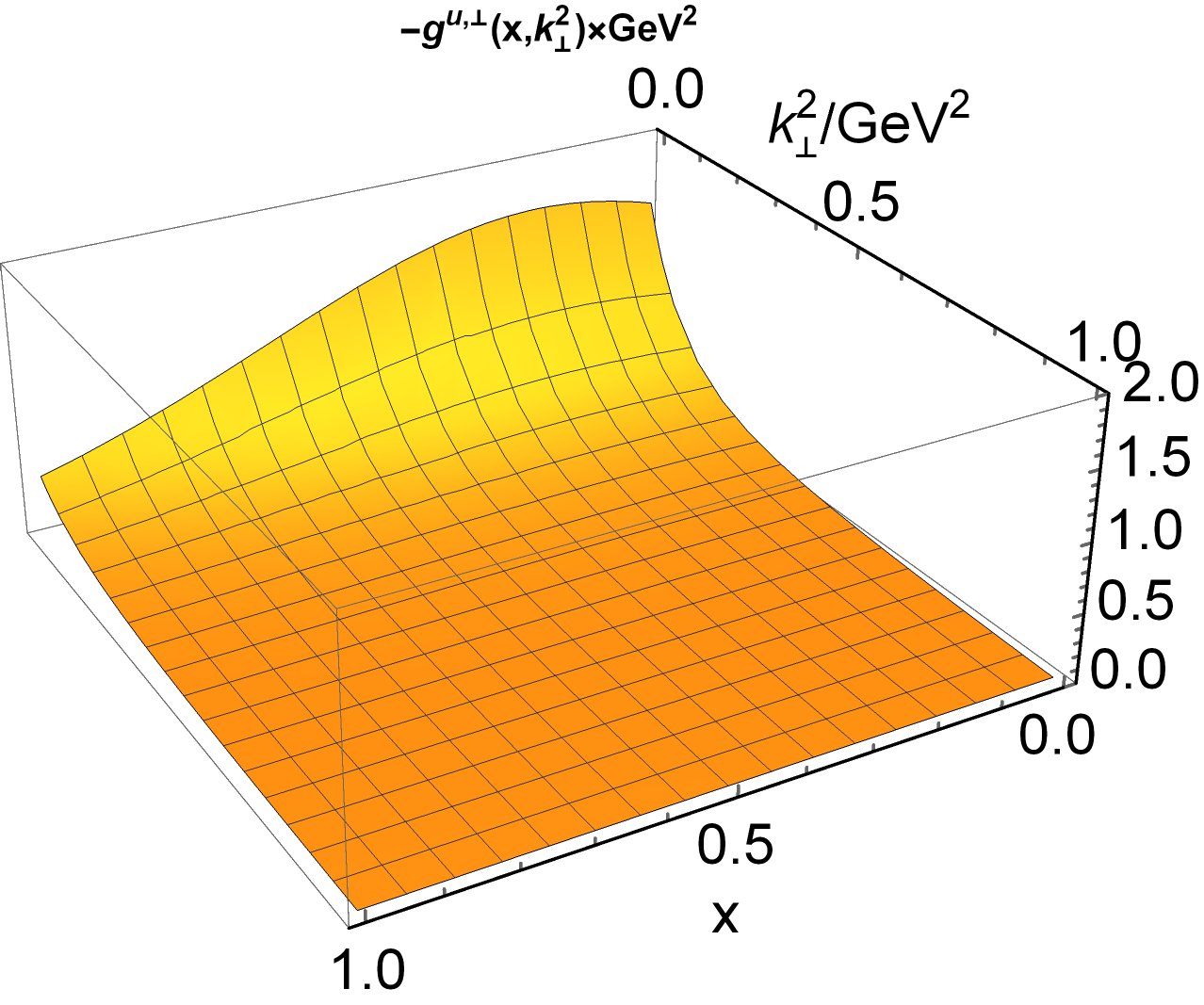}
\qquad
\includegraphics[width=0.47\textwidth]{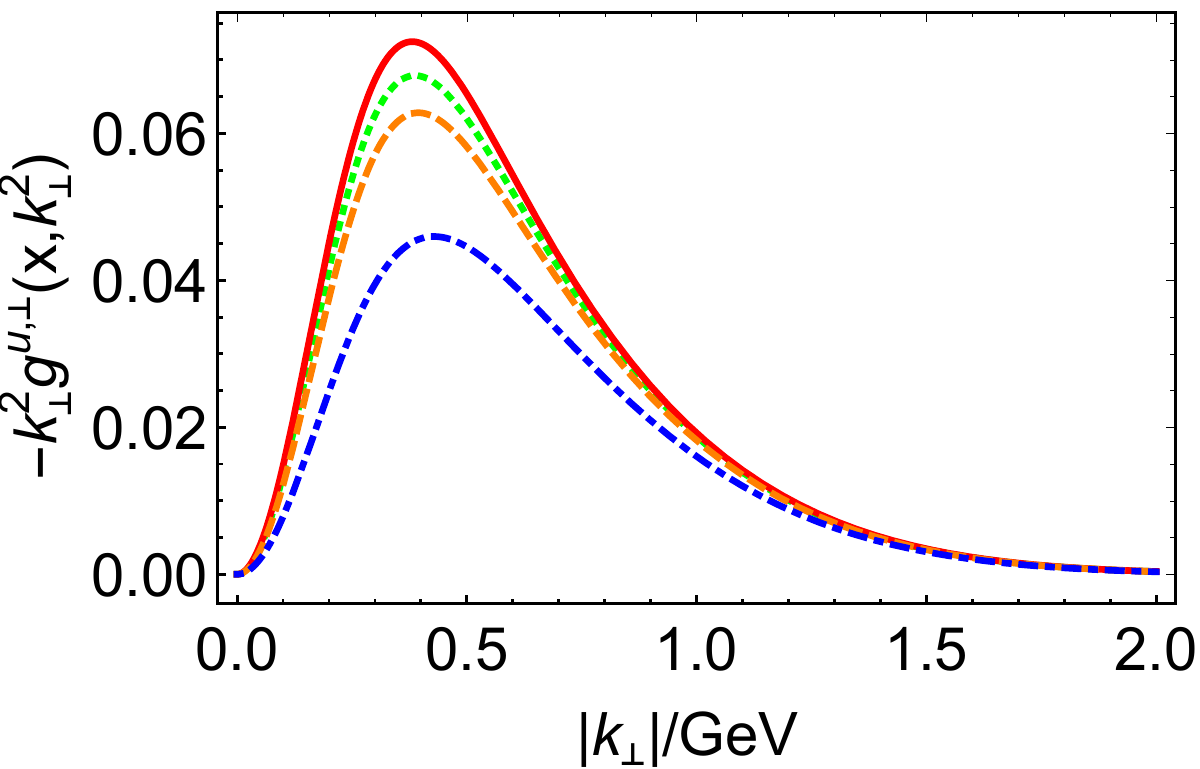}
\caption{\emph{Left panel} -- $g^{u,\perp}\left(x,\bm{k}_{\bot }^2\right)$ kaon $u$ quark twist-two TMD defined in Eq. (\ref{t3tmd3}); \emph{Right panel} -- $\bm{k}_{\bot}^2 g^{u,\perp}\left(x,\bm{k}_{\bot }^2\right)$ at: $x = 0$ - dotted green curve; $x=1/4$ - solid red curve; and $x=1/2$ - dashed orange curve; and $x=3/4$ - dot-dashed blue curve. }\label{tmdp33}
\end{figure}

Another twist-three TMD is chiral even ($r=\bm{k}_{\perp}^2$),
\begin{align}\label{t3tmd2}
f^{u,\perp}(x,r)=&F_2^{u,k}(x,\bm{k}_{\perp}^2,0,0)\nonumber\\
=&\bar{N}^{EF}  \frac{N_c }{\pi^3} \frac{1}{\sigma_1^{r,0}}\bar{\mathcal{C}}_2(\sigma_1^{r,0})+ \frac{N^{EF}N_c }{4\pi^3}(1-x)(m_K^2-\left(M_u-M_s\right)^2) \frac{6\bar{\mathcal{C}}_3(\sigma_1^{r,0})}{[\sigma_1^{r,0}]^2},
\end{align}
$f^{u,\perp}(x,\bm{k}_{\perp}^2)$ is plotted in Fig. \ref{tmdp3}, the r.m.s. value of the kaon $\bm{k}_{\perp}^2$ is $\langle k_{\perp}^2 \rangle_u^{1/2}=0.342$ GeV and $\langle k_{\perp}^2 \rangle_s^{1/2}=0.145$ GeV, for the pion $\langle k_{\perp}^2 \rangle_u^{1/2}=0.086$ GeV. 
\begin{align}\label{t2pdf1}
f^{\perp}(x)=\int d^2\bm{k}_{\perp} f^{\perp}(x,\bm{k}_{\perp}^2),
\end{align}
we obtained
\begin{align}\label{agtmd}
f^{u,\perp}(x)=&\bar{N}^{EF}  \frac{N_c }{2\pi^2}\bar{\mathcal{C}}_1(\sigma_1^{0,0})+ \frac{N^{EF}N_c }{4\pi^2}(1-x)(m_K^2-\left(M_u-M_s\right)^2) \frac{\bar{\mathcal{C}}_2(\sigma_1^{0,0})}{\sigma_1^{0,0}},
\end{align}
The chiral odd TMD
\begin{align}\label{t3tmd3}
g^{u,\perp}(x,r)&=G_2^{u,k}(x,\bm{k}_{\perp}^2,0,0)=-\frac{N^{EF}N_c}{4\pi^3} m_K^2\frac{6\bar{\mathcal{C}}_3(\sigma_1^{r,0})}{[\sigma_1^{r,0}]^2},
\end{align}
which is plotted in Fig. \ref{tmdp33}.

\subsection{Twist-four TMDs}
The kaon twist-four chiral even TMD is ($r=\bm{k}_{\perp}^2$),
\begin{align}\label{t4tmd1}
f_3^u(x,r)=&F_3^u(x,\bm{k}_{\perp}^2,0,0)\nonumber\\
=&-\frac{N^{EE}N_c}{2\pi^3} \frac{\bar{\mathcal{C}}_2(\sigma_1^{r,0})}{\sigma_1^{r,0}}-\frac{\bar{N}^{EF}N_c}{2\pi^3}  (2x-1) \frac{\bar{\mathcal{C}}_2(\sigma_1^{r,0})}{\sigma_1^{r,0}}\nonumber\\
-& \frac{N^{EF}N_c}{4\pi^3} (1-x)^2(m_K^2-\left(M_u-M_s\right)^2) \frac{6\bar{\mathcal{C}}_3(\sigma_1^{r,0})}{[\sigma_1^{r,0}]^2}\nonumber\\
+&  \frac{N^{EF}N_c}{2\pi^3} \frac{1}{m_K^2}(m_K^2-\left(M_u-M_s\right)^2) \frac{\bar{\mathcal{C}}_2(\sigma_1^{r,0})}{\sigma_1^{r,0}}\nonumber\\
-&  \frac{N^{EF}N_c}{4\pi^3} \frac{1}{m_K^2} (1-x)(m_K^2-\left(M_u-M_s\right)^2) \left(M_s^2-M_u^2\right) \frac{6\bar{\mathcal{C}}_3(\sigma_1^{r,0})}{[\sigma_1^{r,0}]^2}.
\end{align}
%
%
%
\begin{figure}
\centering
\includegraphics[width=0.47\textwidth]{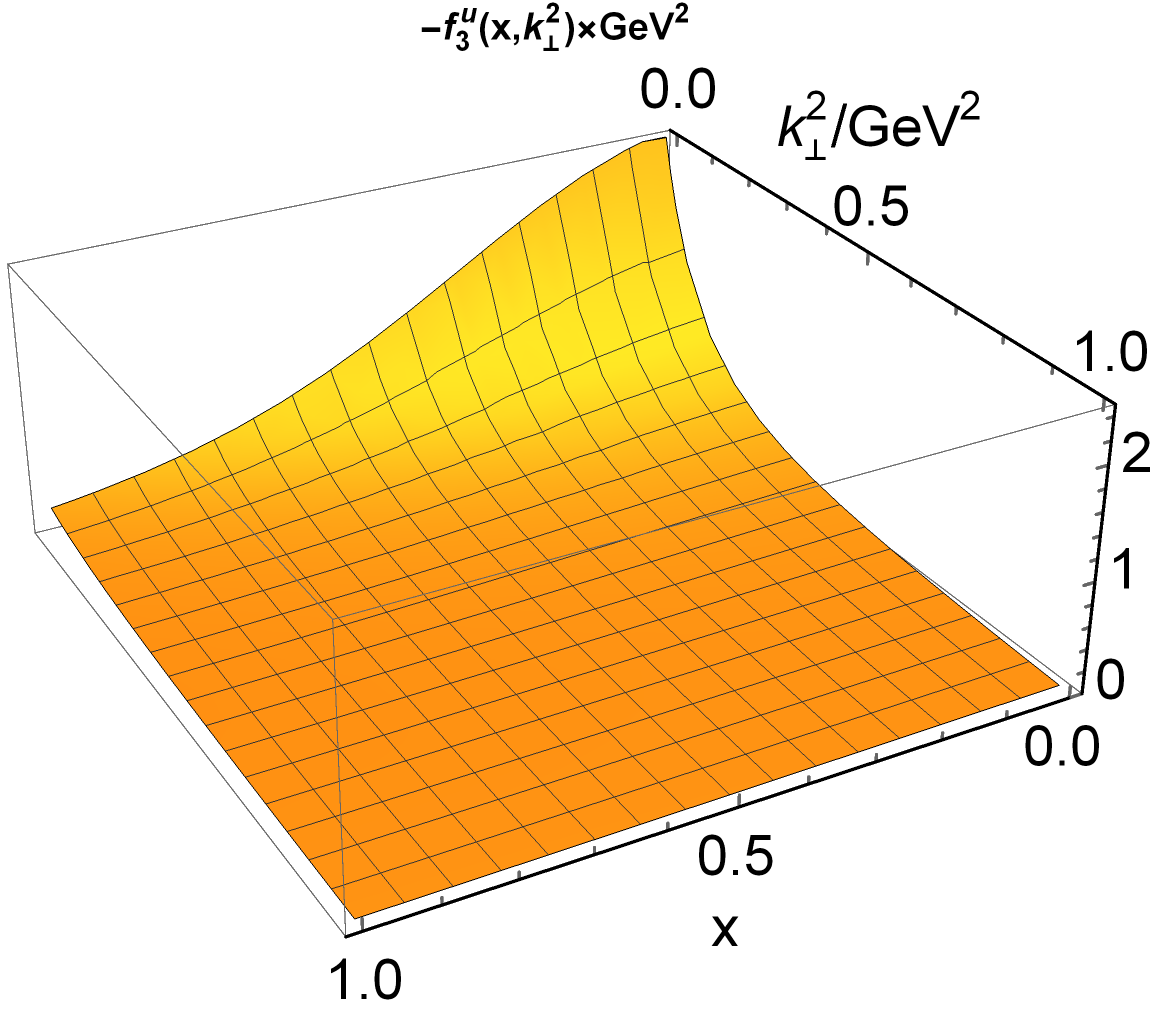}
\qquad
\includegraphics[width=0.47\textwidth]{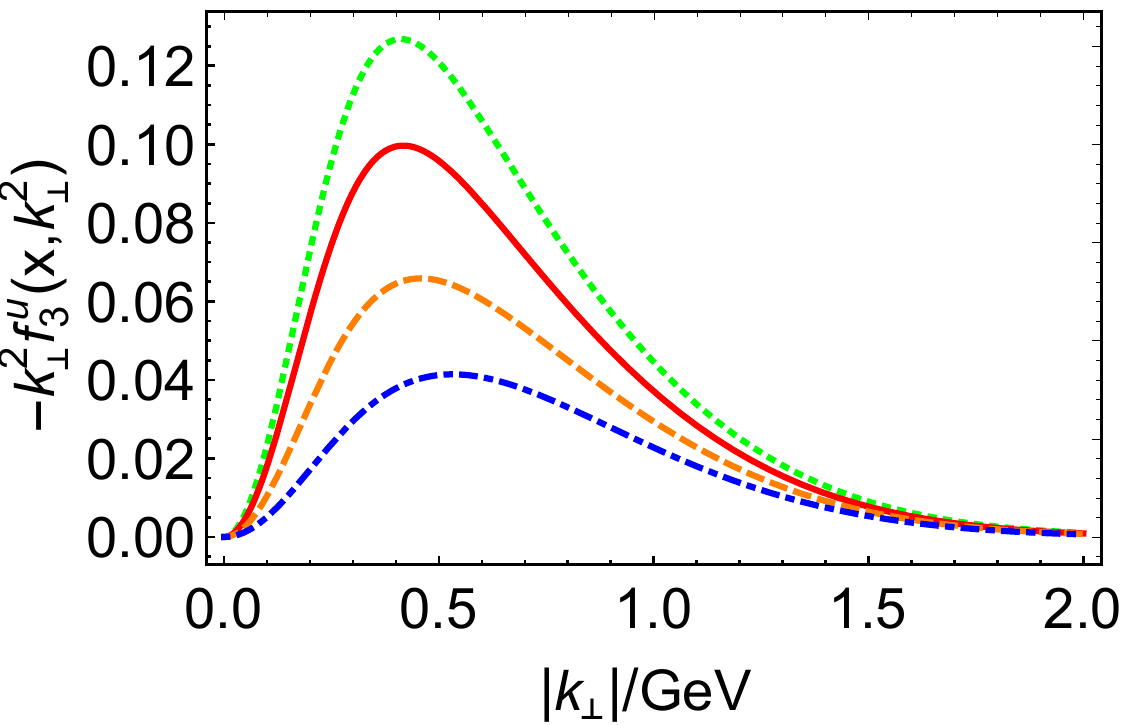}
\caption{\emph{Left panel} -- $-f_3^u\left(x,\bm{k}_{\bot }^2\right)$ kaon $u$ quark twist-two TMD defined in Eq. (\ref{t4tmd1}); \emph{Right panel} -- $-\bm{k}_{\bot}^2 f_3^u\left(x,\bm{k}_{\bot }^2\right)$ at: $x = 0$ - dotted green curve; $x=1/4$ - solid red curve; and $x=1/2$ - dashed orange curve; and $x=3/4$ - dot-dashed blue curve. }\label{tmdp4}
\end{figure}
The r.m.s. value of the kaon $\bm{k}_{\perp}^2$ is $\langle k_{\perp}^2 \rangle_u^{1/2}=0.485$ GeV and $\langle k_{\perp}^2 \rangle_s^{1/2}=0.376$ GeV, for the pion $\langle k_{\perp}^2 \rangle_u^{1/2}=0.324$ GeV. 
\begin{align}\label{t2pdf1}
f_3(x)=\int d^2\bm{k}_{\perp} f_3(x,\bm{k}_{\perp}^2),
\end{align}
we obtain
\begin{align}\label{agtmd}
f_3^u(x)=&-\frac{N^{EE}N_c}{4\pi^2} \bar{\mathcal{C}}_1(\sigma_1^{0,0})-\frac{\bar{N}^{EF}N_c}{4\pi^3}  (2x-1) \bar{\mathcal{C}}_1(\sigma_1^{0,0})\nonumber\\
-& \frac{N^{EF}N_c}{4\pi^2} (1-x)^2(m_K^2-\left(M_u-M_s\right)^2) \frac{\bar{\mathcal{C}}_2(\sigma_1^{0,0})}{\sigma_1^{0,0}}\nonumber\\
+&  \frac{N^{EF}N_c}{4\pi^2} \frac{1}{m_K^2}(m_K^2-\left(M_u-M_s\right)^2)\bar{\mathcal{C}}_1(\sigma_1^{0,0})\nonumber\\
-&  \frac{N^{EF}N_c}{4\pi^2} \frac{1}{m_K^2} (1-x)(m_K^2-\left(M_u-M_s\right)^2)\left(M_s^2-M_u^2\right) \frac{\bar{\mathcal{C}}_2(\sigma_1^{0,0})}{\sigma_1^{0,0}}.
\end{align}
The kaon twist-four chiral odd TMD is ($r=\bm{k}_{\perp}^2$),
\begin{align}\label{t4tmd2}
h_3^{u,\perp}(x,r)=&H_3^u(x,\bm{k}_{\perp}^2,0,0)=- \frac{N^{EF}N_c}{16\pi^3} \frac{(M_s-M_u)}{m_K}  \frac{6\bar{\mathcal{C}}_3(\sigma_1^{r,0})}{[\sigma_1^{r,0}]^2},
\end{align}
which is plotted in Fig. \ref{tmdp4}. 

The formulas of the chiral odd TMDs reveal that without the inclusion of the Wilson line or quark-gluon interaction, the chiral odd TMDs are limited in their ability to provide substantial information about the internal structure of hadrons.
\begin{figure}
\centering
\includegraphics[width=0.47\textwidth]{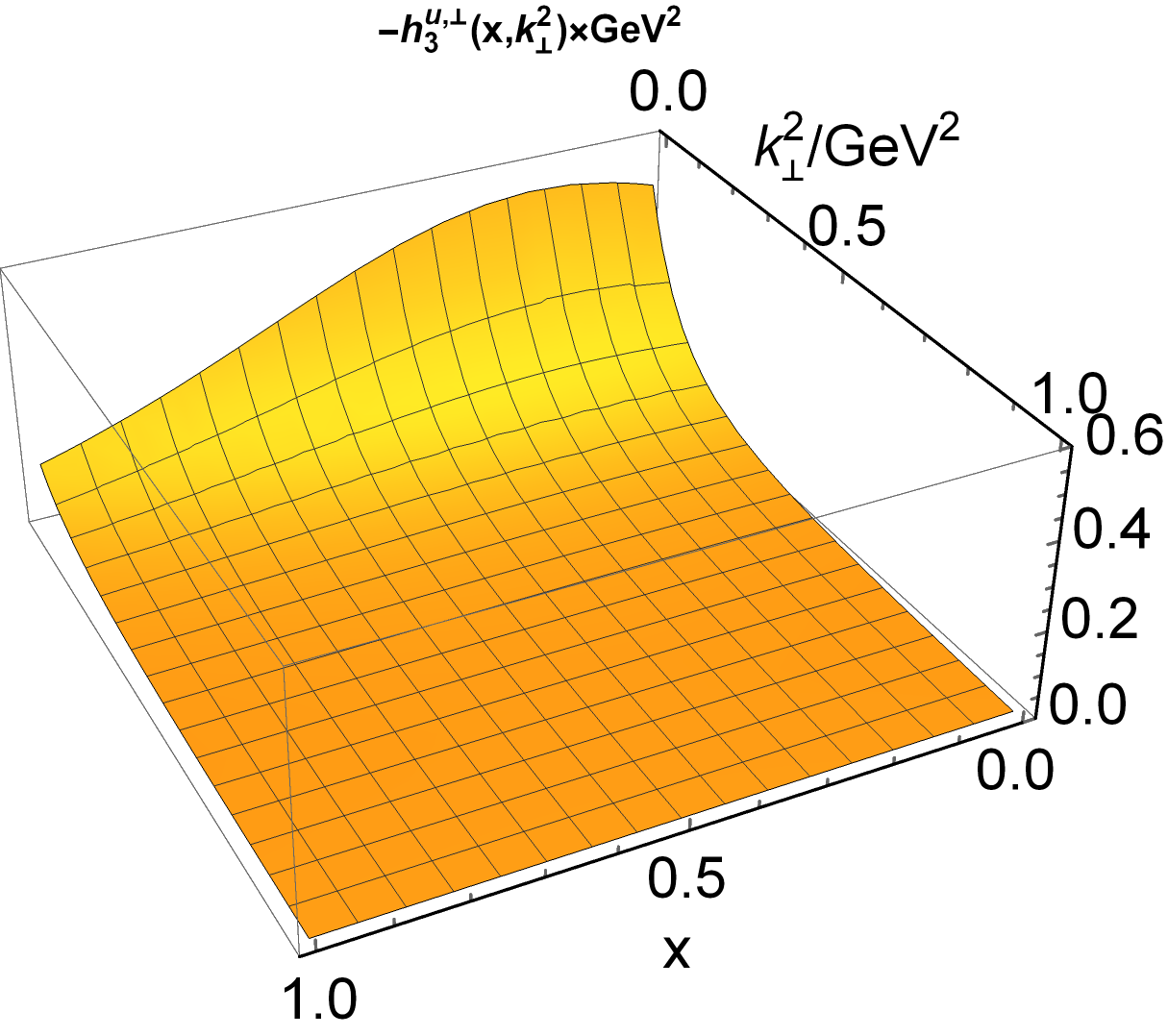}
\qquad
\includegraphics[width=0.47\textwidth]{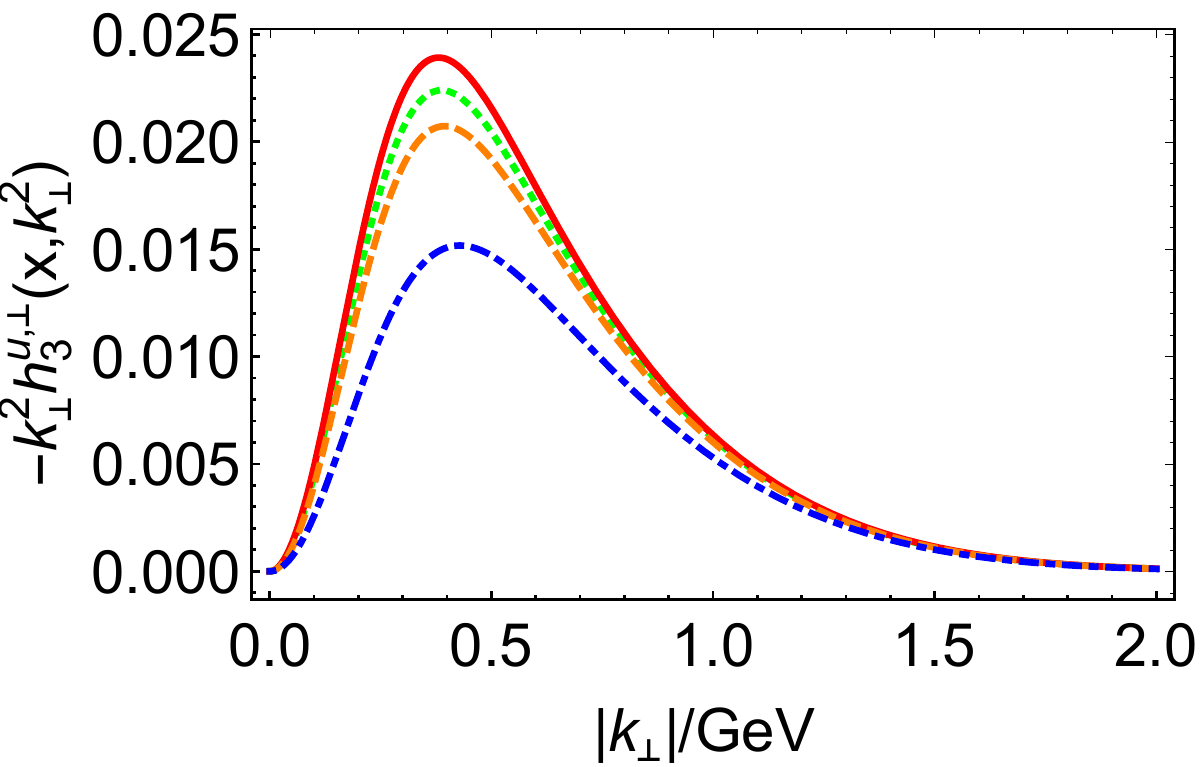}
\caption{\emph{Left panel} -- $-h_3^{u,\perp}\left(x,\bm{k}_{\bot }^2\right)$ kaon $u$ quark twist-two TMD defined in Eq. (\ref{t4tmd2}); \emph{Right panel} -- $-\bm{k}_{\bot}^2 h_3^{u,\perp}\left(x,\bm{k}_{\bot }^2\right)$ at: $x = 0$ - dotted green curve; $x=1/4$ - solid red curve; and $x=1/2$ - dashed orange curve; and $x=3/4$ - dot-dashed blue curve. }\label{tmdp4}
\end{figure}

\subsection{Transverse momentum dependence of TMDs}
Undoubtedly, one of the intriguing aspect of TMDs is their dependence on transverse momentum. This reliance on transverse momentum can be further clarified through $\bm{k}_{\perp}$-weighted moments,
\begin{align}\label{ts}
\langle k_{\perp}^n\rangle_{\alpha}&\equiv \frac{\int_0^1 \mathrm{d}x\int \mathrm{d}^2\bm{k}_{\perp} |\bm{k}_{\perp}|^n \alpha(x,\bm{k}_{\perp}^2) }{\int_0^1 \mathrm{d}x\int \mathrm{d}^2\bm{k}_{\perp} \alpha(x,\bm{k}_{\perp}^2)},
\end{align}
where $n$ represents the order of the moment, $\alpha(x,\bm{k}_{\perp}^2)$ is an arbitrary TMD. The results of $\langle k_{\perp}^n\rangle$ are presented in Table \ref{tsb2}, where we also include the NJL results. For the pion, both $h_1^{\perp}$ and $h_3^{\perp}$ are zero. As for the kaon, the $\langle k_{\perp}^n\rangle$ moments of $h_1^{\perp}$ and $h_3^{\perp}$ for the $u$ quark and $s$ quark are equivalent to those of $g^{\perp}$, therefore they are not listed in the table. In both the NJL model and CI, only quarks are considered as the explicit degrees of freedom, neglecting the presence of gluons in QCD. The chiral odd TMDs exhibit consistent observations, with differing coefficients while maintaining identical distributions; thus, the $\langle k_{\perp}^n\rangle$ moments of the chiral odd GPDs remain the same.

It can be observed from Table \ref{tsb2} that the numerical results in the NJL model are smaller than those in the CI. In the NJL model, the typical transverse momenta of the pion are $\langle k_{\perp}\rangle_{\alpha}\approx 0.36-0.42$ GeV and $\langle k_{\perp}^2\rangle_{\alpha}\approx 0.18-0.23$ GeV, while in the CI model, they are $\langle k_{\perp}\rangle_{\alpha}\approx 0.4-0.5$ GeV and $\langle k_{\perp}^2\rangle_{\alpha}\approx 0.25-0.38$ GeV. For the $K$ meson, in the NJL model, the typical transverse momenta are $\langle k_{\perp}\rangle_{\alpha}\approx 0.34-0.52$ GeV and $\langle k_{\perp}^2\rangle_{\alpha}\approx 0.16-0.33$ GeV, whereas in the CI model, they are $\langle k_{\perp}\rangle_{\alpha}\simeq 0.38\sim 0.53$ GeV and $\langle k_{\perp}^2\rangle_{\alpha}\simeq0.21\sim 0.39$ GeV.

\begin{center}
\begin{table}
\caption{The $\langle k_{\perp}\rangle$ and $\langle k_{\perp}^2\rangle$ moments of the various TMDs defined by Eq. (\ref{ts}), in units of GeV and GeV$^2$ respectively.}\label{tsb2}
\begin{tabular}{p{1.5cm}p{0.8cm} p{0.8cm} p{0.8cm}p{0.8cm}p{0.8cm}p{0.8cm}p{0.8cm}p{0.8cm}p{0.8cm}p{0.8cm}p{0.8cm}}
\hline\hline
&$\langle k_{\perp}^n\rangle$&$f^u$&$f^s$&$e^u$ &$e^s$&$f^{u,\perp}$&$f^{s,\perp}$&$g^{u,\perp}$&$g^{s,\perp}$&$f_3^{u}$&$f_3^{s}$\\
\hline
NJL ($\pi$)&$\langle k_{\perp}\rangle$&0.415&$-$&0.367&$-$&0.367&$-$&0.367&$-$&0.367&$-$\\
\hline
NJL ($\pi$)&$\langle k_{\perp}^2\rangle$&0.231&$-$&0.183&$-$&0.183&$-$&0.183&$-$&0.183&$-$\\
\hline
DSE ($\pi$)&$\langle k_{\perp}\rangle$&0.516&$-$&0.430&$-$&0.409&$-$&0.409&$-$&0.509&$-$\\
\hline
DSE ($\pi$)&$\langle k_{\perp}^2\rangle$&0.371&$-$&0.266&$-$&0.241&$-$&0.241&$-$&0.363&$-$\\
\hline
NJL ($K$) &$\langle k_{\perp}\rangle$&0.421&0.421&0.346&0.401&0.374&0.389&0.380&0.380&0.381&0.515\\
\hline
NJL ($K$)&$\langle k_{\perp}^2\rangle$&0.236&0.236&0.162&0.217&0.190&0.204&0.196&0.196&0.197&0.331\\
\hline
DSE ($K$)&$\langle k_{\perp}\rangle$&0.518&0.523&0.384&0.477&0.443&0.352&0.422&0.422&0.468&0.532\\
\hline
DSE ($K$)&$\langle k_{\perp}^2\rangle$&0.374&0.378&0.210&0.323&0.282&0.167&0.255&0.255&0.312&0.391\\
\hline\hline
\end{tabular}
\end{table}
\end{center}

\section{Wigner distribution}\label{good3}
\subsection{Twist-two Wigner distributions}
The five dimensional Wigner distributions are functions of impact parameter, longitudinal momentum fraction and transverse momentum of partons. The Wigner distributions are defined as~\cite{Ahmady:2020ynt,Kaur:2019kpi}
\begin{align}\label{ff}
\mathcal{X}^{[\Gamma]}\left(x,\bm{k}_{\perp}, \bm{b}_{\bot }\right)=\int \frac{d^2\bm{\Delta}_{\perp}}{(2 \pi )^2}e^{- i\bm{b}_{\perp}\cdot \bm{\Delta }_{\perp}}X^{[\Gamma]}\left(x,\bm{k}_{\perp}, \bm{\Delta}_{\bot }\right),
\end{align}
where $\Gamma$ represents twist-two vertexs $(i\gamma\cdot n, i\gamma\cdot n\gamma^5, i\sigma_{j\mu}n_{\mu})$, $X\left(x, \bm{\Delta}_{\bot },\bm{k}_{\perp}\right)$ represents $F_1\left(x,\bm{k}_{\perp}^2,0,t\right)$, $\tilde{G }_1\left(x,\bm{k}_{\perp}^2,0,t\right)$, $H_1^{\Delta}(x,\bm{k}_{\perp}^2,0,t)$, $H_1^{k}(x,\bm{k}_{\perp}^2,0,t)$. $\mathcal{X}^{[\Gamma]}\left(x,\bm{k}_{\perp}, \bm{b}_{\bot }\right)$ represents $\mathcal{F}_1\left(x,\bm{k}_{\perp}, \bm{b}_{\bot }\right)$, $\mathcal{G}_1\left(x,\bm{k}_{\perp}, \bm{b}_{\bot }\right)$, $\mathcal{H}_1^k\left(x,\bm{k}_{\perp}, \bm{b}_{\bot }\right)$, $\mathcal{H}_1^{\Delta}\left(x,\bm{k}_{\perp}, \bm{b}_{\bot }\right)$, respectively. The twist-two Wigner distributions of spin-$0$ hadron are defined as
\begin{align}\label{ff}
\rho_{UU}\left(x,\bm{k}_{\perp}, \bm{b}_{\bot }\right)&=\mathcal{F}_1\left(x,\bm{k}_{\perp}, \bm{b}_{\bot }\right)\nonumber\\
\rho_{UL}\left(x,\bm{k}_{\perp}, \bm{b}_{\bot }\right)&=\frac{\varepsilon_{\bot}^{ij}}{m_H^2}\bm{k}_{\bot}^i\frac{\partial }{\partial \bm{b}_{\bot}^j}\mathcal{G}_1\left(x,\bm{k}_{\perp},\bm{b}_{\perp}\right)\nonumber\\
\rho_{UT}\left(x,\bm{k}_{\perp}, \bm{b}_{\bot }\right)&=\frac{\varepsilon_{\bot}^{ij}}{2m_H}\bm{k}_{\bot}^i\mathcal{H}_1^k\left(x,\bm{k}_{\perp}, \bm{b}_{\bot }\right)+\frac{\varepsilon_{\bot}^{ij}}{2m_H}\frac{\partial }{\partial \bm{b}_{\bot}^j}\mathcal{H}_1^{\Delta}\left(x,\bm{k}_{\perp}, \bm{b}_{\bot }\right),
\end{align}
where $m_H$ is the hadron mass. One then can classify the unpolarized, longitudinally polarized and transversely polarized Wigner distributions in kaon as: $\rho_{UU}\left(x,\bm{k}_{\perp}, \bm{b}_{\bot }\right)=\rho^{i\gamma\cdot n}\left(x,\bm{k}_{\perp}, \bm{b}_{\bot }\right)$, $\rho_{UL}\left(x,\bm{k}_{\perp}, \bm{b}_{\bot }\right)=\rho^{i\gamma\cdot n\gamma_5}\left(x,\bm{k}_{\perp}, \bm{b}_{\bot }\right)$ and $\rho_{UT}\left(x,\bm{k}_{\perp}, \bm{b}_{\bot }\right)=\rho^{i\sigma_{j\mu}n_{\mu}}\left(x,\bm{k}_{\perp}, \bm{b}_{\bot }\right)$.

In our model,we can obtain
\begin{align}\label{wigner1}
&\rho_{UU}^u(x,\bm{k}_{\perp},\bm{b}_{\perp})\nonumber\\
=&\frac{\bar{N}_{FF}N_c}{2\pi^3} \int \frac{d^2\bm{\Delta}_{\perp}}{(2 \pi )^2}x(1-x)e^{- i\bm{b}_{\perp}\cdot \bm{\Delta }_{\perp}} \frac{\bar{\mathcal{C}}_2(\sigma_1^{r,0})}{\sigma_1^{r,0}}+ \frac{N^{EE}N_c }{2\pi^3}  \int \frac{d^2\bm{\Delta}_{\perp}}{(2 \pi )^2}e^{- i\bm{b}_{\perp}\cdot \bm{\Delta }_{\perp}}\frac{\bar{\mathcal{C}}_2(\sigma_1^{r,0})}{\sigma_1^{r,0}} \nonumber\\
+&\frac{\bar{N}^{EF}N_c}{2\pi^3}  \int \frac{d^2\bm{\Delta}_{\perp}}{(2 \pi )^2}e^{- i\bm{b}_{\perp}\cdot \bm{\Delta }_{\perp}}(2x-1) \frac{\bar{\mathcal{C}}_2(\sigma_1^{r,0})}{\sigma_1^{r,0}}\nonumber\\
-& \frac{N^{EF}N_c}{8\pi^3}\int \frac{d^2\bm{\Delta}_{\perp}}{(2 \pi )^2}\int _0^{1-x}d\alpha e^{-i\bm{b}_{\perp}\cdot \bm{\Delta }_{\perp}}(\bm{\Delta}_{\perp}^2-x(2(m_K^2-(M_u-M_s)^2)+\bm{\Delta}_{\perp}^2))\frac{6\bar{\mathcal{C}}_3(\sigma_4^r)}{[\sigma_4^r]^2} ,
\end{align}
\begin{align}\label{wigner2}
\rho_{UL}^u\left(x,\bm{k}_{\perp},\bm{b}_{\perp}\right)=\varepsilon_{\bot}^{ij}\bm{k}_{\bot}^i\frac{\partial }{\partial \bm{b}_{\bot}^j}\frac{N^{EF}N_c}{4\pi ^3}\int_0^{1-x}  d\alpha\int \frac{d^2\bm{\Delta}_{\perp}}{(2 \pi )^2} e^{- i\bm{b}_{\perp}\cdot \bm{\Delta }_{\perp}}  \frac{6\bar{\mathcal{C}}_3(\sigma_4^r)}{[\sigma_4^r]^2}  ,
\end{align}
\begin{align}\label{wigner3}
&\rho_{UT}^u\left(x,\bm{k}_{\perp},\bm{b}_{\perp}\right)\nonumber\\
=&\varepsilon_{\bot}^{ij}\bm{k}_{\bot}^i\frac{N^{EF}N_c}{4\pi^3}\int_0^{1-x}  d\alpha\int \frac{d^2\bm{\Delta}_{\perp}}{(2 \pi )^2}e^{- i\bm{b}_{\perp}\cdot \bm{\Delta }_{\perp}}  (M_s-M_u)  \frac{6\bar{\mathcal{C}}_3(\sigma_4^r)}{[\sigma_4^r]^2}\nonumber\\
+&\varepsilon_{\bot}^{ij}\frac{\partial }{\partial \bm{b}_{\bot}^i}\frac{N^{EF}N_c}{4\pi^3}\int_0^{1-x}  d\alpha\int \frac{d^2\bm{\Delta}_{\perp}}{(2 \pi )^2}e^{- i\bm{b}_{\perp}\cdot \bm{\Delta }_{\perp}}  (M_u+(M_s-M_u)\alpha)  \frac{6\bar{\mathcal{C}}_3(\sigma_4^r)}{[\sigma_4^r]^2},
\end{align}
take the first line of Eq. (\ref{wigner1}) as an example,
\begin{align}
\frac{\bar{N}_{FF}N_c}{2\pi ^3}\int \frac{d^2\bm{\Delta}_{\perp}}{(2 \pi )^2} x(1-x) e^{-i\bm{b}_{\perp}\cdot \bm{\Delta}_{\perp}} \frac{\bar{\mathcal{C}}_2(\sigma_1^{r,0})}{\sigma_1^{r,0}} = \frac{\bar{N}_{FF}N_c}{2\pi ^3} \delta^2(\bm{b}_{\bot}) \frac{\bar{\mathcal{C}}_2(\sigma_1^{r,0})}{\sigma_1^{r,0}} ,
\end{align}
the $\delta^2 (\bm{b}_{\bot})$ component - first two lines of Eq. (\ref{wigner1}) - is suppressed.

\begin{figure}
\centering
\includegraphics[width=0.47\textwidth]{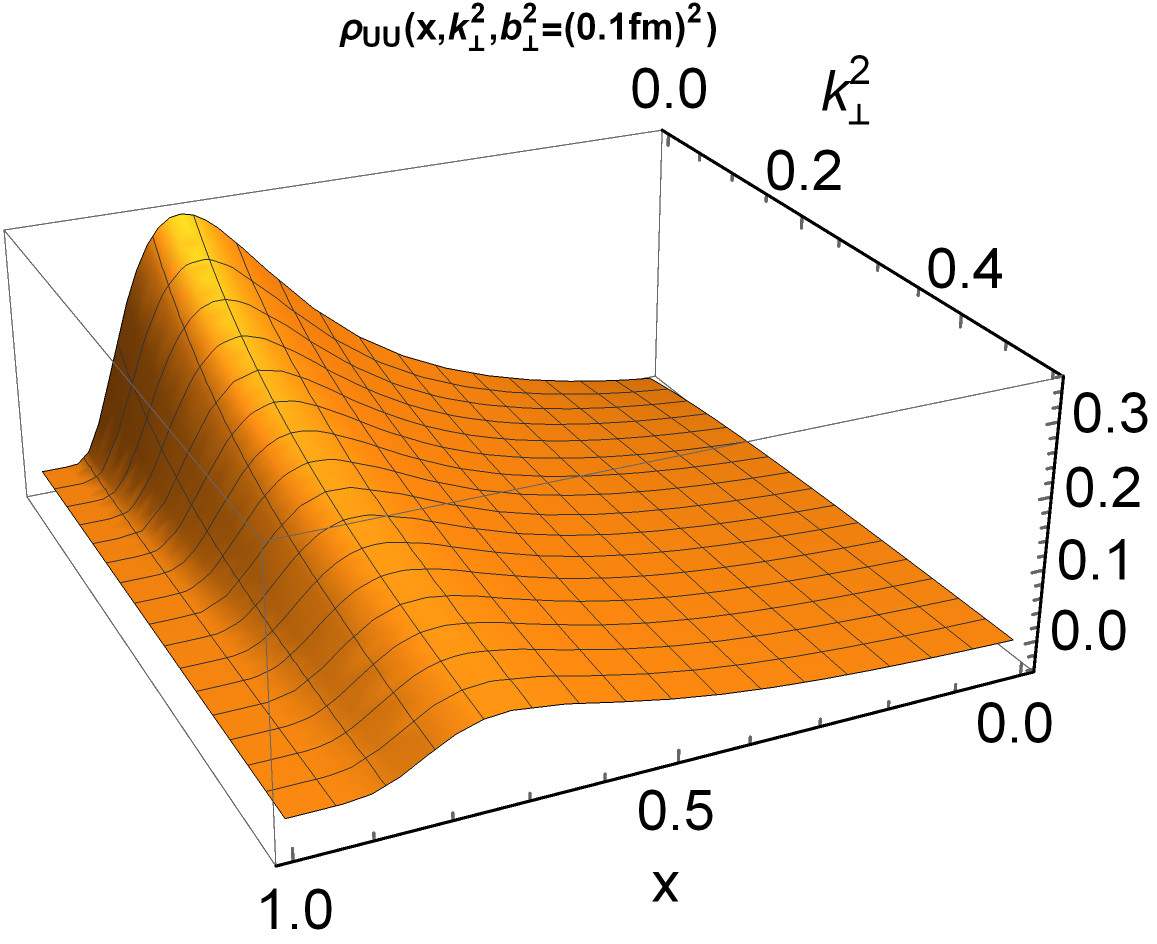}
\qquad
\includegraphics[width=0.47\textwidth]{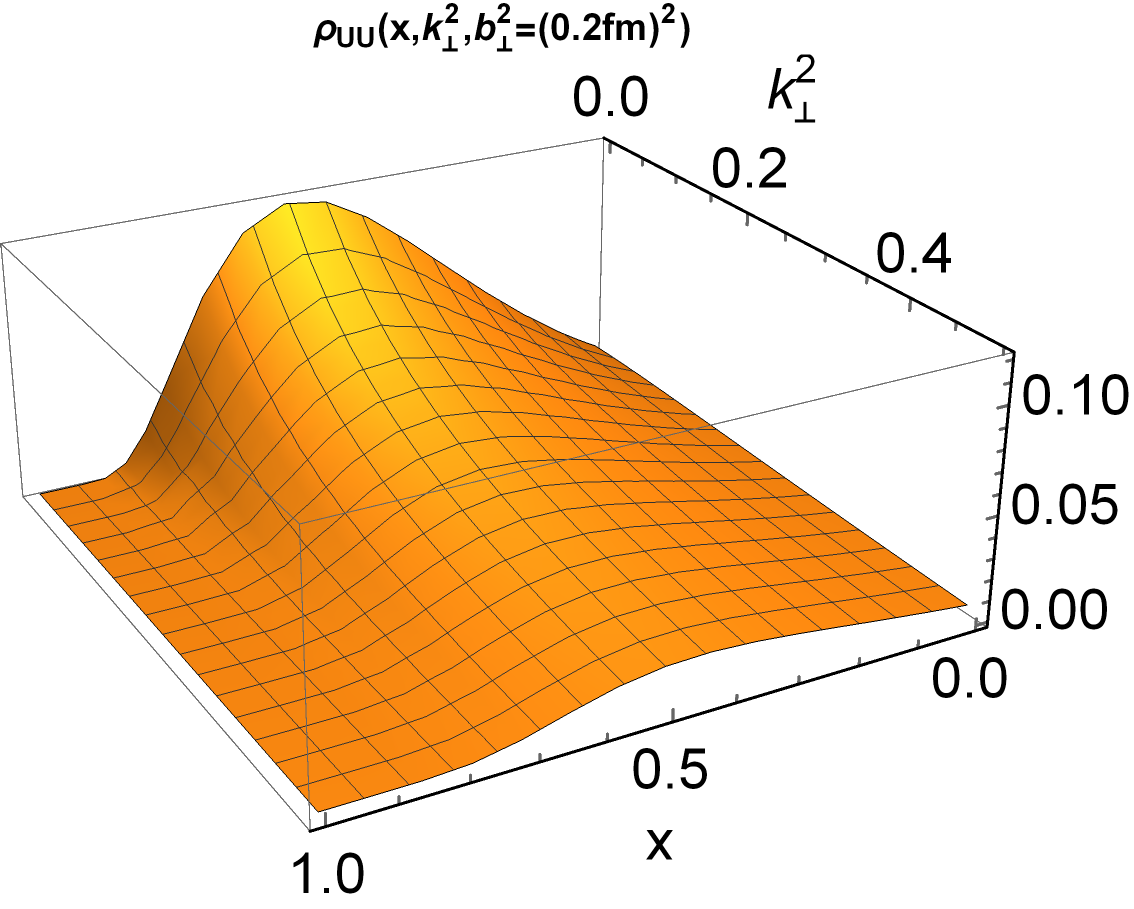}
\caption{The unpolarized Wigner distribution $\rho_{UU}$ at $|\bm{b}_{\perp}|=0.1 $ fm, and $|\bm{b}_{\perp}|=0.2 $ fm. The $\delta^2 (\bm{b}_{\bot})$ component - first two lines of Eq. (\ref{wigner1}) - is suppressed in the image. }\label{wigd}
\end{figure}

These functions are non-zero for $x\in[0,1]$. The unpolarized Wigner distribution is depicted in Fig. \ref{wigd} at different values of $|b_{\perp}|$, specifically $0.1$ fm and $0.2$ fm. It represents the valence $u$ quark Wigner function and exhibits the following properties: Firstly, when $|\bm{b}_{\perp}|\simeq 0$, at $\bm{k}_{\perp}=\bm{0}$, the peak shifts to $x \simeq 1$. As the magnitude of $|\bm{b}_{\perp}|$ increases, the peak moves further away from $x=1$. Secondly, similar to the pion Wigner distribution, the kaon Wigner distribution also demonstrates power-law suppression as both $|\bm{k}_{\perp}|$ and/or $|\bm{b}_{\perp}|$ increase. The transversely polarized distribution differs from that of the pion's $u$ quark due to broken isospin symmetry. It includes a term $\varepsilon_{\bot}^{ij}\bm{k}_{\bot}^i$, which will vanish for a pion with equal up and down quark masses $M_u=M_d$.

The flavor decompositions of valence partons in the kaon are interrelated through a specific relation
\begin{align}\label{relation}
\rho^u\left(x,\bm{k}_{\perp},\bm{b}_{\perp},M_u,M_s\right)=-\rho^s\left(-x,-\bm{k}_{\perp},\bm{b}_{\perp},M_s,M_u\right).
\end{align}

To obtain the purely transverse Wigner distributions, we have conducted an integration over $x$
\begin{align}\label{ff}
\rho_{UX}\left(\bm{k}_{\perp},\bm{b}_{\bot }\right)&=\int_0^1 dx \rho_{UX}\left(x,\bm{k}_{\perp}, \bm{b}_{\bot }\right)
\end{align}
where $X$ stands for the polarization of quark inside kaon.
\begin{figure*}
\centering
\includegraphics[width=0.3\textwidth]{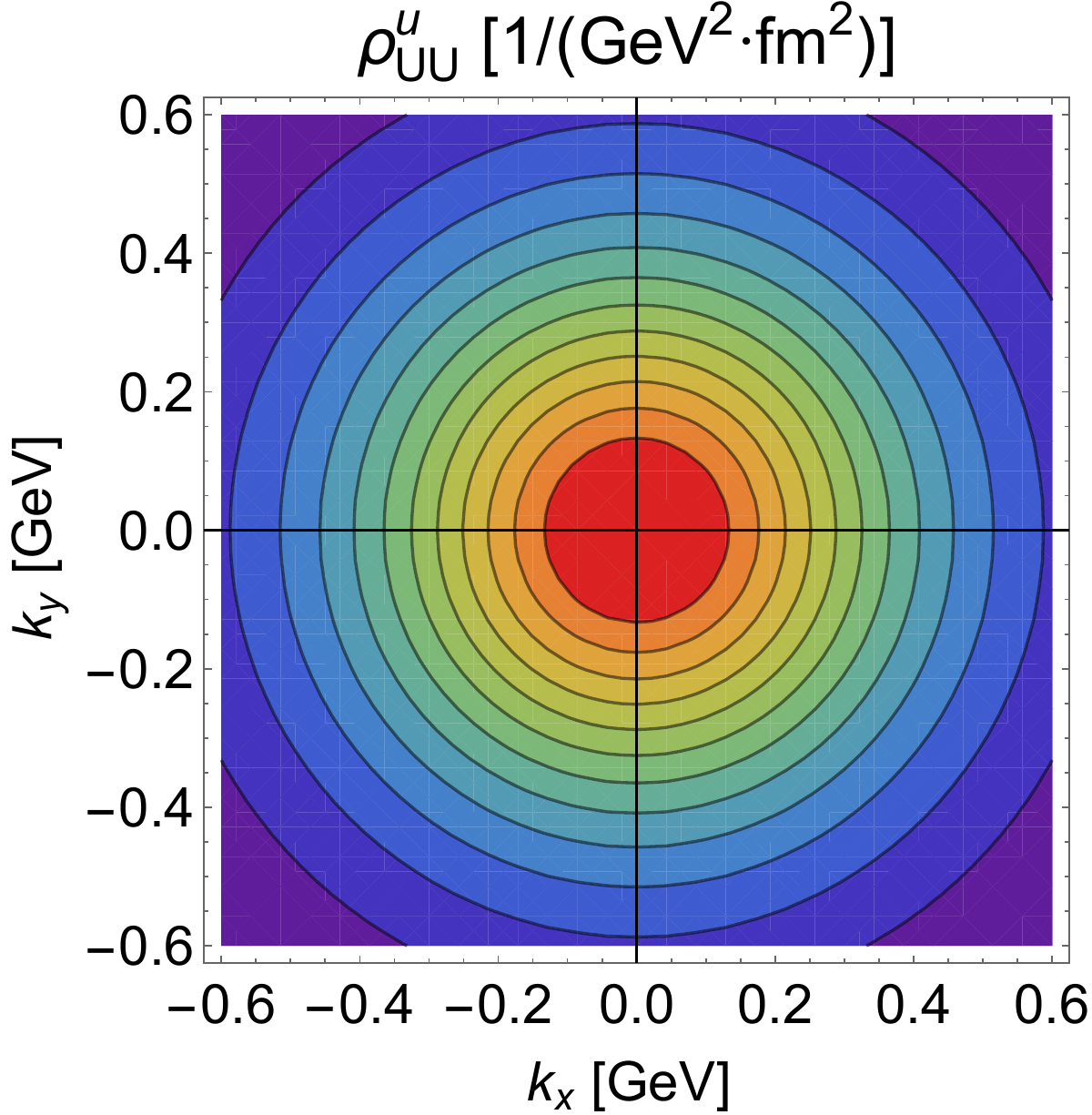}
\qquad
\includegraphics[width=0.045\textwidth]{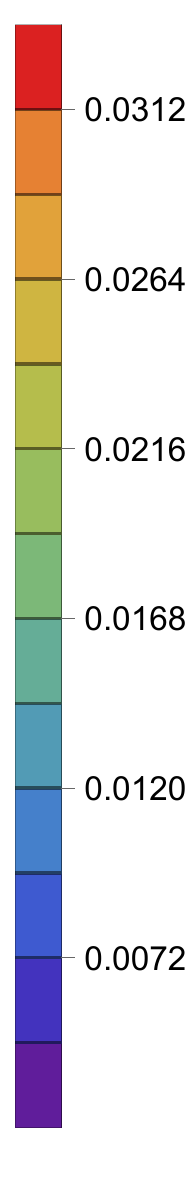}
\qquad
\includegraphics[width=0.3\textwidth]{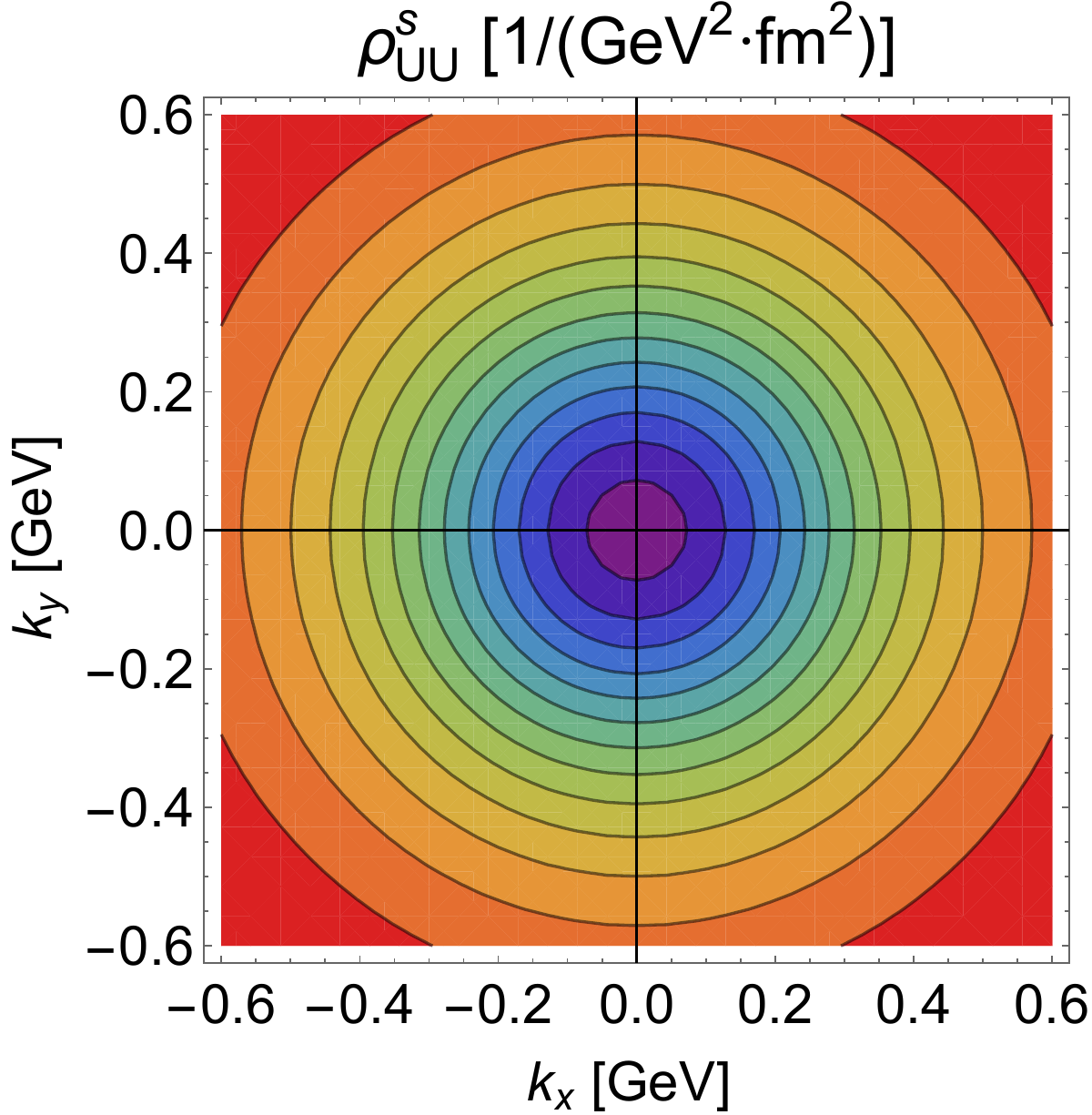}
\qquad
\includegraphics[width=0.05\textwidth]{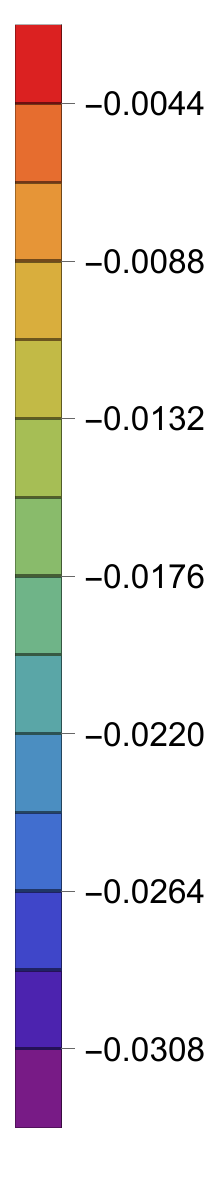}
\qquad
\includegraphics[width=0.3\textwidth]{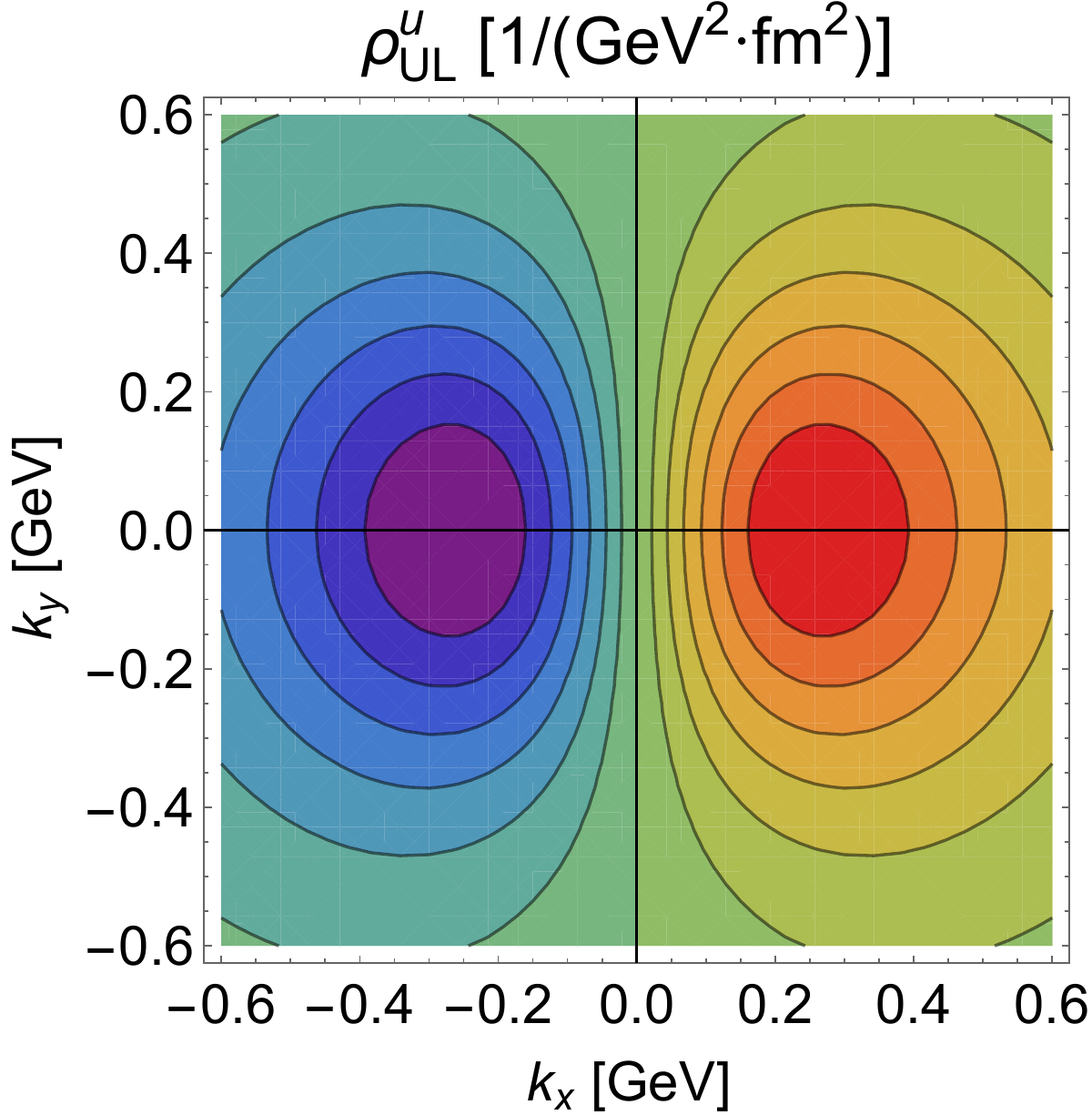}
\qquad
\includegraphics[width=0.05\textwidth]{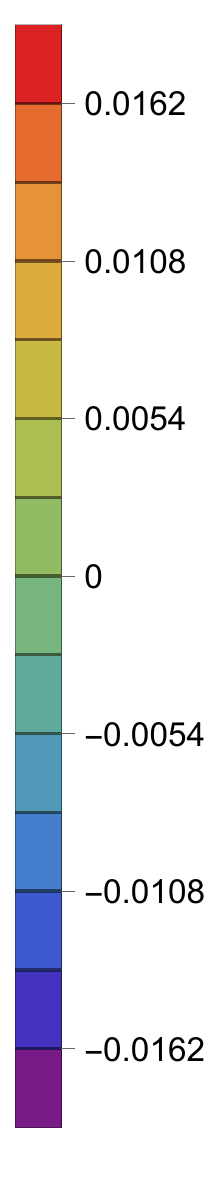}
\qquad
\includegraphics[width=0.3\textwidth]{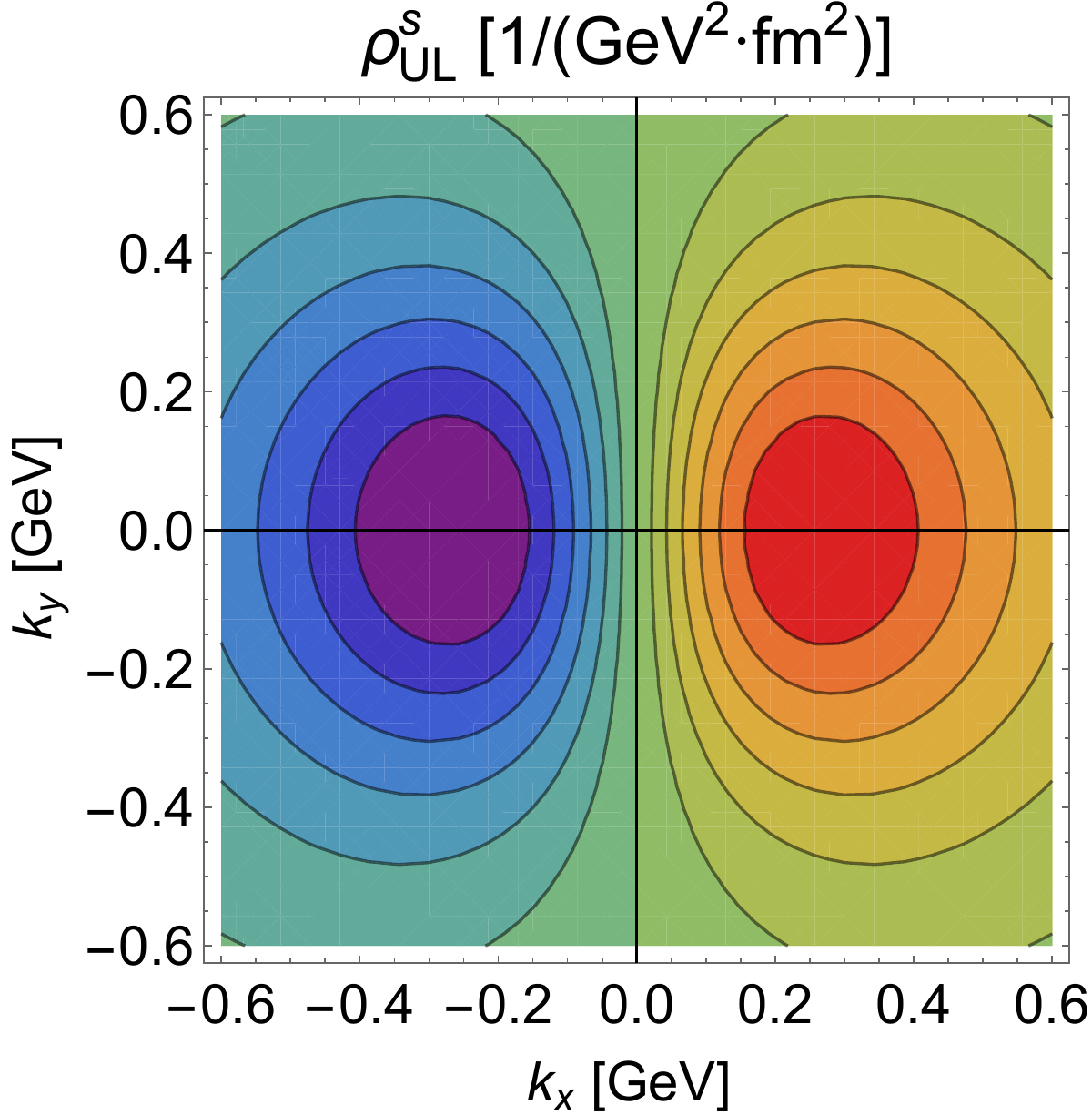}
\qquad
\includegraphics[width=0.045\textwidth]{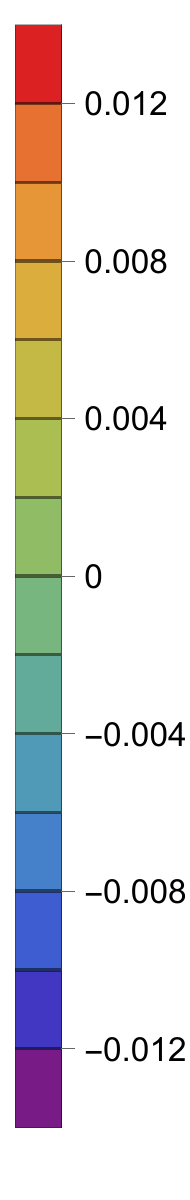}
\qquad
\includegraphics[width=0.3\textwidth]{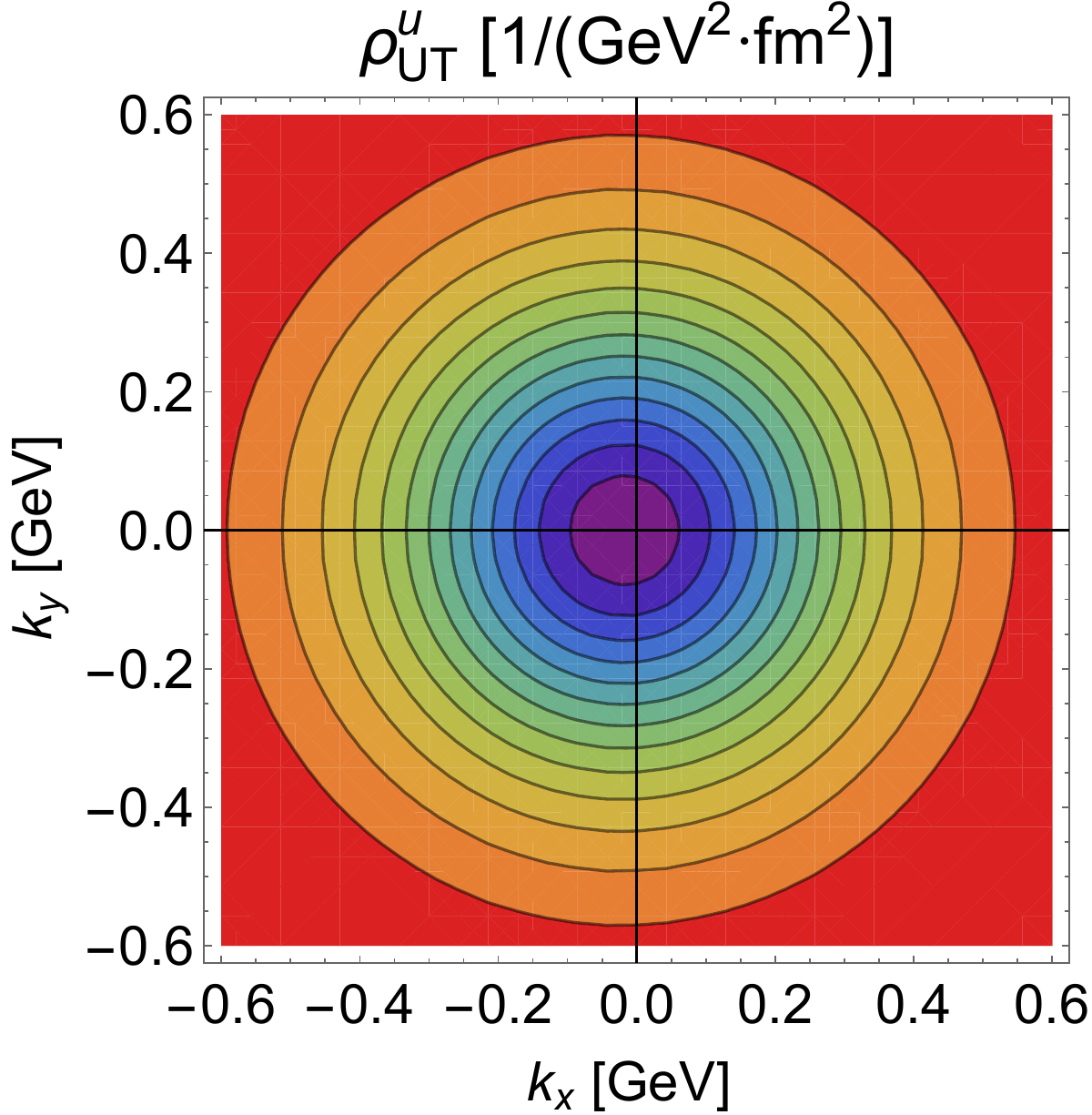}
\qquad
\includegraphics[width=0.05\textwidth]{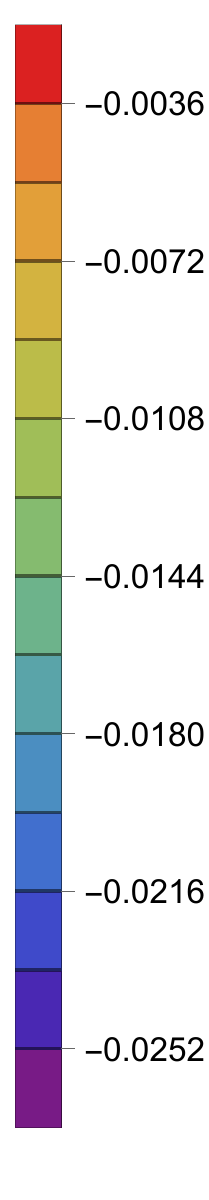}
\qquad
\includegraphics[width=0.3\textwidth]{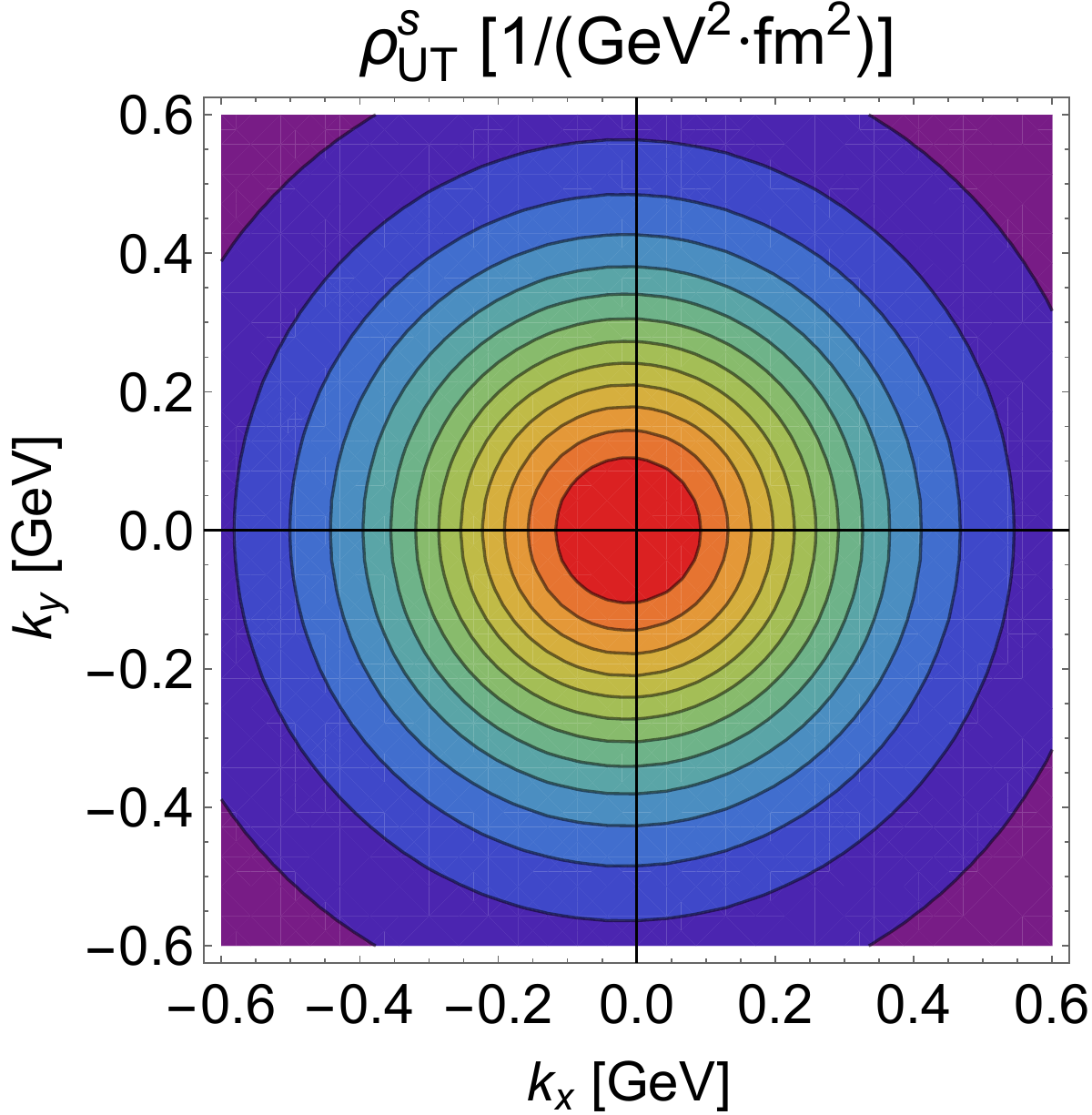}
\qquad
\includegraphics[width=0.045\textwidth]{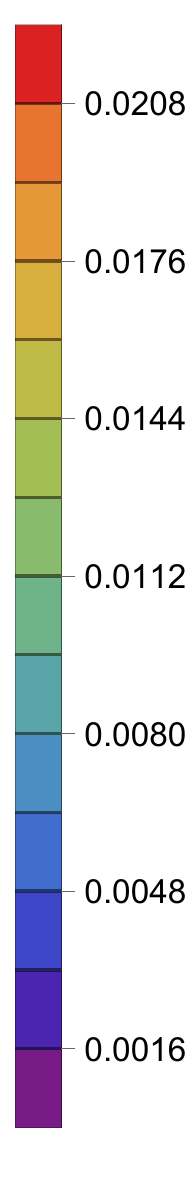}
\caption{The purely Wigner distributions $\rho\left(\bm{b}_{\bot },\bm{k}_{\perp}\right)$ of the unpolarized quark in the kaon (upper panels), the
longitudinal polarized quark (middle panels), and the transversely polarized quark (lower panels). (Left panels) the $u$ quark distributions in the transverse momentum space with $\bm{b}_{\perp}=b_{\bot} \hat{\bm{e}}_y$, $|b_{\bot}|=0.3$ fm. (Right panels) the $s$ quark distributions in the transverse momentum space with $\bm{b}_{\perp}=b_{\bot} \hat{\bm{e}}_y$, $|b_{\bot}|=0.3$ fm. }\label{rho1}
\end{figure*}

In Figs. \ref{rho1}, we present the purely transverse Wigner distributions $\rho_{UU}\left( \bm{k}_{\bot },\bm{b}_{\perp}\right)$ (upper panel), $\rho_{UL}\left(\bm{k}_{\bot },\bm{b}_{\perp}\right)$ (middle panel), $\rho_{UT}\left(\bm{k}_{\bot },\bm{b}_{\perp}\right)$ (lower panel) for the kaon $u$ quark (left panel) and $s$ quark (right panel) in the transverse momentum space with $\bm{b}_{\perp}=b_{\bot} \hat{\bm{e}}_y$, where $|b_{\bot}|=0.3$ fm. In the upper panel, both the unpolarized $u$ quark and $s$ quark distributions exhibit circularly symmetric behavior in transverse momentum space. The longitudinally polarized quark distributions in the middle panel display dipolar distortion patterns, with $\rho_{UL}$ reflecting the spin-orbital correlation. In momentum space, the peaks for both the $u$ quark and $s$ quark are located around $0.2 \quad \text{GeV} <k_x< 0.4 \quad \text{GeV}$, but the center position corresponding to the momentum of the $u$ quark appears to be smaller than that of the $s$ quark. The transversely polarized distributions are depicted in momentum space in the lower panel. For the $u$ quark, it exhibits circular symmetry, but the axis of symmetry no longer resides at $(k_x=0,k_y=0)$; rather, it shifts to $(k_x<0,k_y=0)$. Similarly, for the $s$ quark, it also displays circular symmetry with a shift in the axis of symmetry to $(k_x<0,k_y=0)$. The distribution of polarized quarks inside an unpolarized kaon is described by $\rho_{UT}$. The direction of polarization of the quark is also related to the transverse momentum of the quark, as can be clearly seen from the expression of $\rho_{UT}$ given in Eq. (\ref{wigner3}). This is quite different from the pion's $\rho_{UT}$. For the pion, the first term in Eq. (\ref{wigner3}) is zero, so the direction of polarization of the quark is not related to the transverse momentum of the quark. Therefore, we can consider the direction of transverse momentum along any axis for quark polarization along the $x$-axis.

\begin{figure*}
\centering
\includegraphics[width=0.3\textwidth]{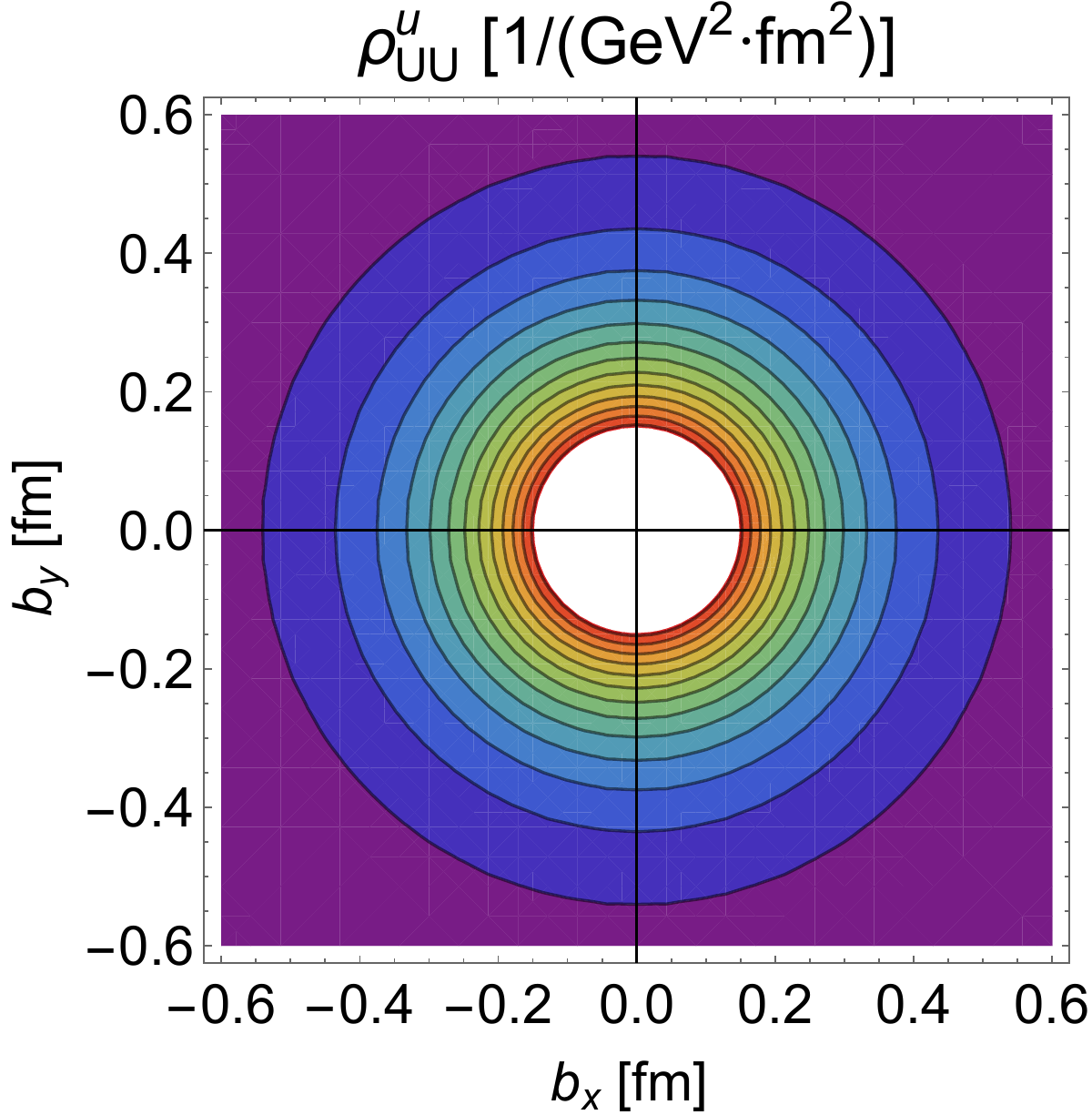}
\qquad
\includegraphics[width=0.045\textwidth]{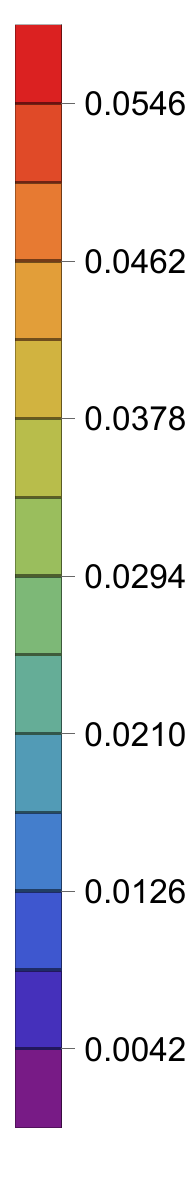}
\qquad
\includegraphics[width=0.3\textwidth]{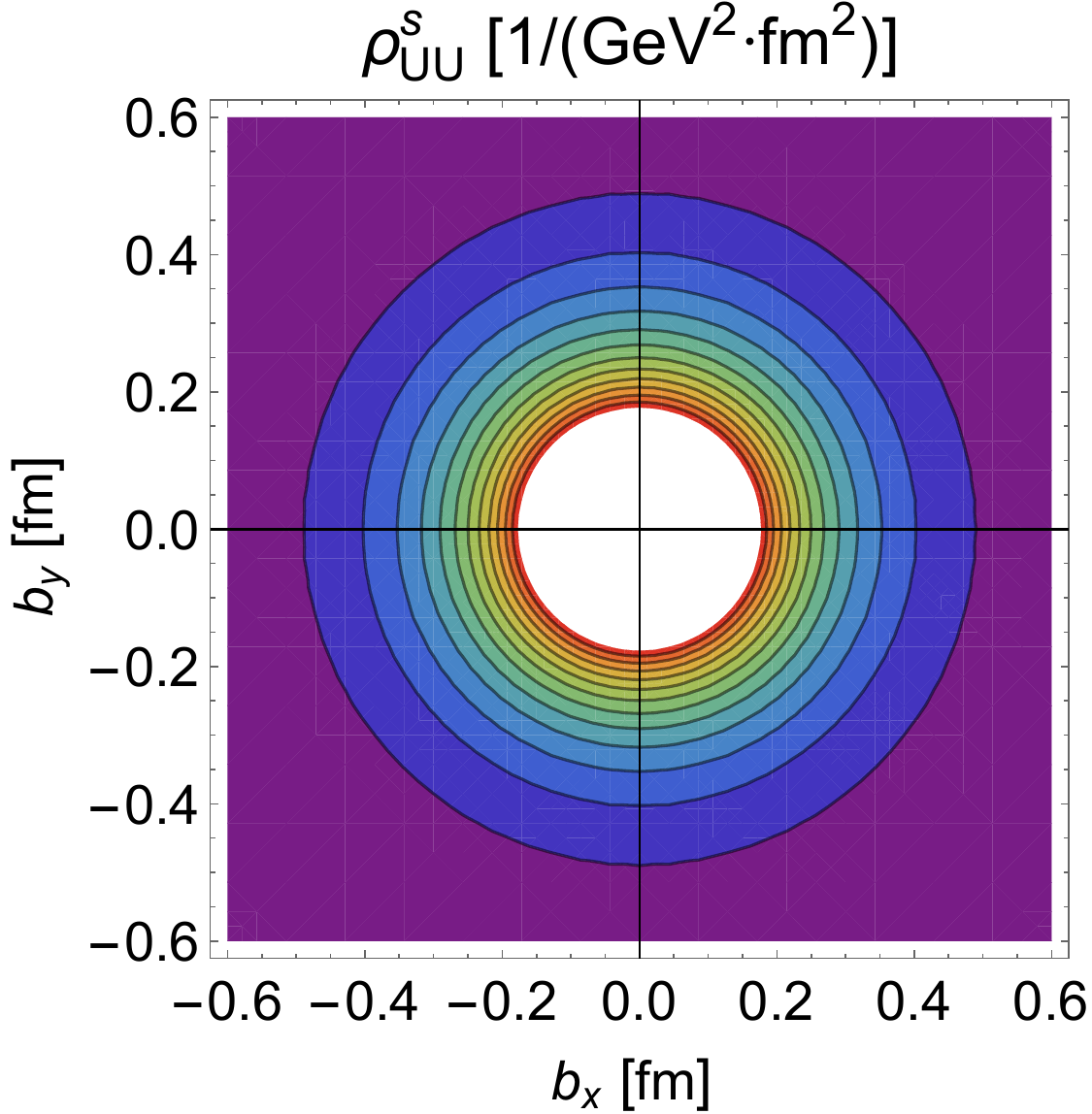}
\qquad
\includegraphics[width=0.04\textwidth]{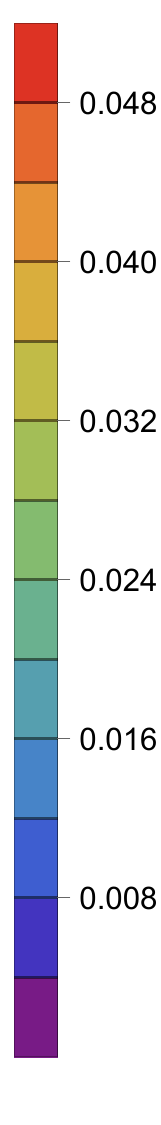}
\qquad
\includegraphics[width=0.3\textwidth]{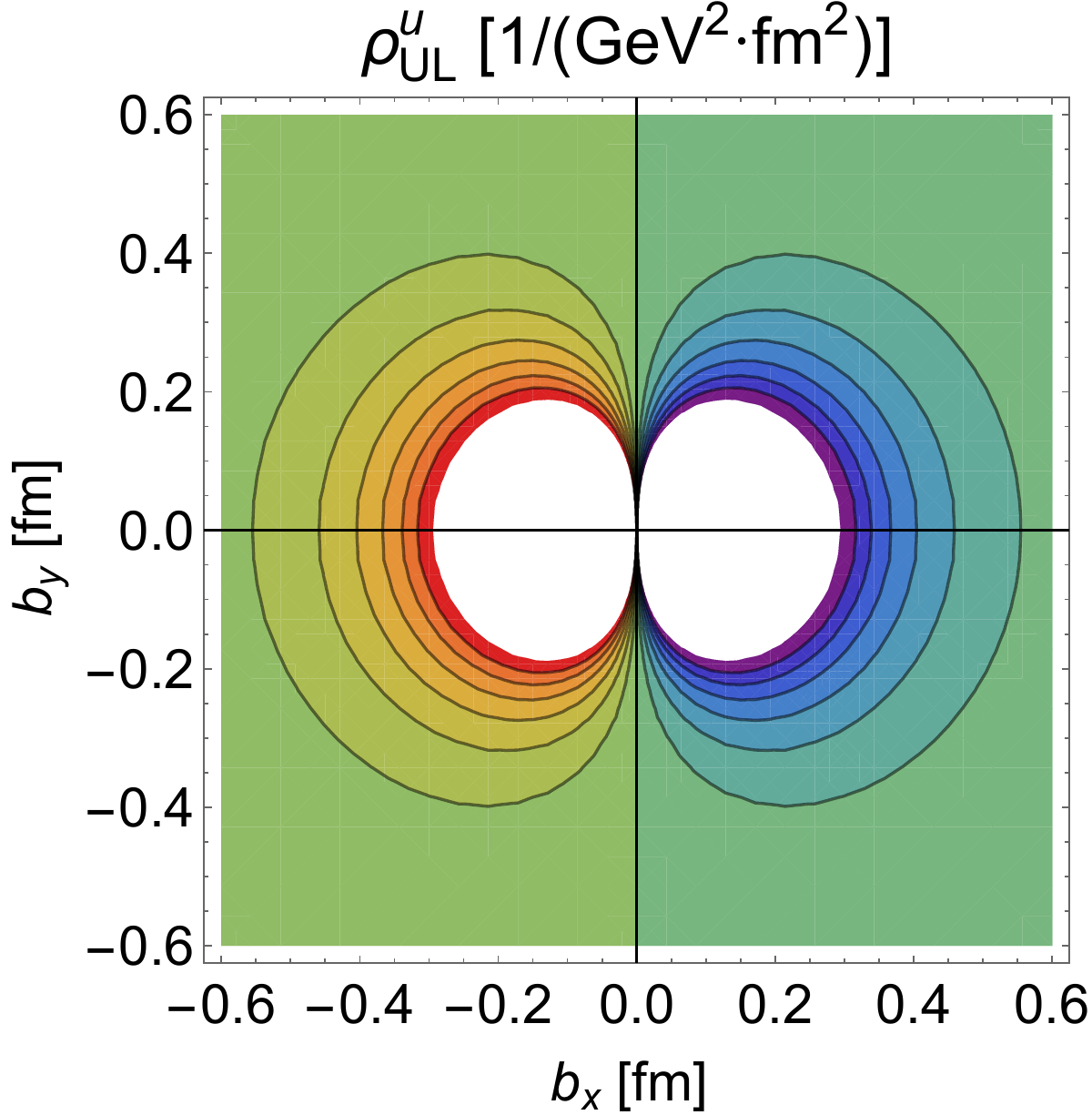}
\qquad
\includegraphics[width=0.06\textwidth]{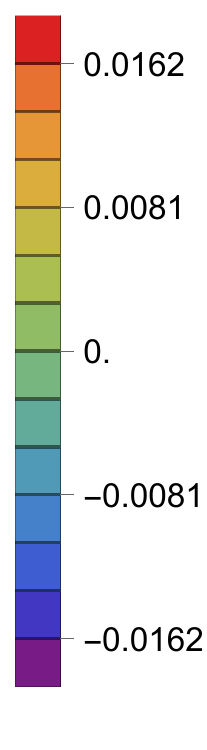}
\qquad
\includegraphics[width=0.3\textwidth]{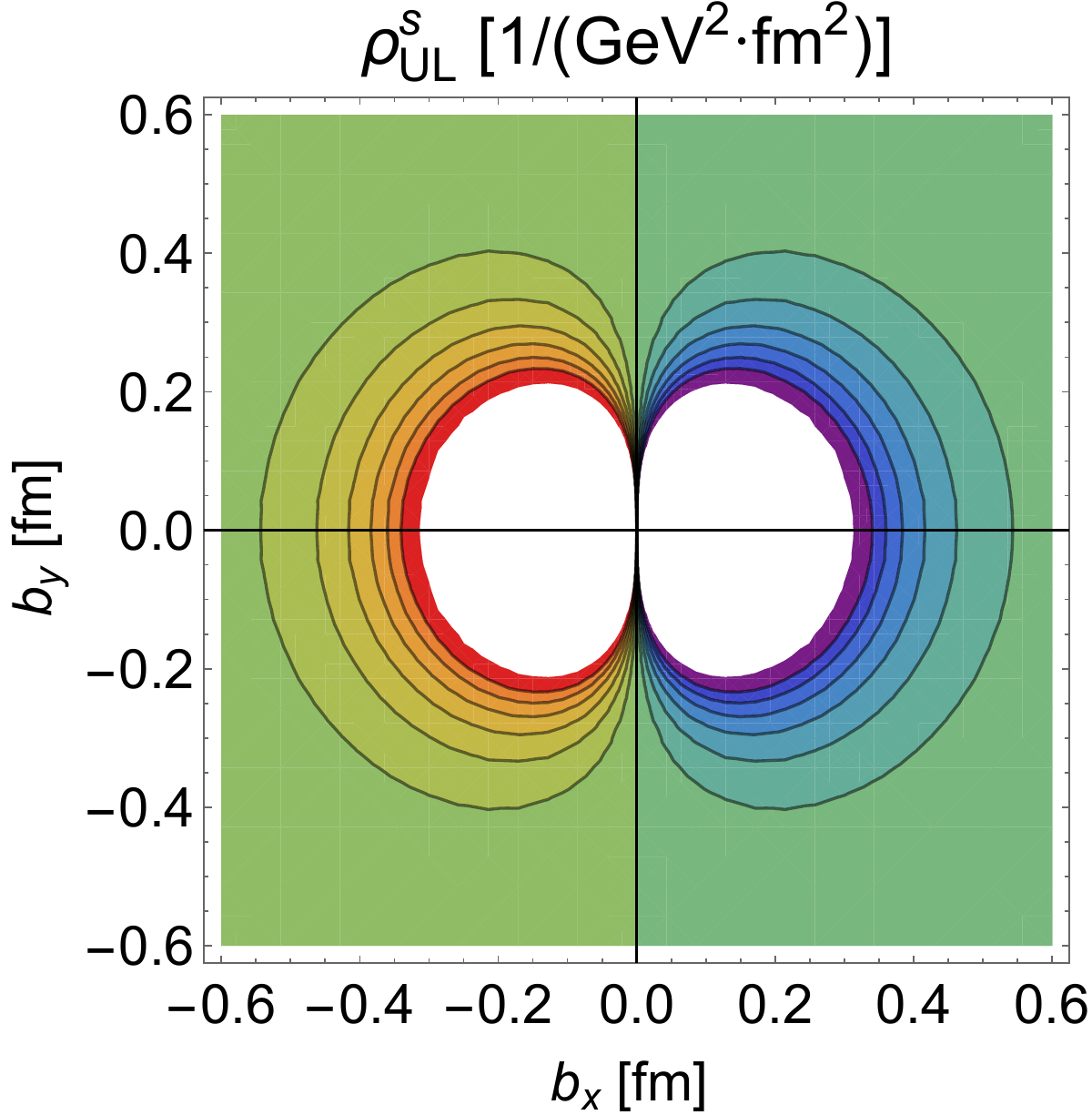}
\qquad
\includegraphics[width=0.06\textwidth]{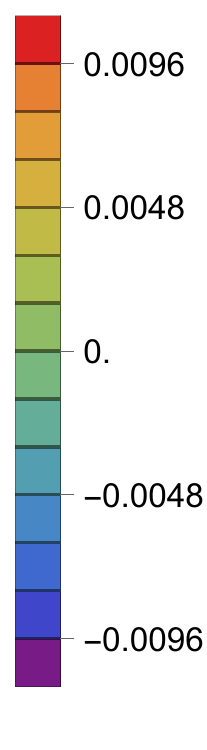}
\qquad
\includegraphics[width=0.3\textwidth]{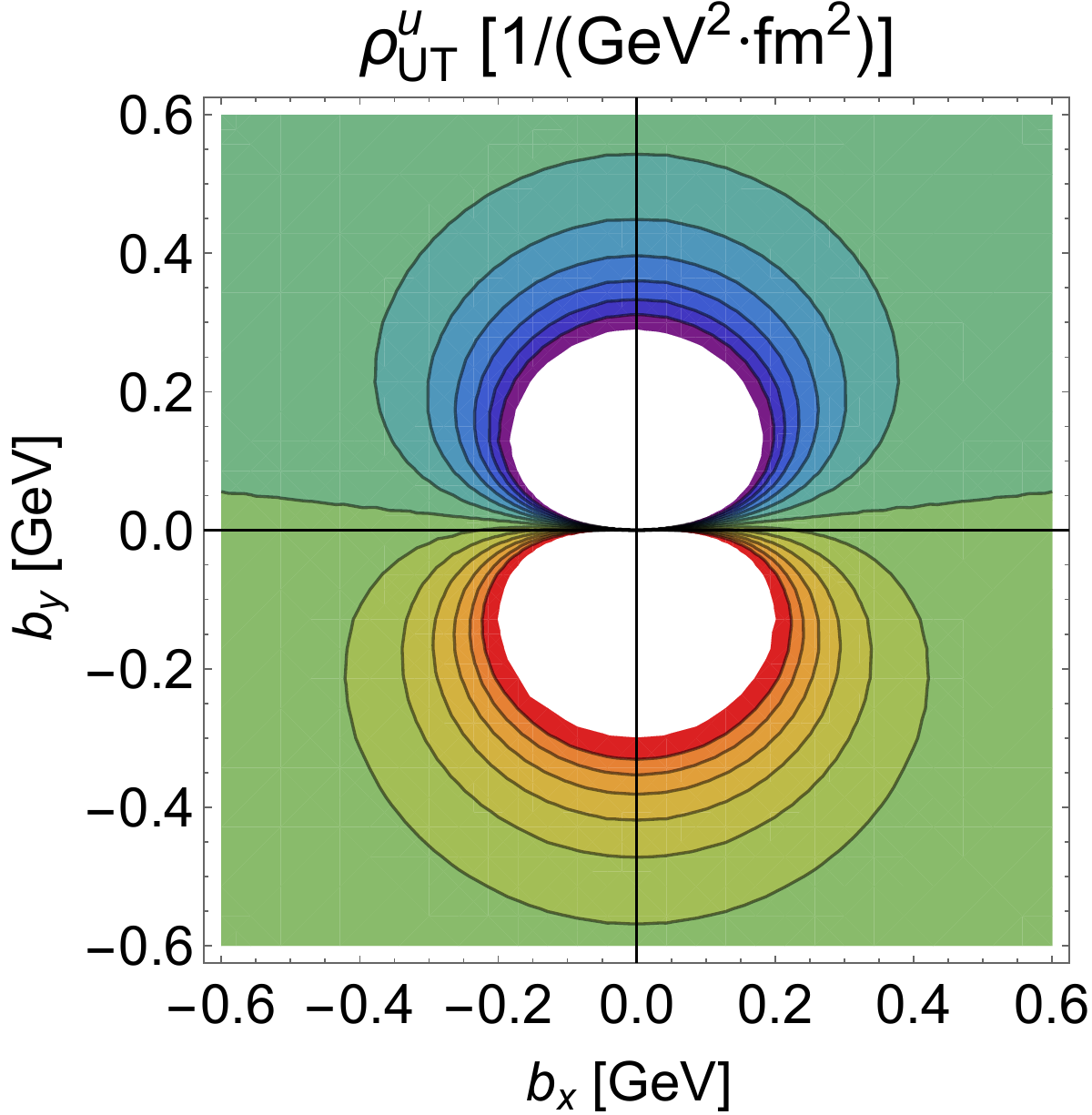}
\qquad
\includegraphics[width=0.06\textwidth]{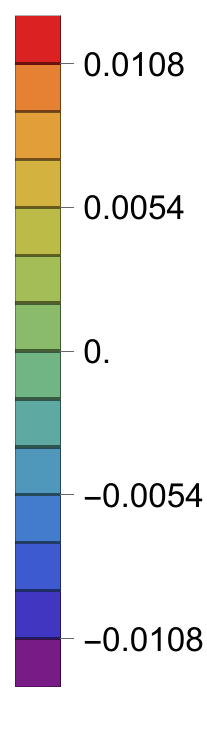}
\qquad
\includegraphics[width=0.3\textwidth]{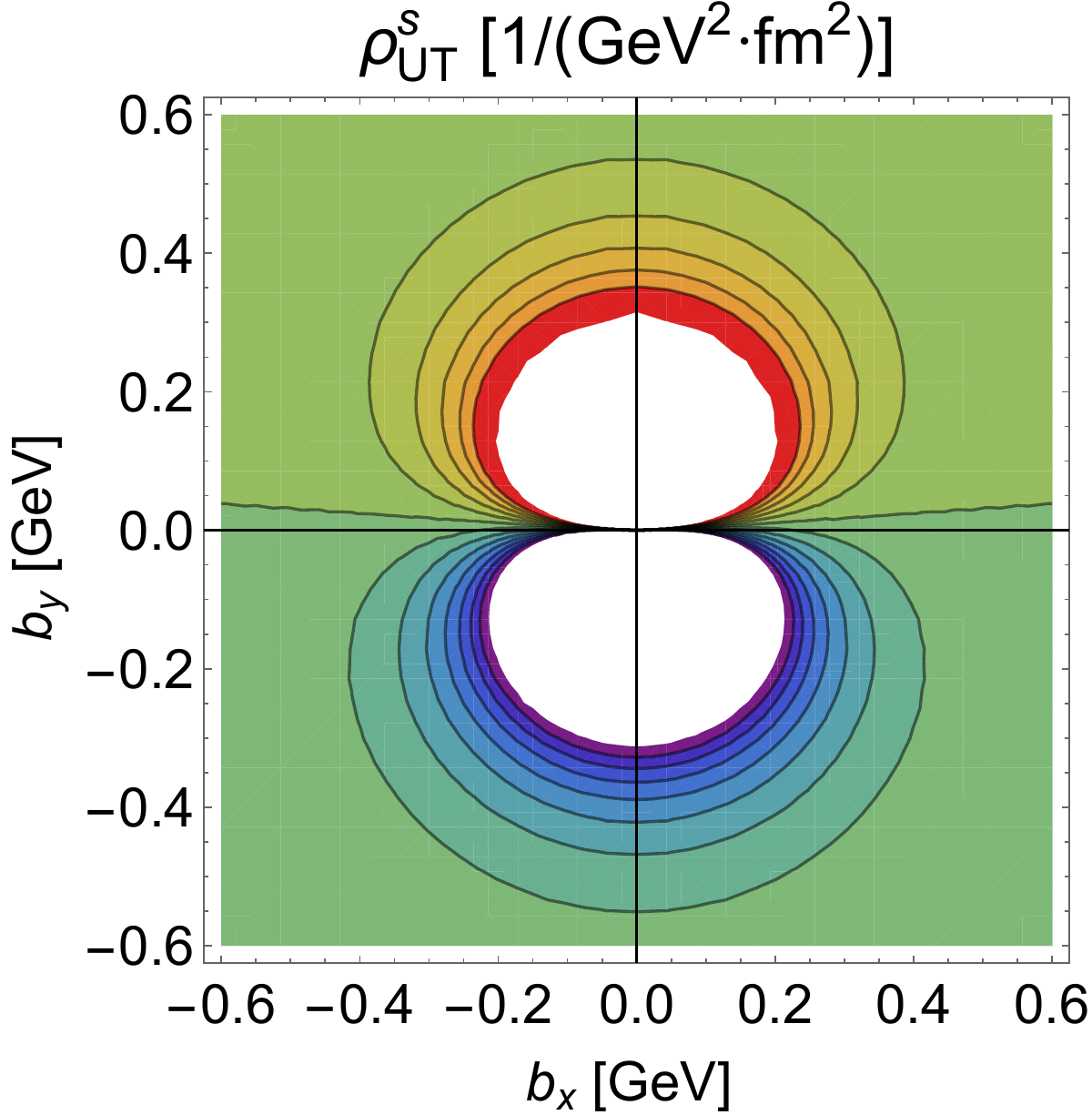}
\qquad
\includegraphics[width=0.06\textwidth]{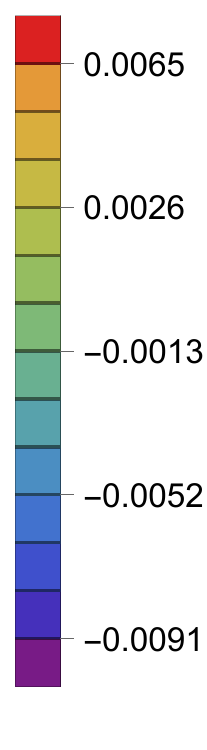}
\caption{The purely Wigner distributions $\rho\left(\bm{b}_{\bot },\bm{k}_{\perp}\right)$ of the unpolarized $u$ quark in the kaon (upper panels), the
longitudinal polarized quark (middle panels), and the transversely polarized quark (lower panels) of pion. (Left panels) the $u$ quark distributions in the impact parameter space with $\bm{k}_{\perp}=k_{\bot} \hat{\bm{e}}_y$, $|k_{\bot}|=0.3$ GeV. (Right panels) the $s$ quark distributions in the impact parameter space with $\bm{k}_{\perp}=k_{\bot} \hat{\bm{e}}_y$, $|k_{\bot}|=0.3$ GeV. }\label{rho2}
\end{figure*}
\begin{figure*}
\centering
\includegraphics[width=0.47\textwidth]{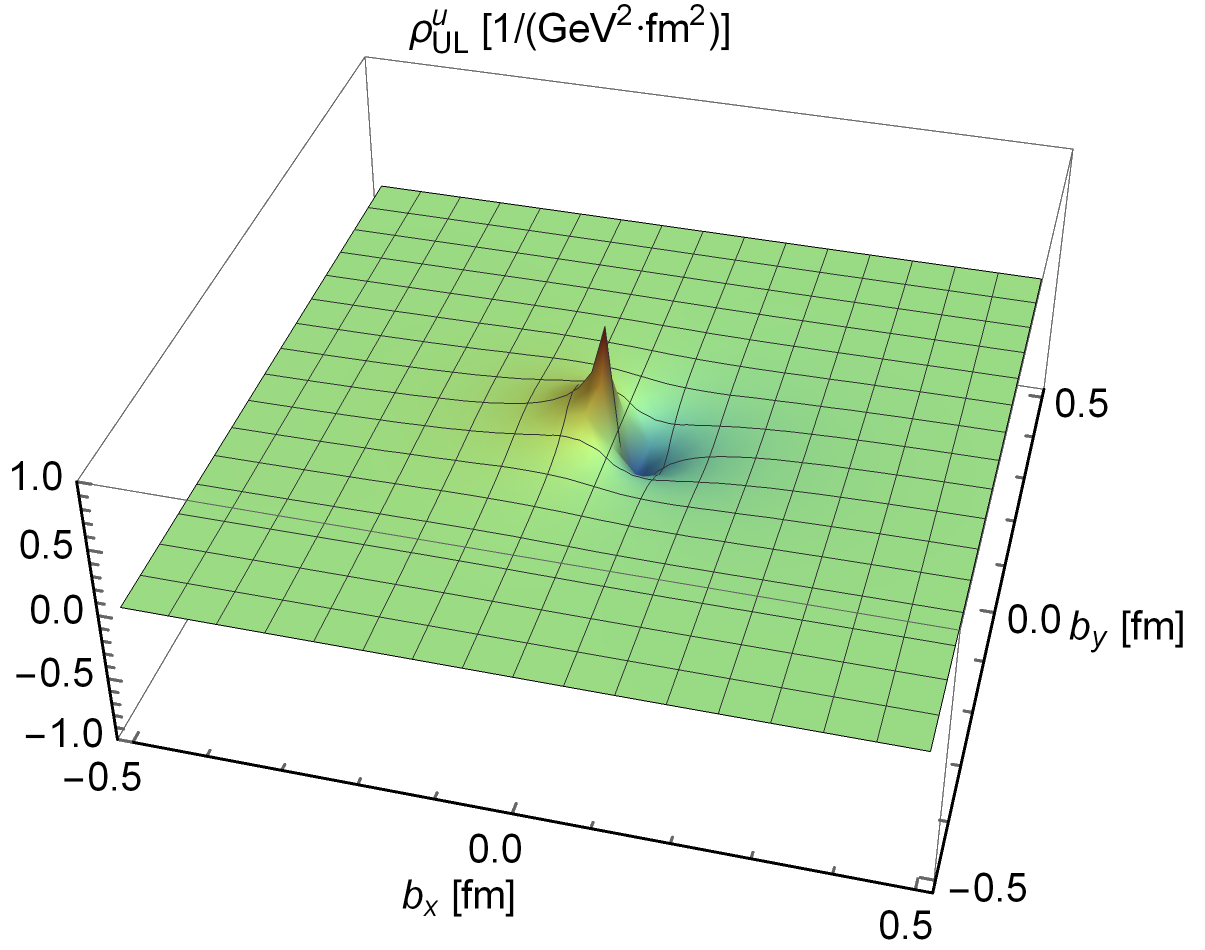}
\qquad
\includegraphics[width=0.47\textwidth]{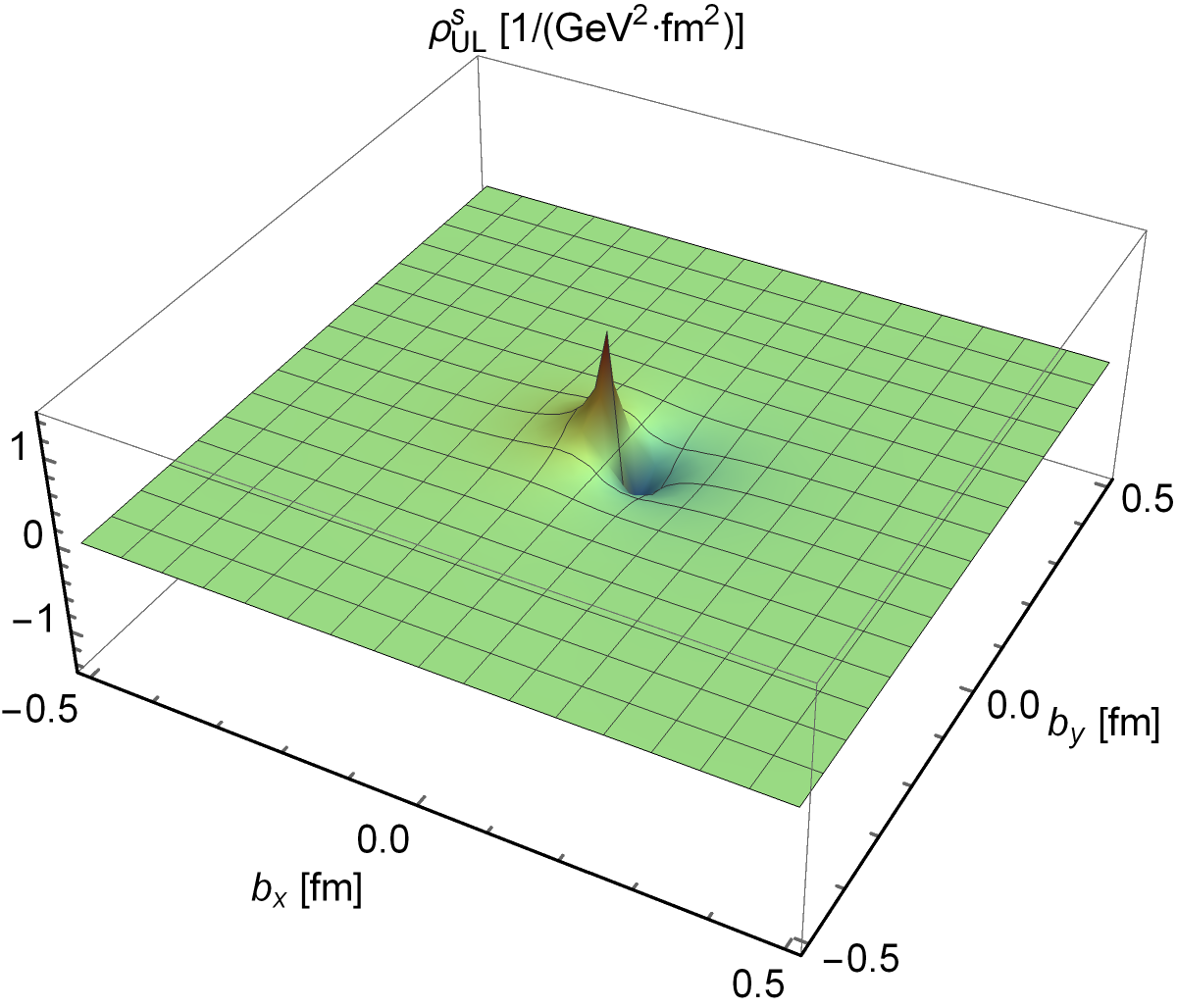}
\qquad
\includegraphics[width=0.47\textwidth]{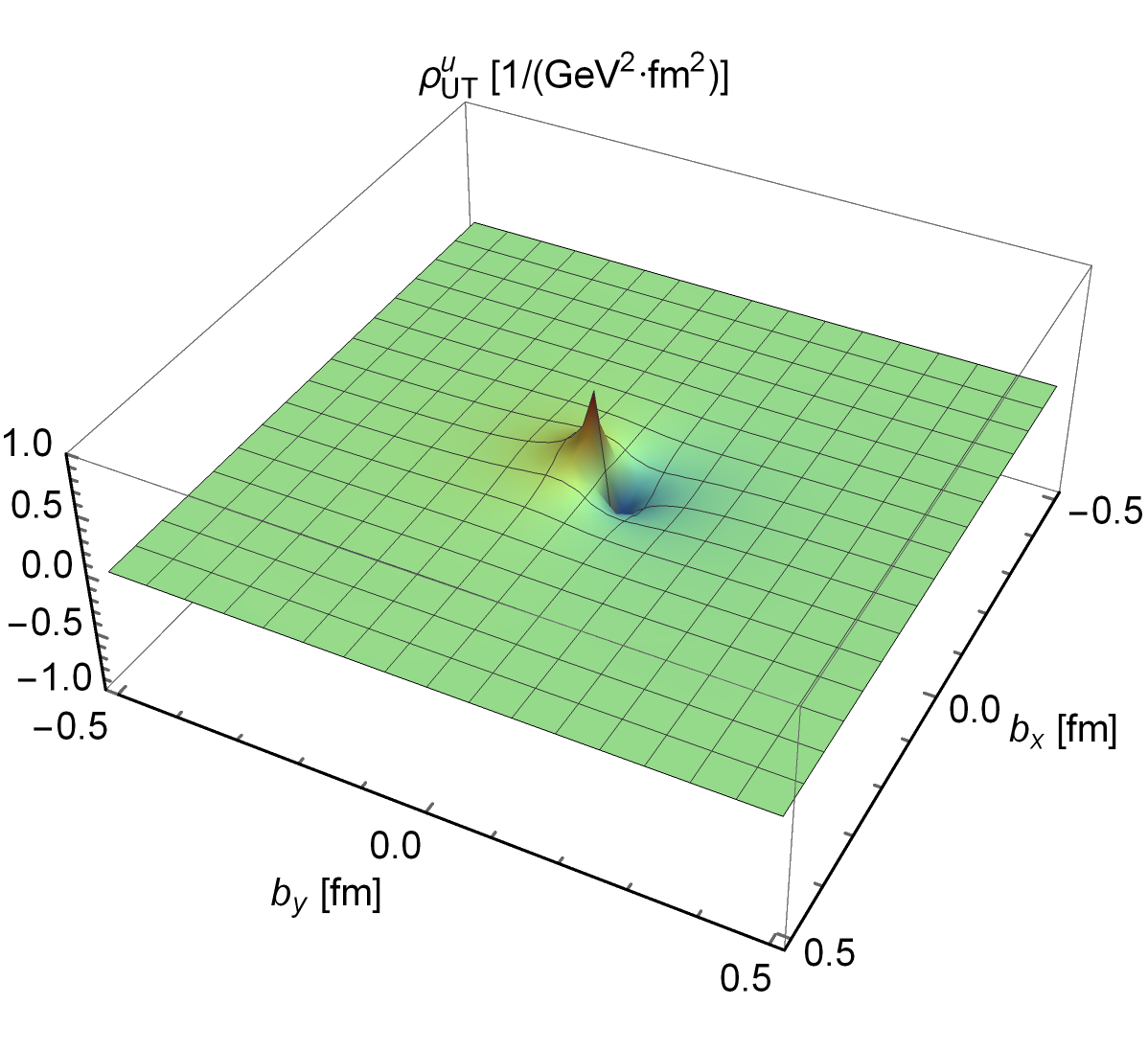}
\qquad
\includegraphics[width=0.47\textwidth]{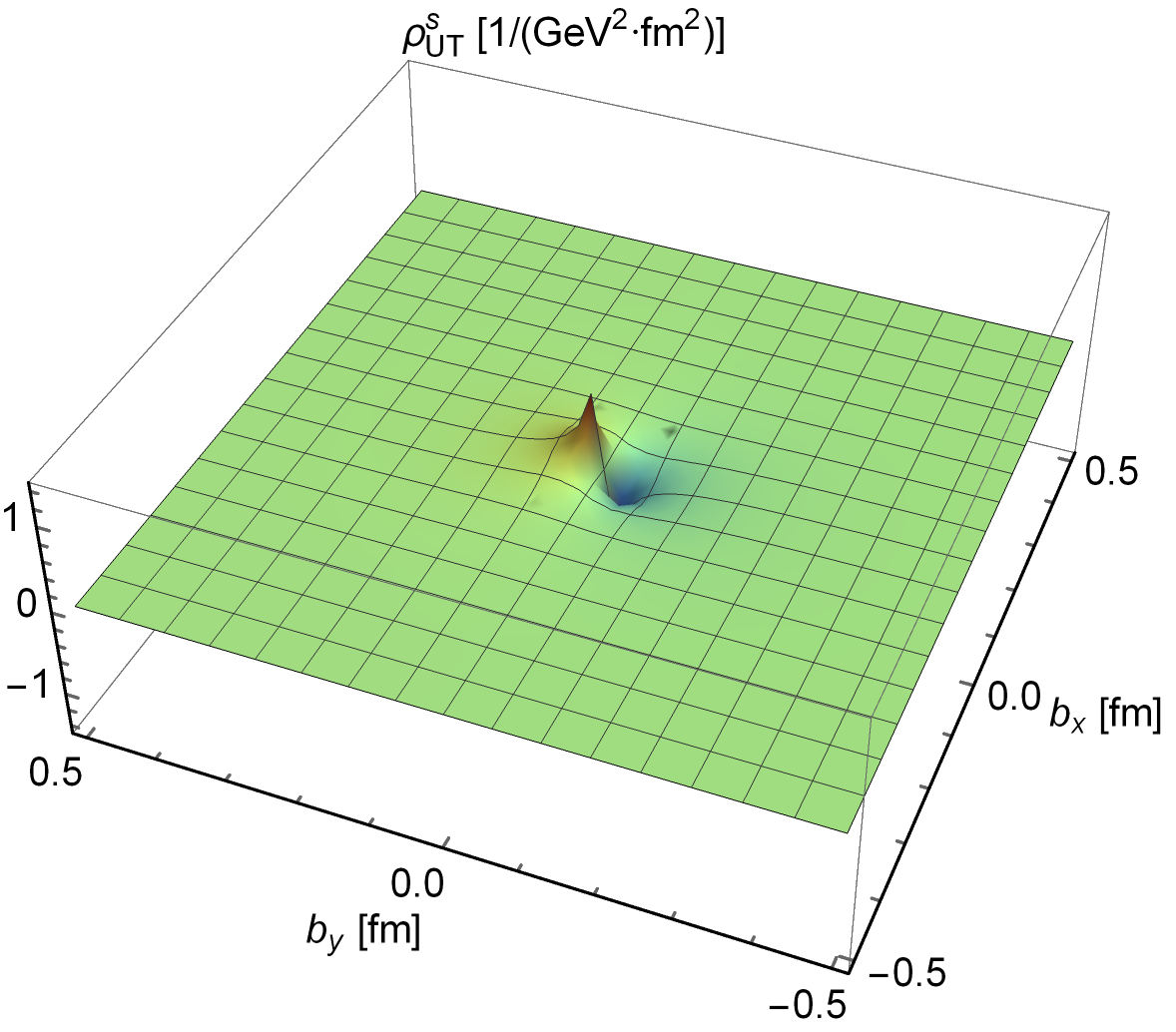}
\caption{The $u$ quark the three-dimensional picture of the longitudinal polarized quark (upper panels), and the transversely polarized quark (lower panels) of kaon with $\bm{k}_{\perp}=k_{\bot} \hat{\bm{e}}_y$, $|k_{\bot}|=0.3$ GeV. }\label{rho22}
\end{figure*}

In Figs. \ref{rho2}, we present the purely transverse Wigner distributions $\rho_{UU}\left(\bm{k}_{\bot },\bm{b}_{\perp}\right)$ (upper panel), $\rho_{UL}\left(\bm{k}_{\bot },\bm{b}_{\perp}\right)$ (middle panel), and $\rho_{UT}\left(\bm{k}_{\bot },\bm{b}_{\perp}\right)$ (lower panel) for the $u$ quark (left panel) and the $s$ quark (right panel) of the kaon in impact parameter space, with $\bm{k}_{\perp}=k_{\bot} \hat{\bm{e}}_y$, and $|k_{\bot}|=0.3$ GeV. The unpolarized distributions of the $u$ quark and $s$ quark exhibit a circularly symmetric behavior in the impact parameter space, which is different from the diagrams in the transverse momentum space. The central area of the contour line appears blank due to a sharp peak in this region, leading to a significant increase in numerical values. The longitudinally polarized quark distributions in the middle panel exhibit dipolar distortion patterns in the impact parameter space, similar to the transverse Wigner distributions. The positions at the poles are left blank due to a sharp increase in numerical value. This presents a challenge in accurately determining the peak of the longitudinally polarized quark. In Fig. \ref{rho22}, we have depicted three-dimensional representations of both the longitudinally polarized quark distributions and the  transversely polarized distributions. It is evident that the peaks of the distributions $\rho_{UL}\left(\bm{k}_{\bot },\bm{b}_{\perp}\right)$ for both $u$ and $s$ quarks fall within the region $0 \quad \text{fm} <|b_x|< 0.1 \quad \text{fm}$. We can also observe that the peaks of the distributions for both $u$ and $s$ quarks in $\rho_{UT}\left(\bm{k}_{\bot },\bm{b}_{\perp}\right)$ are located within the region of $0 \quad \text{fm} <|b_y|< 0.1 \quad \text{fm}$. These findings are consistent with numerical values reported in other references such as Refs.~\cite{Ma:2018ysi,Kaur:2019jow,Ahmady:2020ynt} for pion and Refs.~\cite{Lorce:2011kd,Chakrabarti:2016yuw,Chakrabarti:2017teq} for proton distributions. For the transversely polarized distribution in the lower panel, when the quark is polarized along the $x$-direction, the deformation in $\bm{b}_{\perp}$ space appears in the $y$-direction. The symmetry about the $b_x=0$ axis still holds, but unlike the case of pions, not all values in the region $b_y>0$ are negative and positive in the region $b_y<0$. The dividing line between positive and negative values is no longer at $b_y=0$. This conclusion differs from the calculation in the light-cone quark model using LFWFs as presented in Ref.~\cite{Kaur:2019jow}. We also check the transversely polarized distribution in the NJL model from the LFWFs, the kaon LFWFs have already been studied in Ref.~\cite{Zhang:2021tnr}, the transversely polarized distribution don't have the first item as Eq. (\ref{wigner3}), thus it is still symmetric with respect to $b_y=0$. This shows that the transversely polarized distribution obtained from GTMD are different from those obtained from the LFWFs. This is a question worthy of discussion, and we will dedicate a special study to it and its causes in the future.

Upon comparing the Wigner distributions in the two spaces, it is observed that in the impact parameter space, the peaks are sharper than in the transverse momentum space. This indicates that the twist-two Wigner distributions $\rho_{UX}$ increase much more rapidly with $b_{\perp}$ than with $k_{\bot }$ near the peak.

The gaps in Figs. \ref{rho2} are due to a sharp increase in the numerical value. To obtain the results from Eqs. (\ref{wigner1}) to (\ref{wigner3}), we observe that after the Fourier transform, there will be a factor of $e^{-\bm{b}_{\perp}^2/4\tau \alpha(1-\alpha-x)}$, different from $\bm{k}_{\perp}$, it is a factor of $e^{-\tau \bm{k}_{\perp}^2}$. The rapid changes in numerical values are due to the presence of the factor $4\tau \alpha(1-\alpha-x)$ in the denominator.

To study the regions with gaps, we can refer to Fig. \ref{wigd}, where it is evident that $\rho_{UU}\left(x,\bm{k}_{\perp}, \bm{b}_{\bot }\right)$ is smaller as $x$ approaches $0$ and larger as $x$ approaches $1$. At finite values of $x$, especially when $x$ is small, the numerical values change at a relatively slow rate, allowing us to observe changes in the given space. 

To quantitatively estimate the distortion of the unpolarized quark in the kaon meson, the average quadrupole distortions $Q_b^{ij}(\bm{k}_{\bot})$ and $Q_k^{ij}(\bm{b}_{\bot})$ are defined as 
\begin{align}\label{ff}
Q_b^{ij}(\bm{k}_{\bot})&=Q_b(k_{\bot})(2\hat{\bm{k}}_{\bot}^i\hat{\bm{k}}_{\bot}^j-\delta^{ij})=\frac{\int d^2\bm{b}_{\bot}(2\bm{b}_{\bot}^i\bm{b}_{\bot}^j-\delta^{ij}\bm{b}_{\bot}^2) \rho_{UU}\left(\bm{k}_{\perp}, \bm{b}_{\bot }\right)}{\int d^2\bm{b}_{\bot}\bm{b}_{\bot}^2\rho_{UU}\left(\bm{k}_{\perp}, \bm{b}_{\bot }\right)},
\end{align}
\begin{align}\label{ff}
Q_k^{ij}(\bm{b}_{\bot})&=Q_k(b_{\bot})(2\hat{\bm{b}}_{\bot}^i\hat{\bm{b}}_{\bot}^j-\delta^{ij})=\frac{\int d^2\bm{k}_{\bot}(2\bm{k}_{\bot}^i\bm{k}_{\bot}^j-\delta^{ij}\bm{k}_{\bot}^2) \rho_{UU}\left(\bm{k}_{\perp}, \bm{b}_{\bot }\right)}{\int d^2\bm{k}_{\bot}\bm{k}_{\bot}^2\rho_{UU}\left(\bm{k}_{\perp}, \bm{b}_{\bot }\right)},
\end{align}
where $\hat{\bm{b}}_{\bot}=\bm{b}_{\bot}/b_{\bot}$ and $\hat{\bm{k}}_{\bot}=\bm{k}_{\bot}/k_{\bot}$, $i,j=x,y$. When utilizing the CI, it is observed that the average quadrupole distortion is determined to be zero, as indicated in Refs.~\cite{Chakrabarti:2016yuw,Chakrabarti:2017teq}, due to the even nature of $\rho_{UU}\left(\bm{k}_{\perp}, \bm{b}_{\bot }\right)$ with respect to $\bm{k}_{\perp}$ and $\bm{b}_{\perp}$.

\begin{figure*}
\centering
\includegraphics[width=0.3\textwidth]{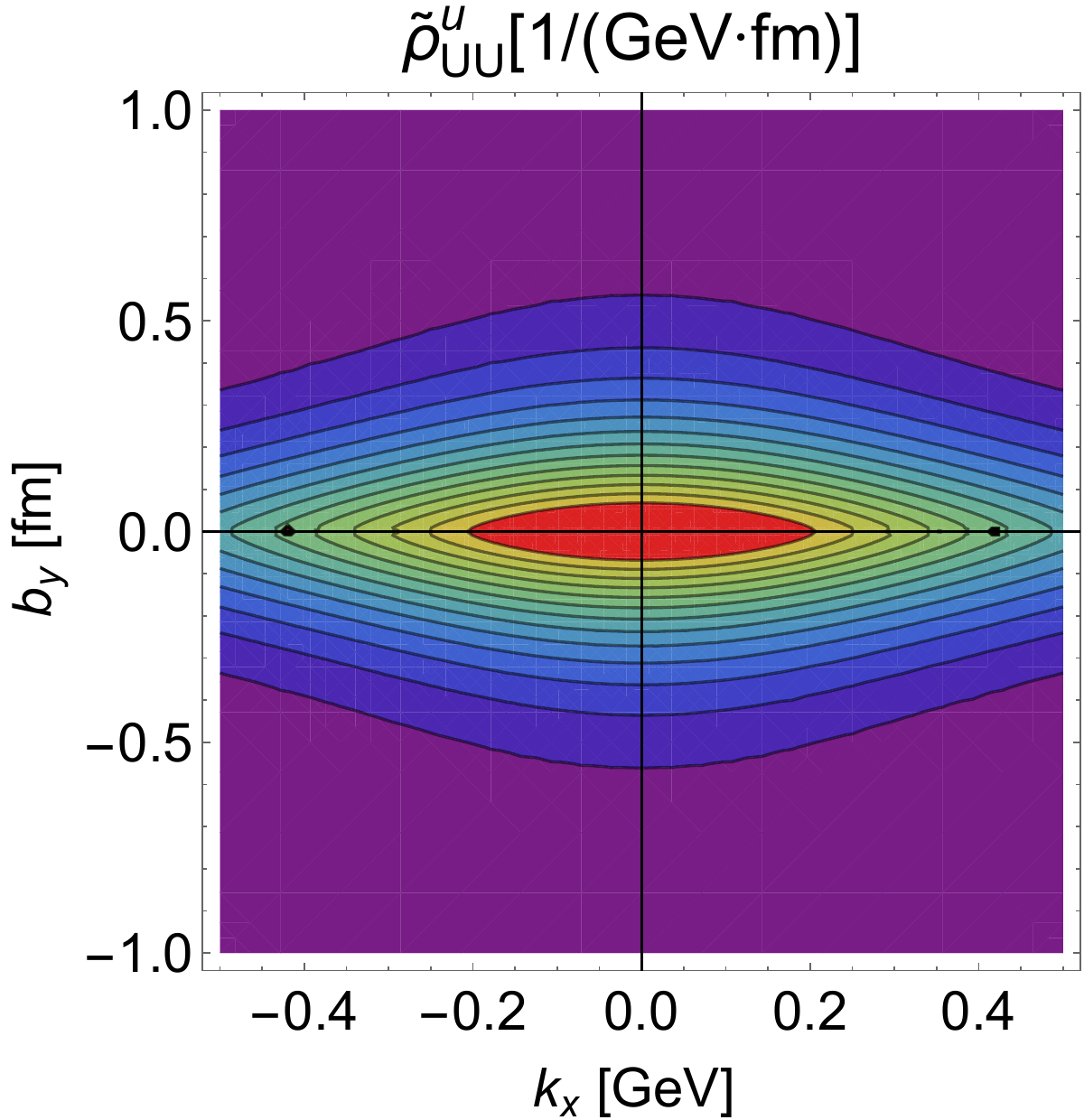}
\qquad
\includegraphics[width=0.045\textwidth]{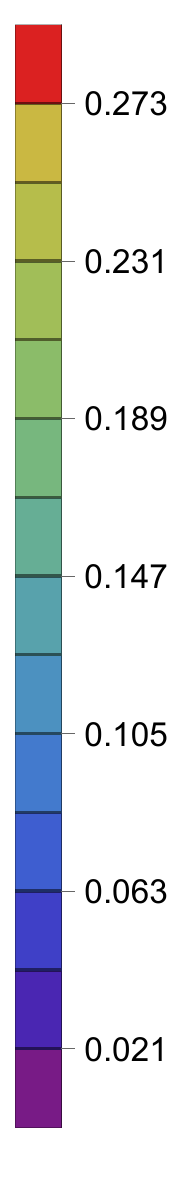}
\qquad
\includegraphics[width=0.3\textwidth]{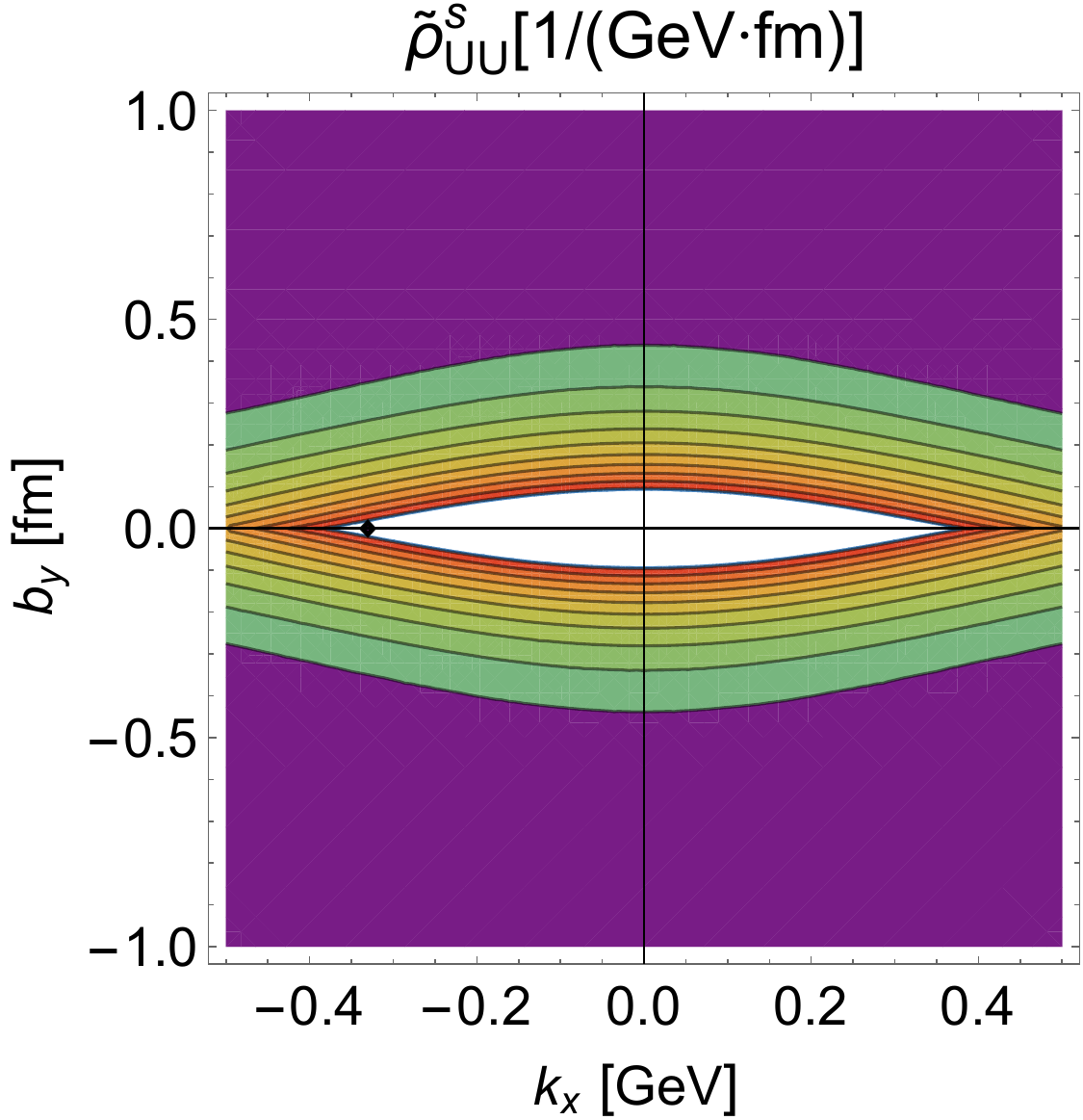}
\qquad
\includegraphics[width=0.045\textwidth]{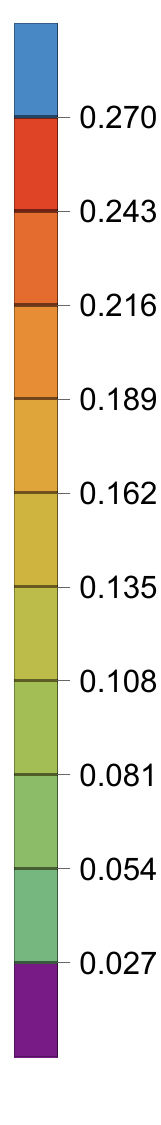}

\caption{The mixed transverse Wigner distributions of the $u$ quark in the kaon: $\tilde{\rho}_{UU}^u\left(k_x, b_y\right)$ -- left panel, $\tilde{\rho}_{UU}^s\left(k_x, b_y\right)$ -- right panel. }\label{urho11}
\end{figure*}

\begin{figure*}
\centering
\includegraphics[width=0.3\textwidth]{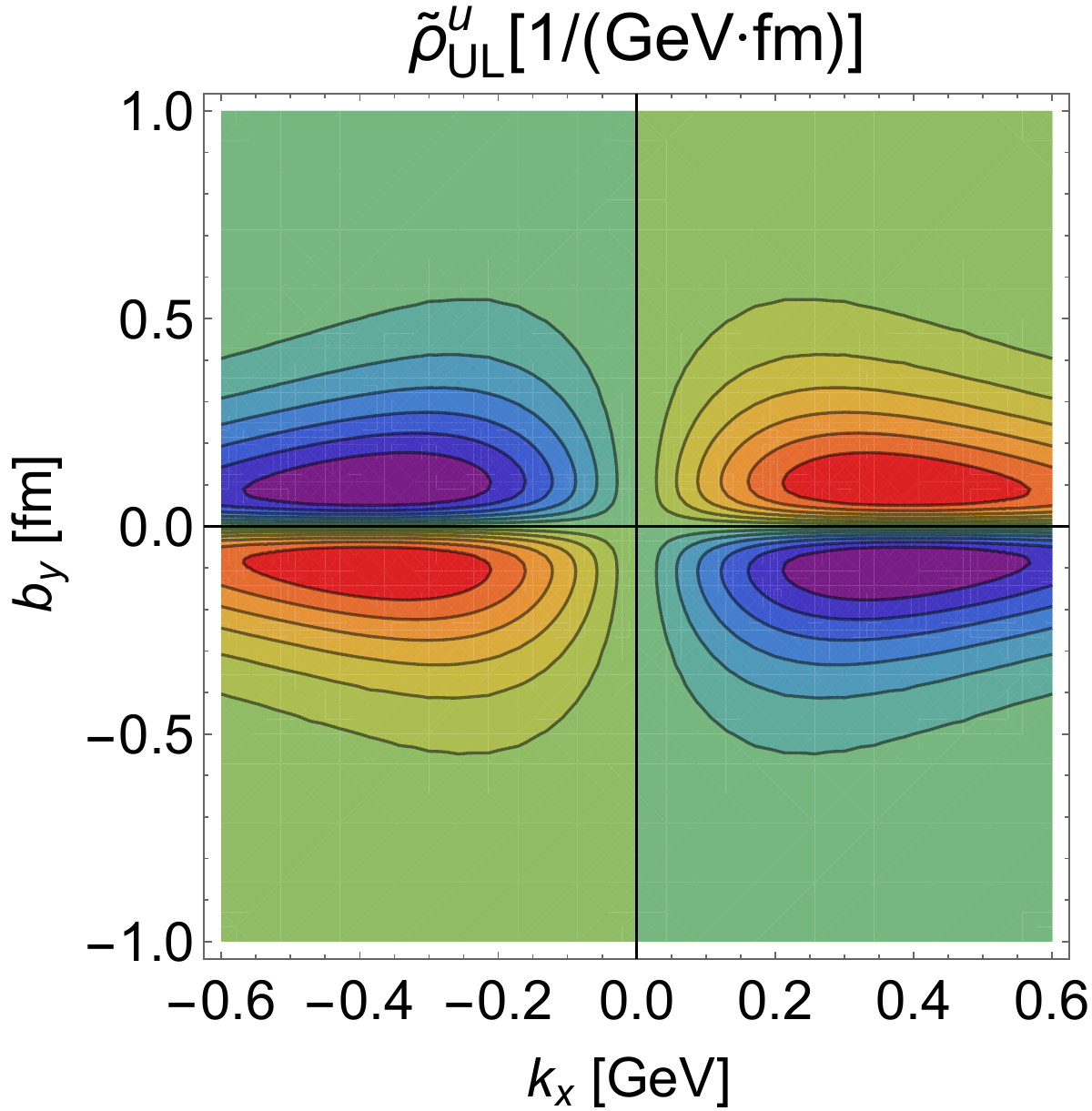}
\qquad
\includegraphics[width=0.04\textwidth]{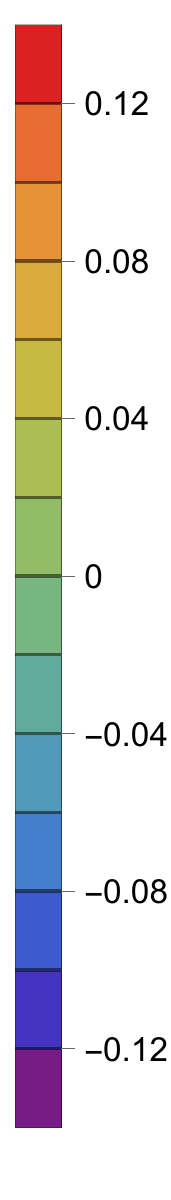}
\qquad
\includegraphics[width=0.3\textwidth]{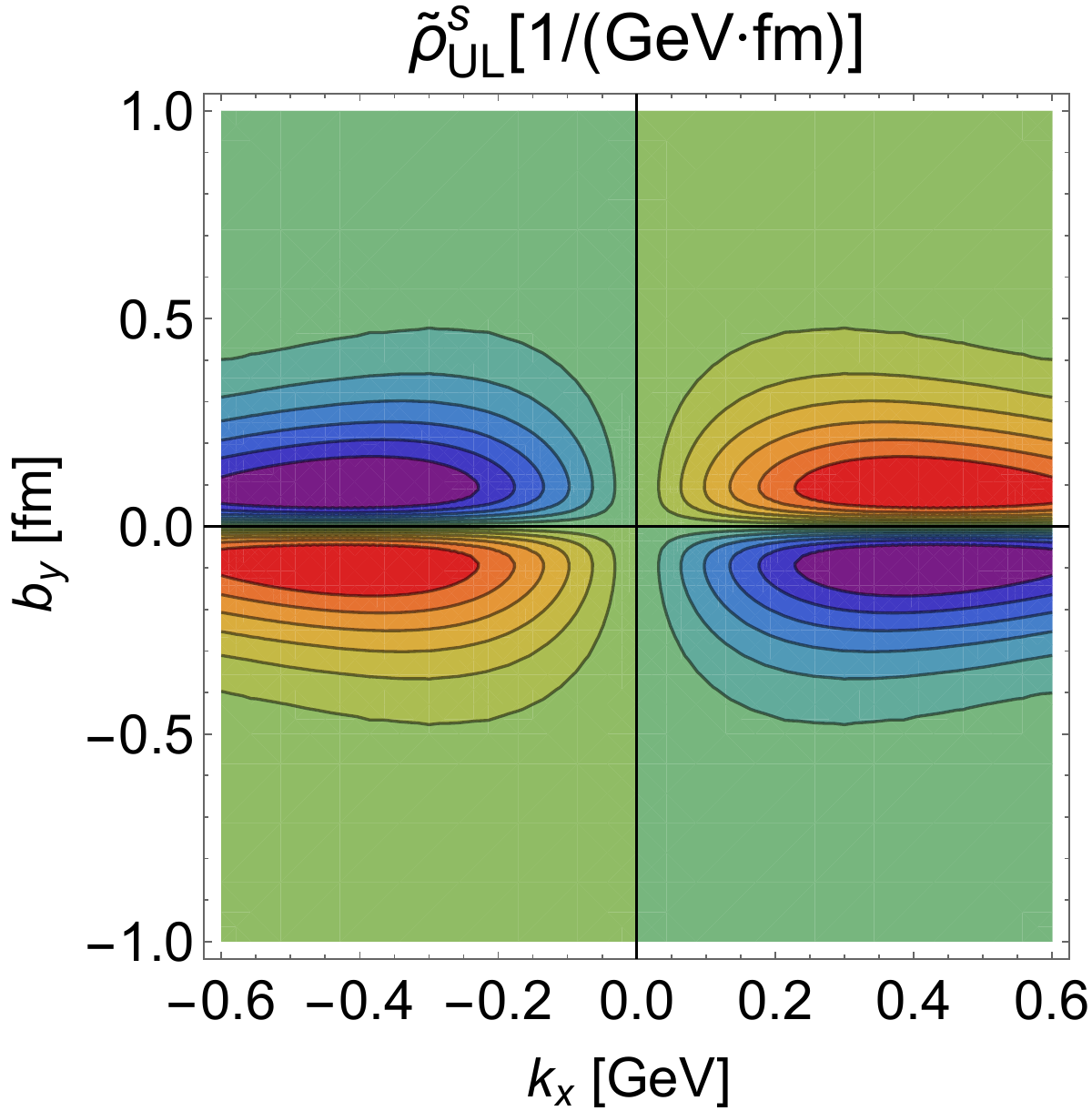}
\qquad
\includegraphics[width=0.045\textwidth]{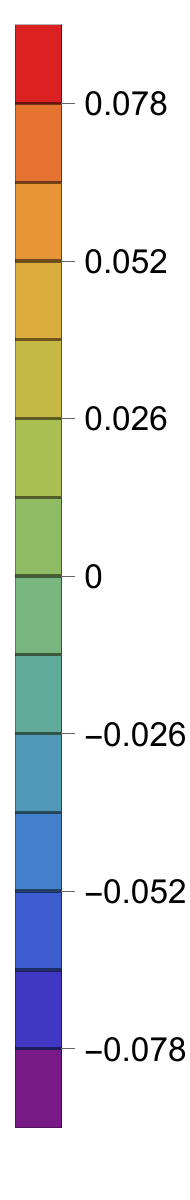}
\qquad
\includegraphics[width=0.3\textwidth]{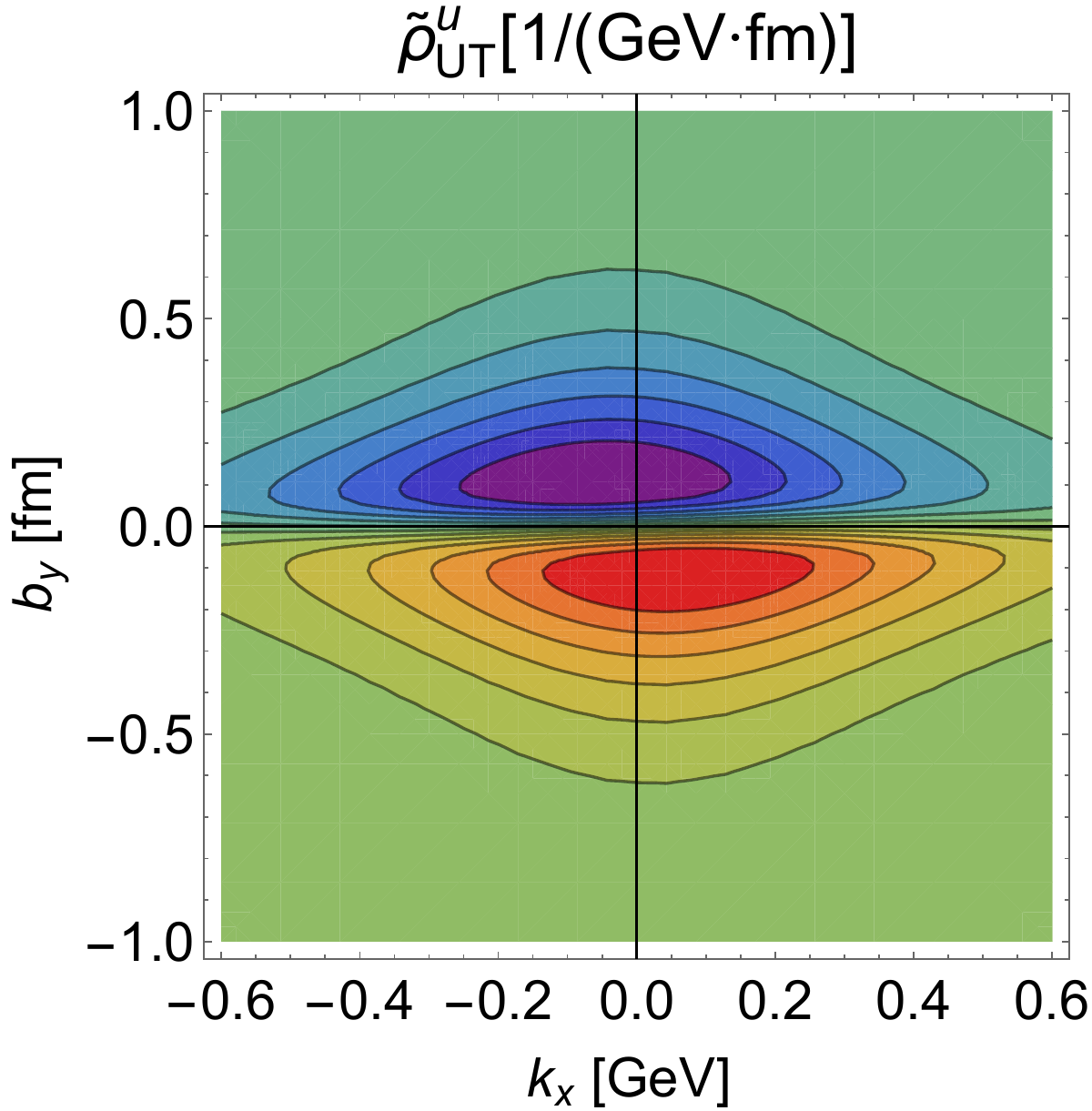}
\qquad
\includegraphics[width=0.045\textwidth]{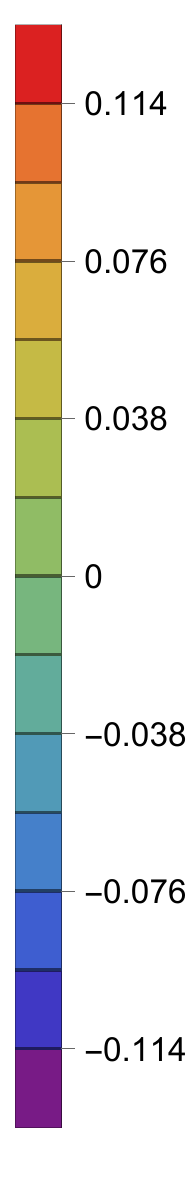}
\qquad
\includegraphics[width=0.3\textwidth]{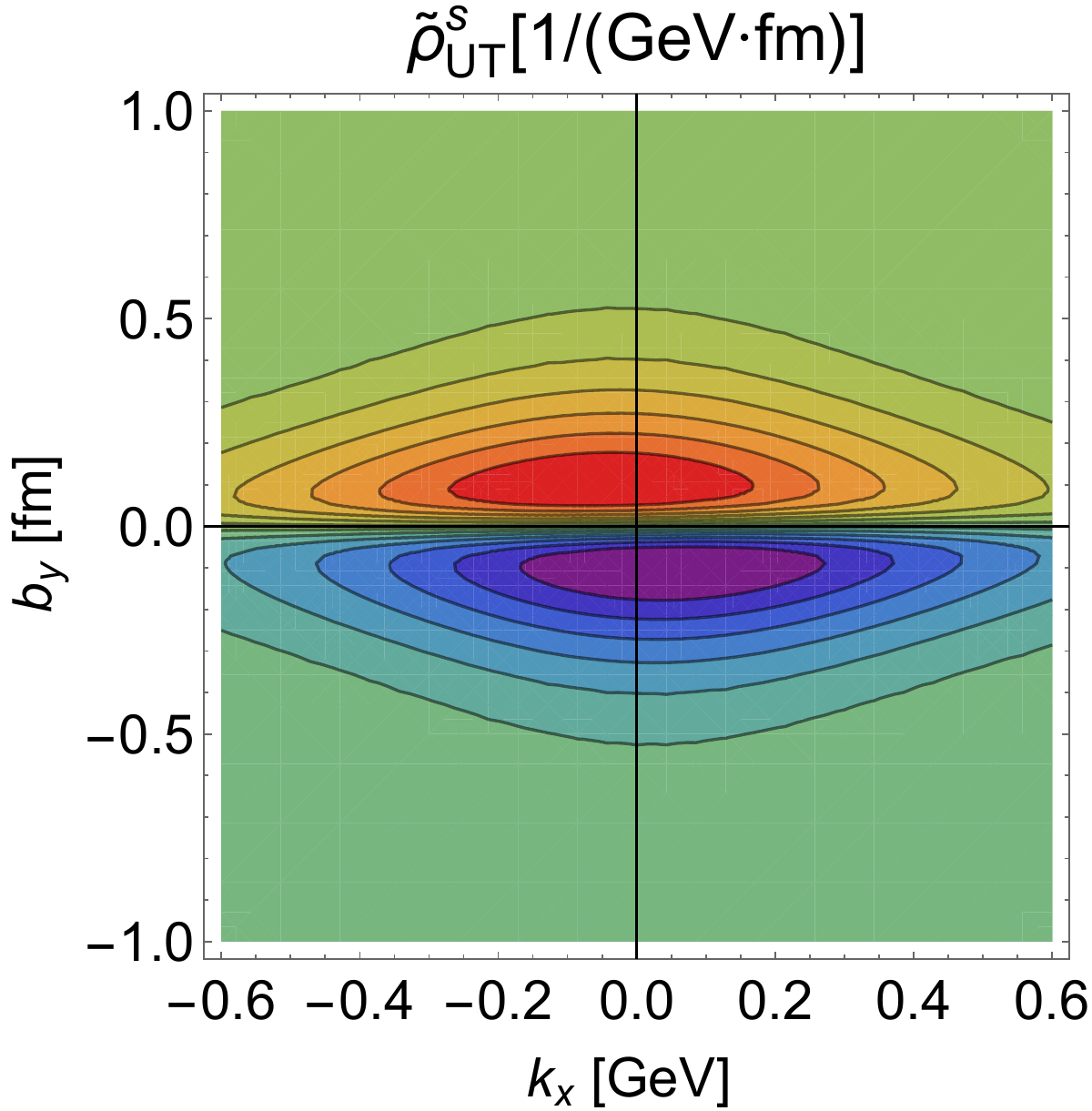}
\qquad
\includegraphics[width=0.04\textwidth]{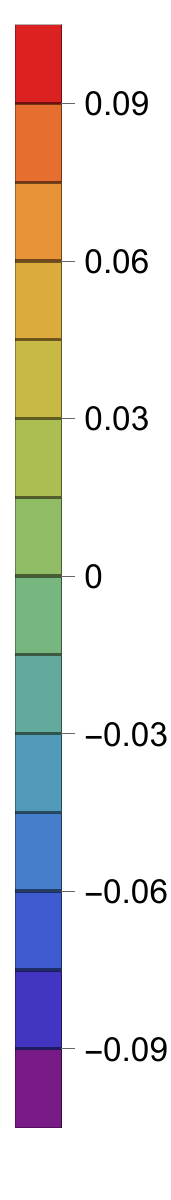}
\caption{The mixed transverse Wigner distributions of the $u$ quark in the kaon: $\tilde{\rho}_{UL}\left(\pm0.1,k_x, b_y\right)$ -- upper panel, $\tilde{\rho}_{UL}\left(\pm0.1,k_x, b_y\right)$ -- lower panel. }\label{urho112}
\end{figure*}

The mixed transverse Wigner distributions~\cite{Lorce:2011kd} are defined as,
\begin{align}\label{mixw}
\tilde{\rho}^{\Gamma}\left(k_x, b_y\right)=\int db_x \int dk_y\rho^{\Gamma}\left(\bm{k}_{\perp}, \bm{b}_{\bot }\right),
\end{align}
in which $b_x$ and $k_y$ are integrated out. From the mixed transverse Wigner distributions, we can extract more information. The $\tilde{\rho}^{\Gamma}$ has a real probability explanation because the left variables $b_y$ and $k_x$ are commutable. Thus, $\tilde{\rho}^{\Gamma}\left(k_x, b_y\right)$ represents the probability density of the quark in the $(k_x, b_y)$ plane. For the $\tilde{\rho}_{UU}\left(k_x, b_y\right)$ of the $u$ quark and $s$ quark in the kaon, we have plotted in Fig. \ref{urho11}. It is observed that for the $u$ quark $\tilde{\rho}_{UU}\left(k_x, b_y\right)$, the maximum values occur at the origin of coordinates $\left(k_x=0, b_y=0\right)$ and decrease with the spread of coordinates. The behavior of the $s$ quark is similar, but there is a blank central area due to its more rapid changes compared to those of the $u$ quark.
For the longitudinally polarized quark distribution $\tilde{\rho}_{UL}\left(k_x, b_y\right)$ and the transversely polarized quark distribution $\tilde{\rho}_{UT}\left(k_x, b_y\right)$, there are blank areas in the plots. Therefore, we plot the $\tilde{\rho}_{UL}\left(x, k_x, b_y\right)$ and $\tilde{\rho}_{UT}\left(x,k_x, b_y\right)$ without integrating out $x$. In Fig. \ref{urho112}, we plot $\tilde{\rho}_{UL}\left(\pm 0.1, k_x, b_y\right)$ and $\tilde{\rho}_{UT}\left(\pm 0.1,k_x, b_y\right)$. The symbol $+$ represents the $u$ quark and $-$ represents the $s$ quark.

For $\tilde{\rho}_{UL}\left(\pm 0.1, k_x,b_y\right)$, the maximum values occur around $\left(|k_x|=0.4, |b_y|=0.1\right)$. In the first and third quadrants of the plot, the helicity of quarks is positive; in the second and fourth quadrants it is negative.

The $\tilde{\rho}_{UT}\left(\pm 0.1,k_x,b_y\right)$ for the $u$ quark is not symmetric at axis $k_x=0$; its maximum values occur at $(k_{x}=\pm 0.1,b_{y}=\mp 0.1)$. It is negative when $b_{y}> 0$, and positive when $b_{y}< 0$. Conversely, the behavior of the $s$ quark is opposite to that of the $u$ quark in this regard.

The kaon's behavior in $\tilde\rho _{UT}(\pm 0.1, k_x, b_y )$ is very different from that of pion's behavior. In the isospin limit where $M_u=M_d$, $\tilde\rho _{ UT }(\pm 0.1,k_x,b_y )$ is still axisymmetric about $k_ { x }=0$, but this symmetry no longer exists for $K$ meson.

\subsection{The spin-orbit correlations}
When integrating over $\bm{b}_{\perp}$ and $\bm{k}_{\perp}$ for $\rho_{UU}\left( \bm{k}_{\bot },\bm{b}_{\perp}\right)$, the resulting value is zero, which is expected as there is no corresponding TMD for $\tilde{G}_{1}$. This implies that the total quark spin in a pseudoscalar meson is zero. Additionally, we perform numerical calculations to determine the distribution of quark orbital angular momentum in a kaon meson using the following expression:
\begin{align}\label{bk1}
\int d^2\bm{b}_{\perp}d^2\bm{k}_{\perp}(\bm{b}_{\perp}\times \bm{k}_{\perp})_z\rho_{UU}\left(\bm{k}_{\bot },\bm{b}_{\perp}\right),
\end{align}
which is also equal to zero. This is understandable because the kaon is a spin-$0$ hadron; that is, it has no net spin and orbital angular momentum from the quarks and antiquarks.

Studying the spin-OAM correlation of the quark inside the kaon meson, as defined by~\cite{Lorce:2011kd,Chakrabarti:2016yuw,Chakrabarti:2017teq} is also an intriguing area of research,
\begin{align}\label{bk2}
C_z=\int d^2\bm{b}_{\perp}d^2\bm{k}_{\perp}(\bm{b}_{\perp}\times \bm{k}_{\perp})_z\rho_{UL}\left(\bm{k}_{\bot },\bm{b}_{\perp}\right),
\end{align}
for $C_z>0$, the alignment of quark spin and OAM is observed, while for $C_z<0$, they tend to be anti-aligned. 

In terms of GTMD, the quark spin-orbit correlator is defined as follows:
\begin{align}\label{1bk2}
C_z=\int dx \int d^2\bm{k}_{\perp}\frac{\bm{k}_{\perp}^2}{m_K^2}\tilde{G}_{1}(x,\bm{k}_{\perp}^2,0,0).
\end{align}
In Ref.~\cite{Tan:2021osk}, they give another GPD method to calculate the spin-OAM correlation, which is defined as 
\begin{align}\label{2bk2}
C_z^q=\int dx \left(\frac{M_q}{m_K}E_T^{q}(x,0,0)-\frac{1}{2}H^q(x,0,0)\right), 
\end{align}
where $E_T$ is the tensor GPD, $H$ is the vector GPD. 

The numerical results of $C_z$ for quarks in the kaon and pion in the DSE and other models are presented in Table \ref{tb42}. The table shows that the absolute value of $C_z$ in the NJL model are greater than those in DSE. The findings indicate that the OAM of the $u$ quark is oriented in the opposite direction to its spin, while the OAM of the $s$ quark aligns with its spin. The table shows that the absolute value of of the $u$ quark in the kaon is smaller to that of the $u$ quark in the pion. Furthermore, the absolute value of the spin-OAM correlation for the $s$ quark is observed to be smaller than that of the $u$ quark in the kaon. This finding is a little larger compare with Ref.~\cite{Kaur:2019jow}, where they report $C_{z,K}^u=-0.234$ and $C_{z,K}^s=0.176$. In Ref.~\cite{Ma:2018ysi}, a value of $C_{z,\pi}=-0.159$ is reported, which is slightly smaller in absolute value than our result. In Refs.~\cite{Acharyya:2024enp,Tan:2021osk}, values of $C_{z,\pi}=-0.32$ and $C_{z,\pi}=-0.272$ are given, respectively, both of which have smaller absolute values compared to our results. Additionally, in Ref.~\cite{Acharyya:2024enp}, a value of $C_{z,K}^u=-0.251$ is reported. We also present the results obtained from the GPD method in the last row. It appears that the results derived from the GPDs are comparable to those of other methods. These findings show that our results are significantly correlated with existing literature and are consistent with the results for $u$ and $d$ quarks in the proton calculated in Refs.~\cite{Chakrabarti:2016yuw,Chakrabarti:2017teq}.

\begin{figure}
\centering
\includegraphics[width=0.47\textwidth]{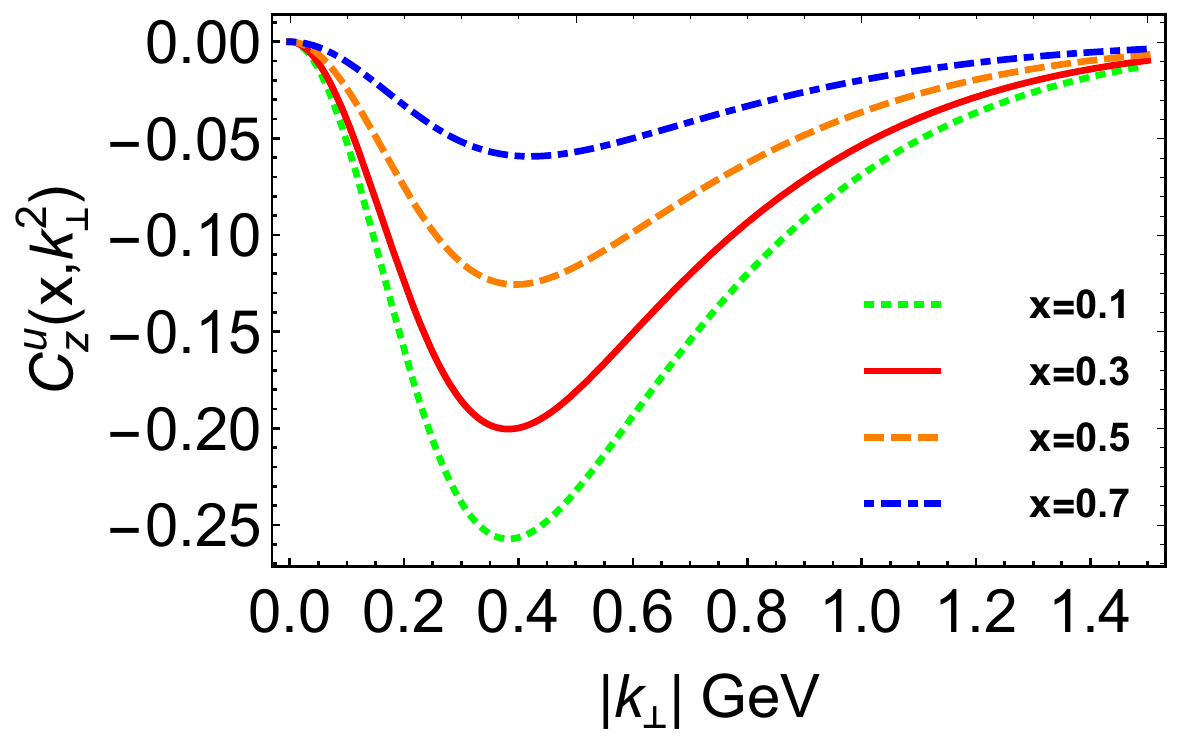}
\qquad
\includegraphics[width=0.47\textwidth]{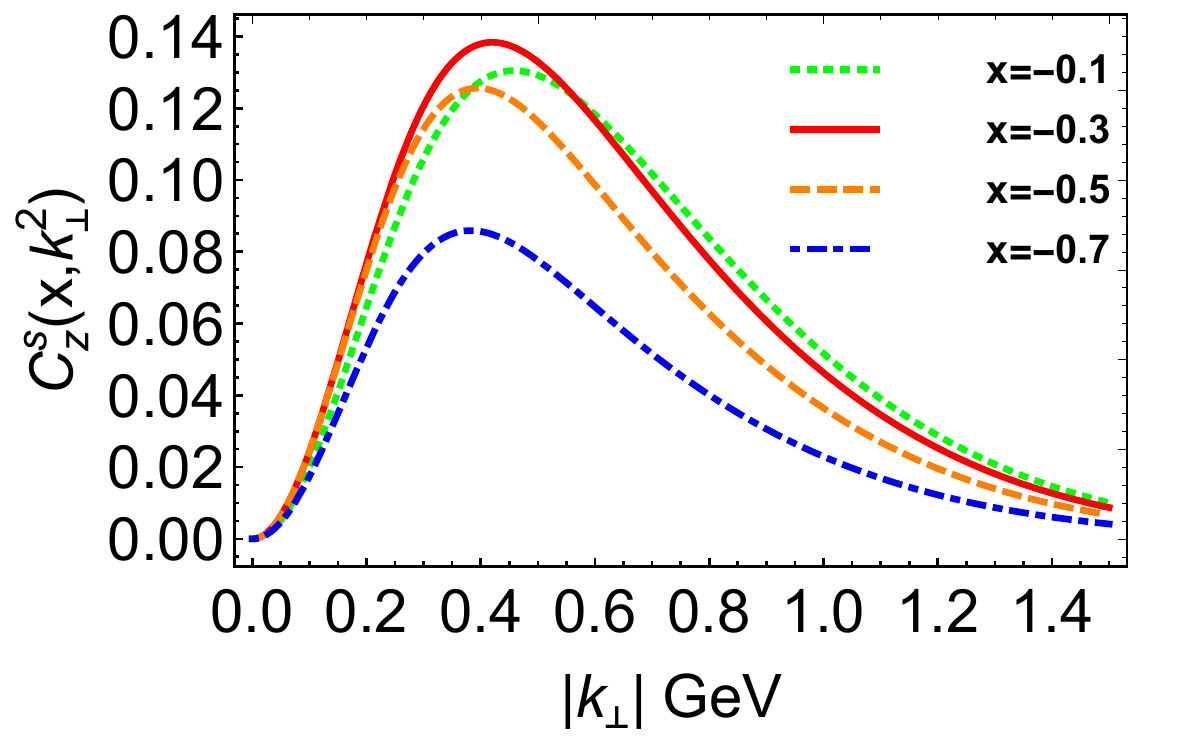}
\caption{Left panel: The $\bm{k}_{\perp}$-dependence of the unintegrated quark correlation $C_z^u(x,\bm{k}_{\perp}^2)$; Right panel: $C_z^s(x,\bm{k}_{\perp}^2)$.}\label{soc}
\end{figure}

In order to show the contribution of quark spin-orbit correlation at different transverse momenta, Fig. \ref{soc} illustrates the $\bm{k}_{\perp}$-dependence of the unintegrated quark correlation $C_z^u(x,\bm{k}_{\perp}^2)$, where $C_z$ is kept unintegrated for both $x$ and $\bm{k}_{\perp}$. The negativity of $C_{z,K}^u(x,\bm{k}_{\perp}^2)$ across the entire region aligns with findings from the GTMD approach in Ref.~\cite{Tan:2021osk}. It is evident that the distribution peak for $C_{z,K}^{u}(x,\bm{k}_{\perp}^2)$ occurs at a transverse momentum of approximately $k_{\perp}\simeq0.38$ GeV, while for $C_{z,K}^{s}(x,\bm{k}_{\perp}^2)$ it occurs at around  $k_{\perp}\simeq0.4$ GeV.

\begin{center}
\begin{table}
\centering
\caption{Comparison of the quark $C_z$ for pion and kaon in the DSE with other models.}\label{tb42}
\begin{tabular}{p{3.0cm}p{1.7cm} p{1.5cm} p{1.5cm} }
\hline\hline
 &$C_{z,K}^{u}$&$C_{z,K}^{s}$&$C_{z,\pi}^{u}$\\
\hline
LFQM ~\cite{Acharyya:2024enp} &$-$0.251&&$-$0.272\\
LCQM ~\cite{Ma:2018ysi} &&&$-$0.159\\
LCQM ~\cite{Tan:2021osk} &&&$-$0.32\\
LCQM ~\cite{Kaur:2019jow} &$-$0.234&0.176&\\
NJL(GTMD) &$-$0.492&0.370&$-$0.491\\
DSE (GTMD) &$-$0.336&0.242&$-$0.374\\
DSE (GPD) &$-$0.289&0.241&$-$0.303\\
\hline\hline
\end{tabular}
\end{table}
\end{center}

\begin{figure}
\centering
\includegraphics[width=0.47\textwidth]{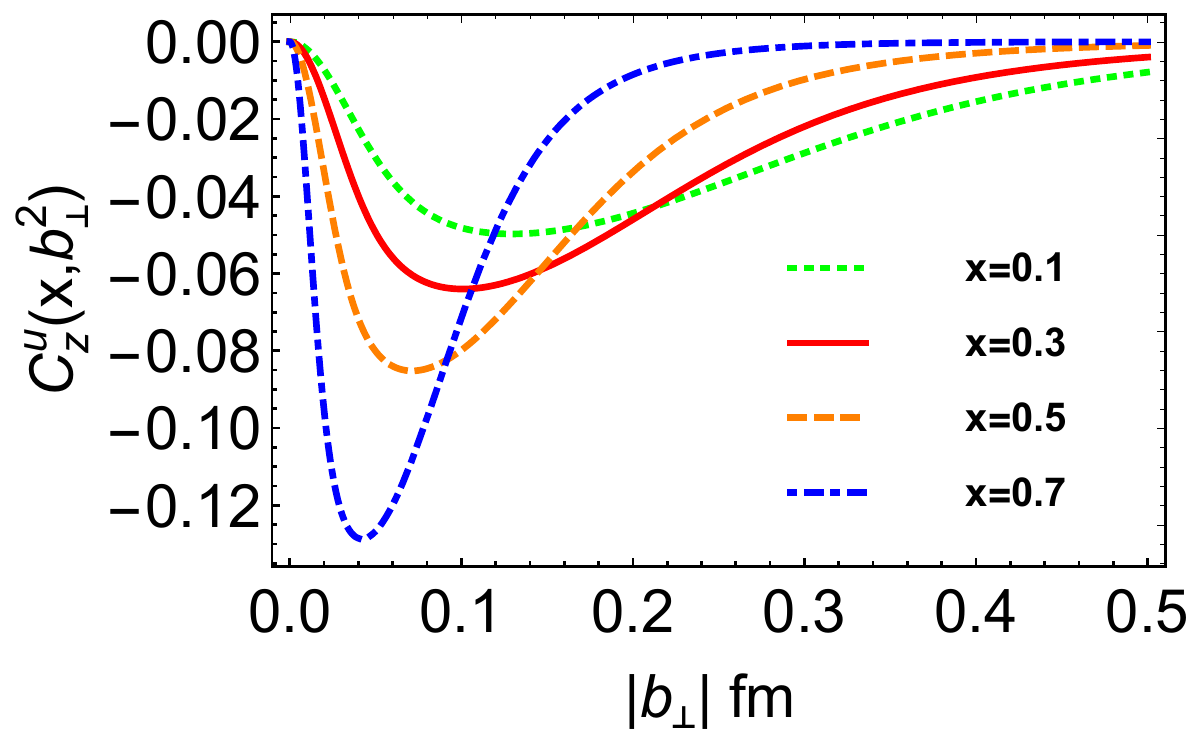}
\qquad
\includegraphics[width=0.47\textwidth]{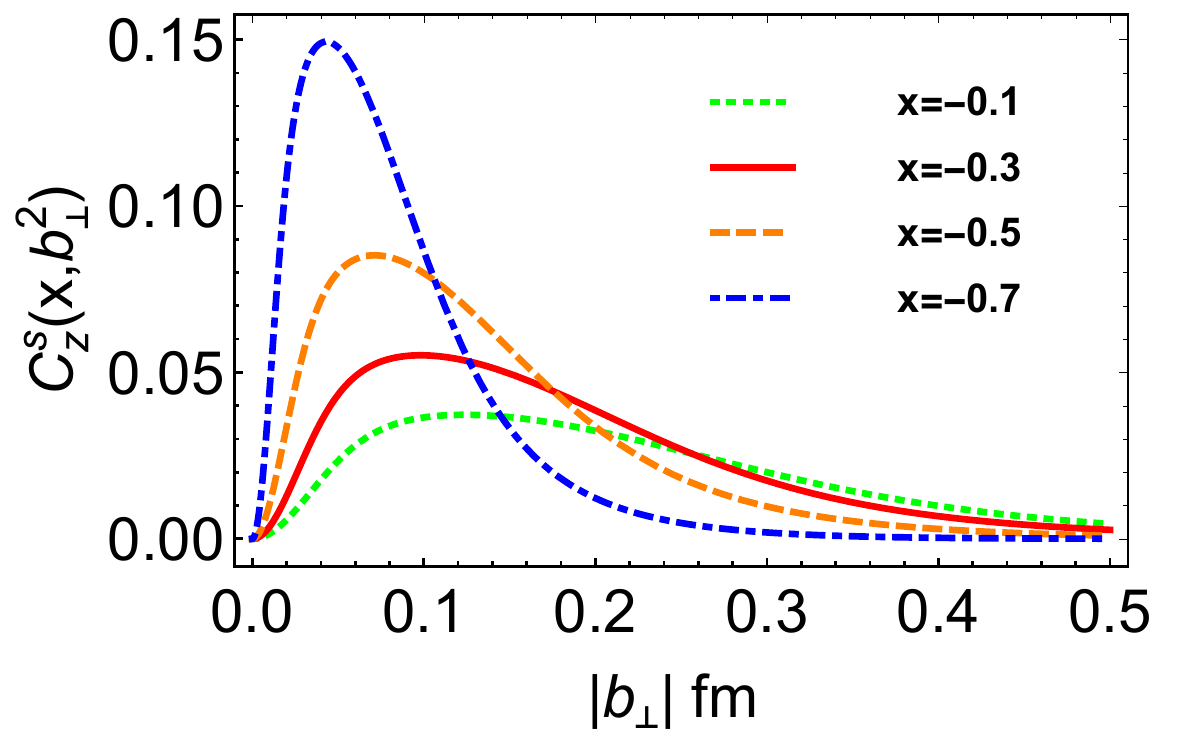}
\caption{Left panel: The $\bm{b}_{\perp}$-dependence of the unintegrated quark correlation $C_z^u(x,\bm{b}_{\perp}^2)$; Right panel: $C_z^s(x,\bm{b}_{\perp}^2)$.}\label{sbc}
\end{figure}

Furthermore, we aim to investigate the quark spin-orbit correlation at different impact parameter $b_{\perp}$, Fig. \ref{sbc} illustrates the $\bm{b}_{\perp}$-dependence of the unintegrated quark correlation $C_z^u(x,\bm{b}_{\perp}^2)$, where $C_z$ is kept unintegrated for both $x$ and $\bm{b}_{\perp}$. From the diagram, it can be observed that as the absolute value of $x$ increases for both up ($u$) and strange ($s$) quarks, the peak corresponding to $|\bm{b}_{\perp}|$, which contributes the most, decreases in magnitude while its height increases. This suggests an increasing contribution from this particular position with respect to $|\bm{b}_{\perp}|$. It is evident that the quark spin-orbit correlation at different impact parameters $b_{\perp}$ differs from transverse momentum $k_{\perp}$. 

\subsection{Impact parameter space PDF}
\begin{figure}
\centering
\includegraphics[width=0.47\textwidth]{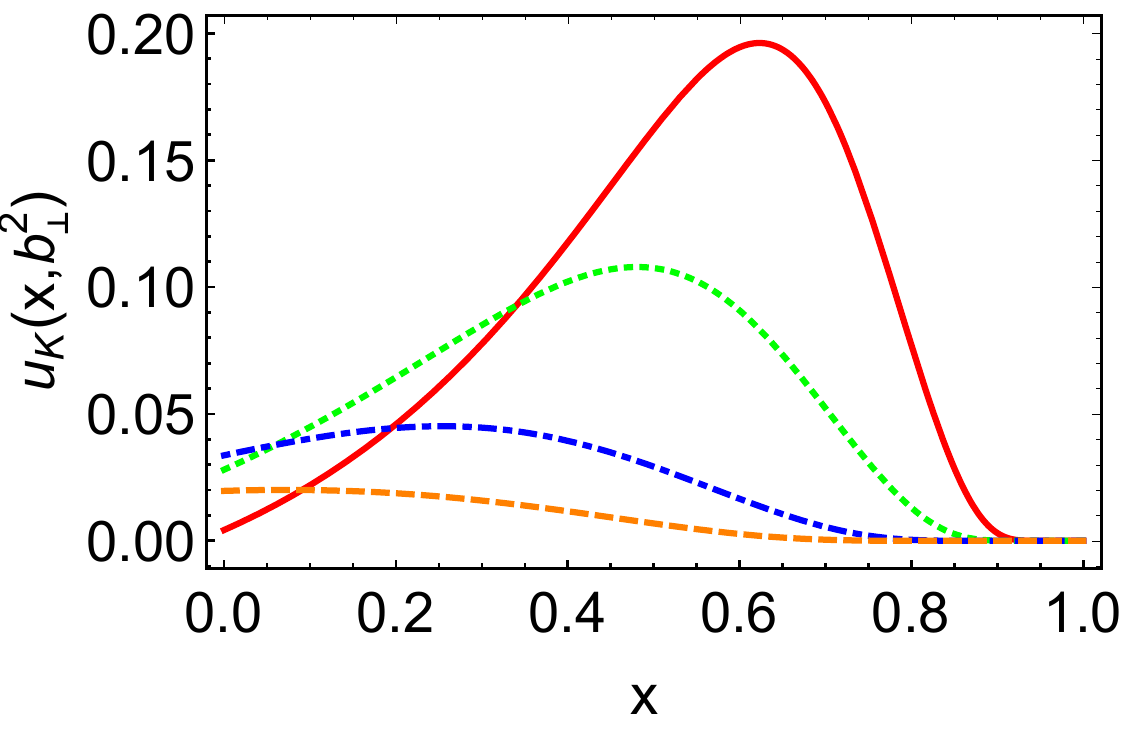}
\qquad
\includegraphics[width=0.47\textwidth]{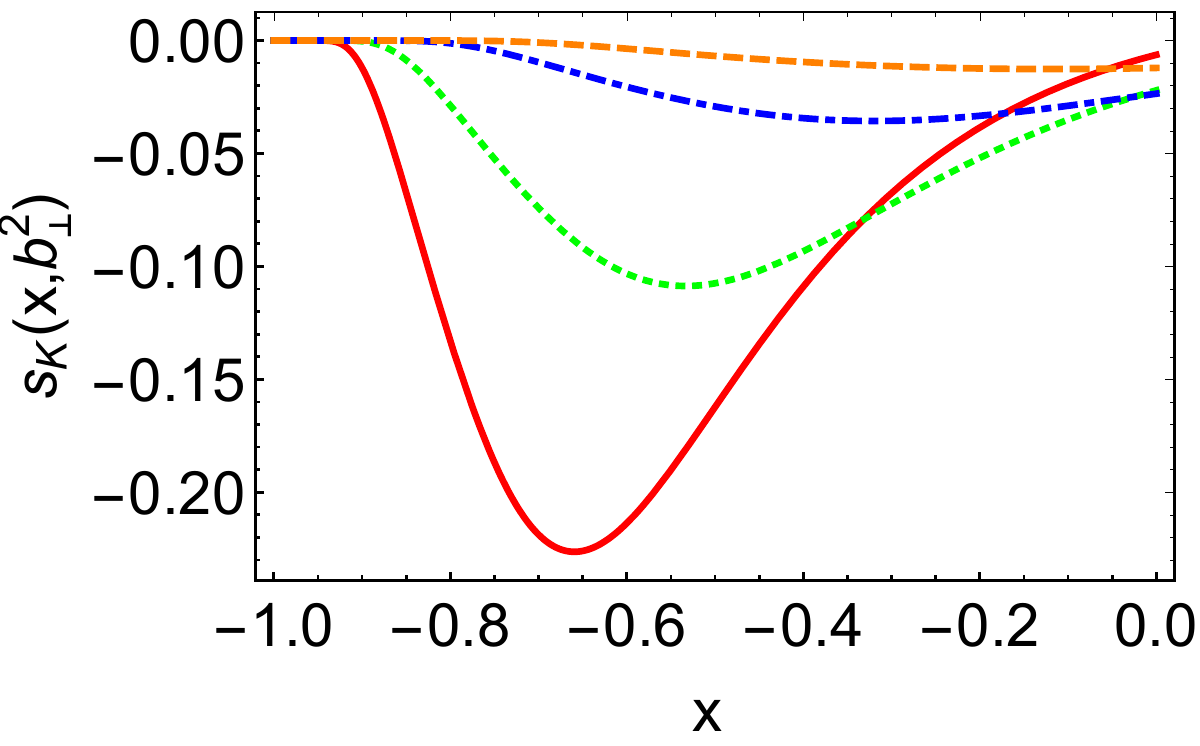}
\caption{IPS PDFs : left panel -- $u_K\left(x,\bm{b}_{\perp}^2\right)$ and right panel -- $s_K\left(x,\bm{b}_{\perp}^2\right)$ both panels with $\bm{b}_{\perp}^2=0.5$ GeV$^{-2}$ --- red solid curve, $\bm{b}_{\perp}^2=1$ GeV$^{-2}$ --- green dotted curve, $\bm{b}_{\perp}^2=2.5$ GeV$^{-2}$ --- blue dot-dashed curve, $\bm{b}_{\perp}^2=5$ GeV$^{-2}$ --- orange dashed curve. }\label{qxb}
\end{figure}

\begin{figure}
\centering
\includegraphics[width=0.47\textwidth]{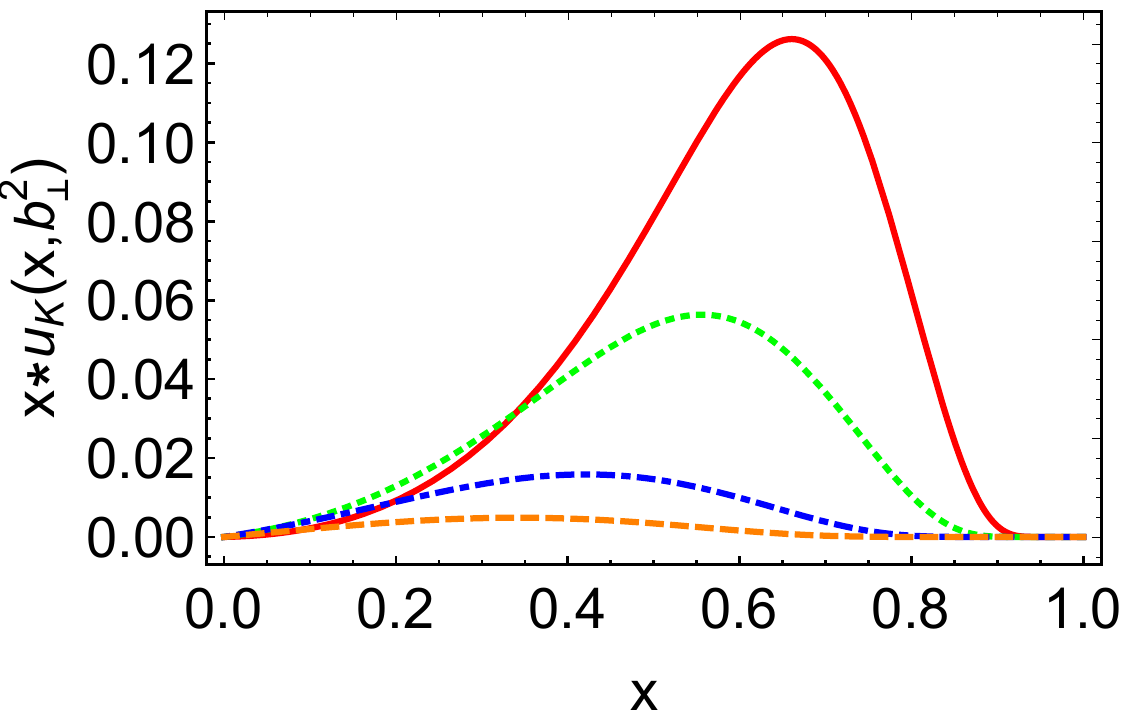}
\qquad
\includegraphics[width=0.47\textwidth]{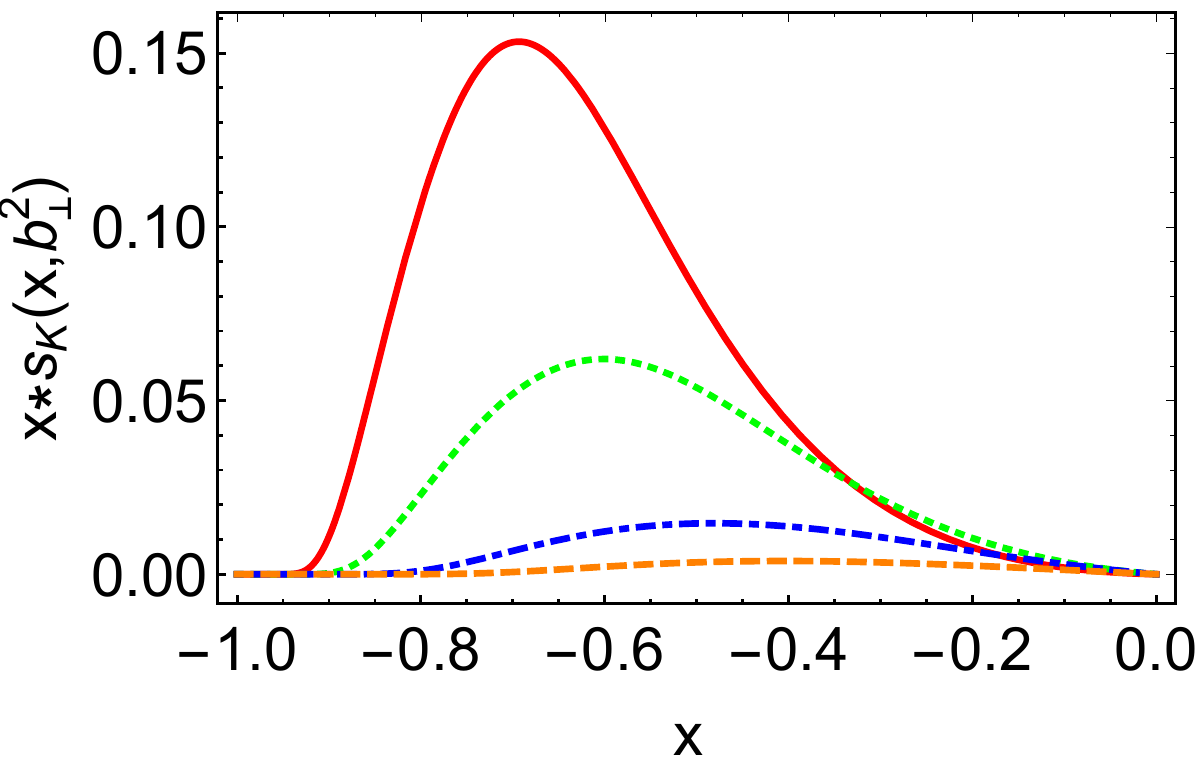}
\caption{IPS PDFs : left panel -- $xu_K\left(x,\bm{b}_{\perp}^2\right)$ and right panel -- $xs_K\left(x,\bm{b}_{\perp}^2\right)$ both panels with $\bm{b}_{\perp}^2=0.5$ GeV$^{-2}$ --- red solid curve, $\bm{b}_{\perp}^2=1$ GeV$^{-2}$ --- green dotted curve, $\bm{b}_{\perp}^2=2.5$ GeV$^{-2}$ --- blue dot-dashed curve, $\bm{b}_{\perp}^2=5$ GeV$^{-2}$ --- orange dashed curve.}\label{qxxb}
\end{figure}

The impact parameter space PDF is defined as,
\begin{align}\label{aG9}
q\left(x,\bm{b}_{\perp}^2\right)=\int \frac{d^2\bm{k}_{\perp}}{(2 \pi )^2}\rho_{UU}\left(x,\bm{k}_{\perp}, \bm{b}_{\bot }\right),
\end{align}
this density represents the probability of finding a dressed quark within the light front at a transverse position $\bm{b}_{\perp}$ from the kaon's center of transverse momentum (CoTM). Then one obtains:
\begin{align}\label{kipspdf}
u_K(x,\bm{b}_{\perp}^2)=&\frac{\bar{N}^{FF}N_c}{4\pi^2} \int \frac{d^2\bm{\Delta}_{\perp}}{(2 \pi )^2}x\bar{x}e^{- i\bm{b}_{\perp}\cdot \bm{\Delta }_{\perp}} \bar{\mathcal{C}}_1(\sigma_1^{0,0})+ \frac{N^{EE}N_c }{4\pi^2}  \int \frac{d^2\bm{\Delta}_{\perp}}{(2 \pi )^2}e^{- i\bm{b}_{\perp}\cdot \bm{\Delta }_{\perp}}\bar{\mathcal{C}}_1(\sigma_1^{0,0}) \nonumber\\
+&\frac{\bar{N}^{EF}N_c}{4\pi^2}  \int \frac{d^2\bm{\Delta}_{\perp}}{(2 \pi )^2}e^{- i\bm{b}_{\perp}\cdot \bm{\Delta }_{\perp}}(2x-1) \bar{\mathcal{C}}_2(\sigma_1^{0,0})\nonumber\\
-& \frac{N^{EF}N_c}{4\pi^2}\int \frac{d^2\bm{\Delta}_{\perp}}{(2 \pi )^2}\int _0^{1-x}d\alpha e^{-i\bm{b}_{\perp}\cdot \bm{\Delta }_{\perp}}(\bar{x}\bm{\Delta}_{\perp}^2-2x(m_K^2-(M_u-M_s)^2))\frac{\bar{\mathcal{C}}_2(\sigma_4^0)}{\sigma_4^0} .
\end{align}
A CI treatment of the kaon does not exhibit strong $x$-$t$ correlations when compared with realistic interactions. The impact parameter space (IPS) PDFs $q_K(x,\bm{b}_{\perp}^2)$ and $xq_K(x,\bm{b}_{\perp}^2)$ for the kaon's $u$ quark and $s$ quark are depicted in Figs. \ref{qxb} and \ref{qxxb}. From the diagrams, it is evident that the peak corresponding to the position of the $u$ quark in terms of $x$ decreases as $\bm{b}_{\perp}^2$ increases. Conversely, for the $s$ quark, the peak increases as $\bm{b}_{\perp}^2$ increases, due to the negative value of $x$.

\subsection{The width distribution}
\begin{figure}
\centering
\includegraphics[width=0.47\textwidth]{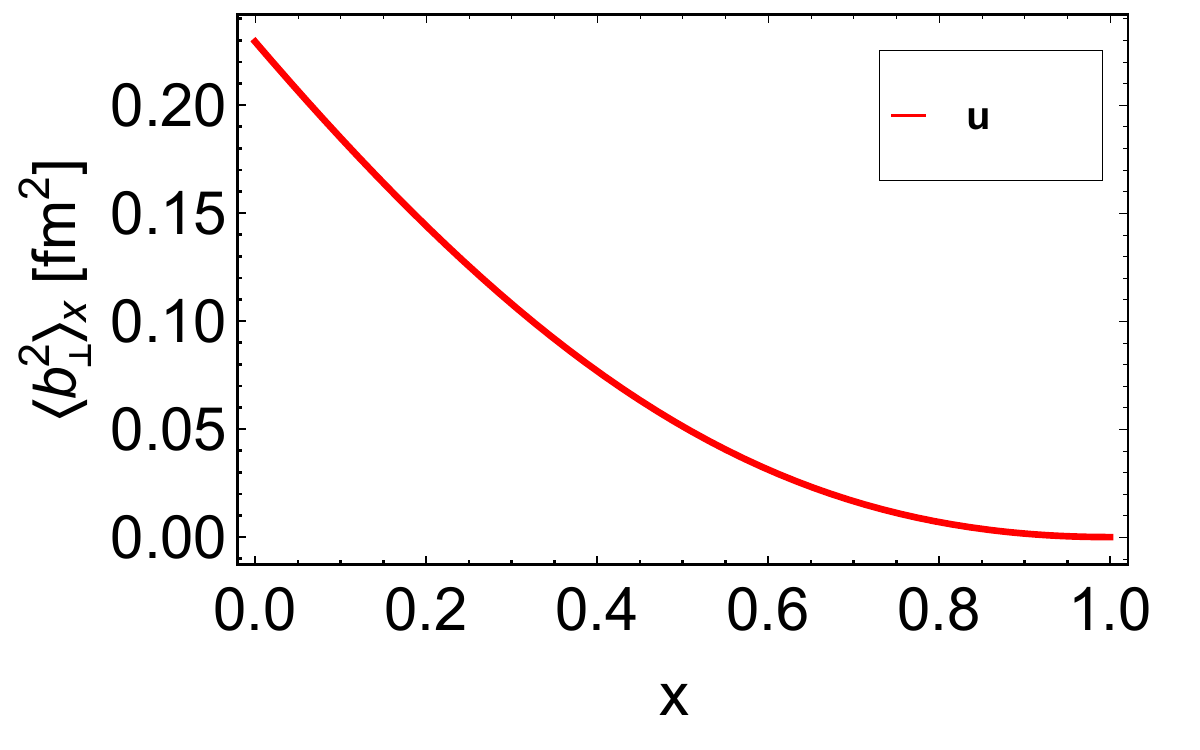}
\includegraphics[width=0.47\textwidth]{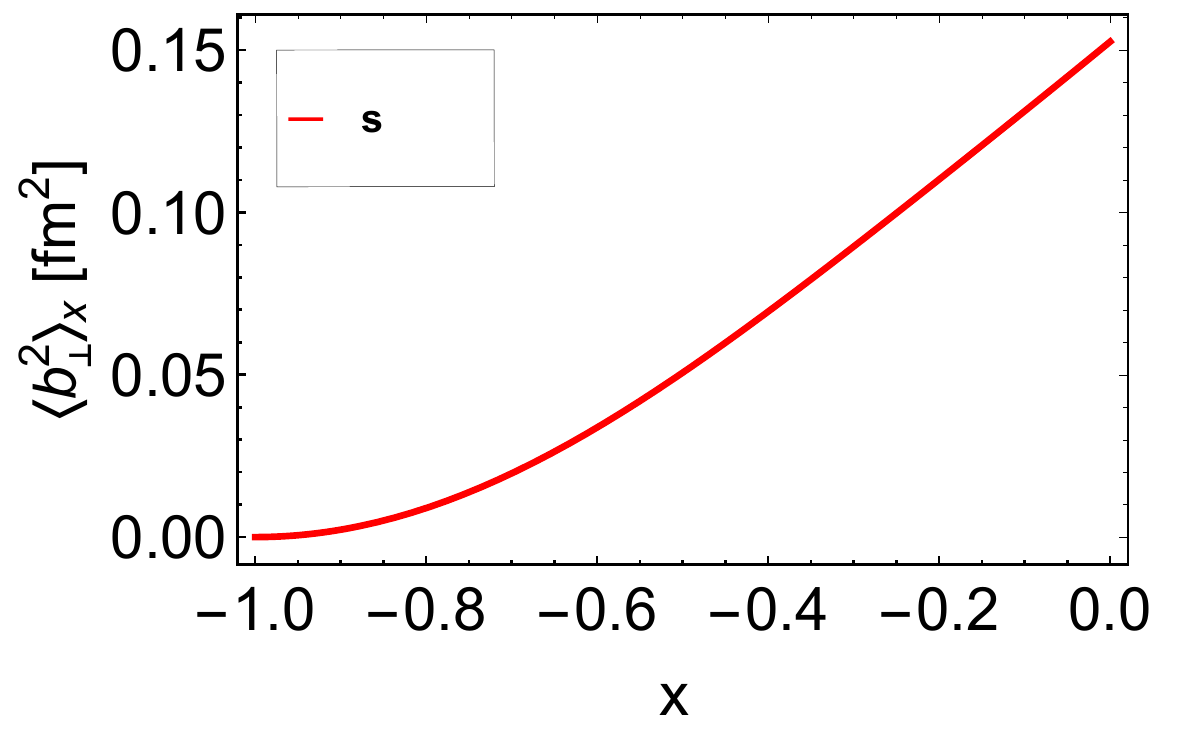}
\caption{The width distribution of partons in the kaon for a given momentum fraction $x$ defined in Eq. (\ref{tmmf}): left panel $-$ $u$ quark; right panel $-$ $s$ quark.}\label{witf}
\end{figure}

The width distribution of partons in the kaon for a given momentum fraction $x$ is
\begin{align}\label{tmmf}
\langle \bm{b}_{\bot}^2\rangle_x^q &=\frac{\int d^2 \bm{b}_{\bot}\bm{b}_{\bot}^2q_K(x,\bm{b}_{\bot}^2)}{\int d^2 \bm{b}_{\bot}q_K(x,\bm{b}_{\bot}^2)},
\end{align}
when $x\rightarrow 1$, the impact parameter should approach zero. This is because the struck quark moves closer to the center of momentum as its momentum increases. As a result, for our findings
\begin{align}\label{tmmff}
\langle \bm{b}_{\bot}^2\rangle_x^u =\frac{1}{u_K(x)}\left(\frac{N^{EF}N_c}{\pi^2}\bar{x}^2 \frac{\bar{\mathcal{C}}_2(\sigma_1^{0,0})}{\sigma_1^{0,0}}+ \frac{N^{EE} }{2\pi^2}  x\bar{x}^3(m_K^2-(M_u-M_s)^2)\frac{6\bar{\mathcal{C}}_3(\sigma_1^{0,0})}{[\sigma_1^{0,0}]^2}\right),
\end{align}
where $u_K(x)$ represents the valence $u$ quark PDF as defined in Eq. (\ref{kapdf}). In the equation above, both terms contain a factor of $(1-x)$, resulting in the value of the equation being zero at $x=1$. The width distribution is depicted in Fig. \ref{witf}. These results are consistent with the $x$-dependence of $\langle \bm{b}_{\bot}^2\rangle_x$ for quarks in the proton as shown in Fig. 24 of Ref. ~\cite{Dupre:2017hfs}. Specifically, at $x=0$, the values of $\langle \bm{b}_{\bot}^2\rangle_0$ for quarks in pions are smaller than those for quarks in protons.

The squared radius independent of $x$ is calculated from $\langle \bm{b}_{\bot}^2\rangle_x^q$ by taking the following average over $x$:
\begin{align}\label{tmmf1}
\langle \bm{b}_{\bot}^2\rangle^q &=\frac{1}{N_q}\int_0^1 dx q_K(x)\langle \bm{b}_{\bot}^2\rangle_x^q,
\end{align}
with the integrated number of valence quarks $N_u=1$ and $N_s = 1$, for the $K^+$, we present the $x$-independent squared radius of kaon and pion in the NJL model and DSE in Table \ref{tb41}. The table reveals that the $u$ quark's $x$-independent squared radius of kaon and pion are similar. However, inside the kaon, the $u$ quark's squared radius is larger than that of the $s$ quark in both the NJL model and DSE. Furthermore, it is noteworthy that the results obtained from the NJL model are larger than those from DSE.
\begin{center}
\begin{table}
\centering
\caption{Comparison of the $x$-independent squared radius for quarks $\langle b_{\bot}^2\rangle^q$ of the kaon and pion in the NJL model and DSE in units of fm$^2$.}\label{tb41}
\begin{tabular}{p{1.2cm}p{1.2cm} p{1.2cm} p{1.2cm} p{1.2cm}p{1.2cm}}
\hline\hline
 &$\langle b_{\bot}^2\rangle^{u,K}$&$\langle b_{\bot}^2\rangle^{u,K}$&$\langle b_{\bot}^2\rangle^{u,\pi}$&$\langle b_{\bot}^2\rangle^{K}$&$\langle b_{\bot}^2\rangle^{\pi}$\\
\hline
NJL &0.152&0.088&0.140&0.131&0.140\\
DSE &0.079&0.053&0.080&0.070&0.080\\
\hline\hline
\end{tabular}
\end{table}
\end{center}

The Dirac squared radius $\langle \bm{b}_{\bot}^2\rangle$ is then obtained as the charge-weighted sum over the valence quarks
\begin{align}\label{tmmf2}
\langle \bm{b}_{\bot}^2\rangle &=e_u\langle \bm{b}_{\bot}^2\rangle^u+e_s\langle \bm{b}_{\bot}^2\rangle^s,
\end{align}
with quark electric charges of $e_u=+2/3$ and $e_s=1/3$, it is observed from Table \ref{tb41} that the Dirac squared radius $\langle \bm{b}_{\bot}^2\rangle^K$ for the kaon is smaller than $\langle \bm{b}_{\bot}^2\rangle^{\pi}$ for the pion in both the NJL model and the DSE.
For our results, $\int d^2 \bm{b}_{\bot}u_K(x,\bm{b}_{\bot}^2)$ is $u$ quark PDF in Eq. (\ref{kapdf}), when $x\rightarrow 1$,
\begin{align}
u_K(1)=\frac{N^{EE}N_c}{4\pi^2} \bar{\mathcal{C}}_1(M_s^2)+\frac{\bar{N}^{EF}N_c}{4\pi^2}\bar{\mathcal{C}}_1(M_s^2),
\end{align}
it is a definite value. For the integral $\int d^2 \bm{b}_{\bot}\bm{b}_{\bot}^2u_K(x,\bm{b}_{\bot}^2)$, pertaining to the first term of Eq. (\ref{kipspdf}),
\begin{align}
&\quad \int d^2 \bm{b}_{\bot}\bm{b}_{\bot}^2\frac{\bar{N}^{FF}N_c}{4\pi ^2}\int \frac{d^2\bm{\Delta}_{\perp}}{(2 \pi )^2} x\bar{x}e^{-i\bm{b}_{\perp}\cdot \bm{\Delta}_{\perp}} \bar{\mathcal{C}}_1(\sigma_1^{0,0})\nonumber\\
&= \frac{\bar{N}^{FF}N_c}{4\pi ^2} \int d^2 \bm{b}_{\bot}\bm{b}_{\bot}^2\delta^2(\bm{b}_{\bot})x\bar{x}\bar{\mathcal{C}}_1(\sigma_1^{0,0})=0,
\end{align}
which means this term is zero, it is $x$ independent. The integration of the final term in Eq. (\ref{kipspdf}) is defined as Eq. (\ref{tmmff}) multiplied by $u_K(x)$. As $x$ approaches $1$, the integration will vanish, resulting in $\int d^2 \bm{b}_{\bot}\bm{b}_{\bot}^2u_K(x,\bm{b}_{\bot}^2)$ also vanishing. Our findings confirm that $\langle \bm{b}_{\bot}^2\rangle_{x\rightarrow 1}=0$. This observation is further supported by Fig. \ref{witf}, where it can be seen that as $x$ approaches $1$, the width distribution of partons in the kaon $\langle \bm{b}_{\bot}^2\rangle_x$ becomes zero.

In small momentum fraction region $u$ quark is wider than $s$ quark, with the increasing of $|x|$, the two lines almost coincide, which means the two quarks have almost the same width. This result is coincide with the NJL result in Ref.~\cite{Zhang:2021tnr}.

\subsection{Light-front transverse-spin distribution}
The light-front transverse-spin distribution of $u$ quark within the kaon defined in IPS~\cite{QCDSF:2007ifr}:
\begin{align}\label{lftsd}
\rho_n\left(\bm{b}_{\bot },\bm{s}_{\perp}\right)
&=\frac{1}{2}\int_0^1  dx \int d^2\bm{k}_{\perp}x^{n-1} [\rho_{UU}\left(x,\bm{k}_{\perp}, \bm{b}_{\bot }\right)-\bm{s}_{\perp}^i\rho_{UT}\left(x,\bm{k}_{\perp}, \bm{b}_{\bot }\right)],
\end{align}
For a quark polarized in the $+x$ direction and $\hat{s}_{\perp}\cdot \hat{b}_{\perp}=\cos \phi_{\perp}$, the expression $\varepsilon_{\perp}^{ij}s_{\perp}^ib_{\perp}^j=|b_{\perp}|\sin\phi_{\perp}$ holds true. In the context of the CI treatment, all kaon form factors are considered to be hard.
\begin{figure}
\centering
\includegraphics[width=0.47\textwidth]{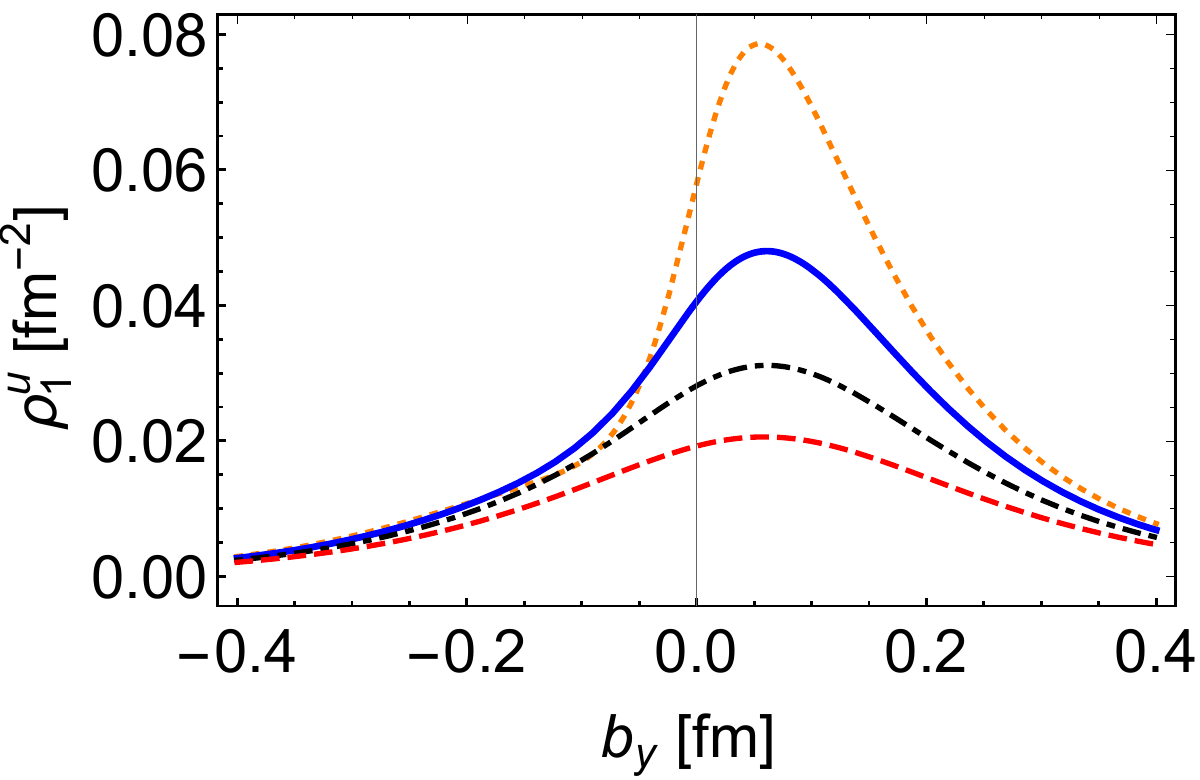}
\qquad
\includegraphics[width=0.47\textwidth]{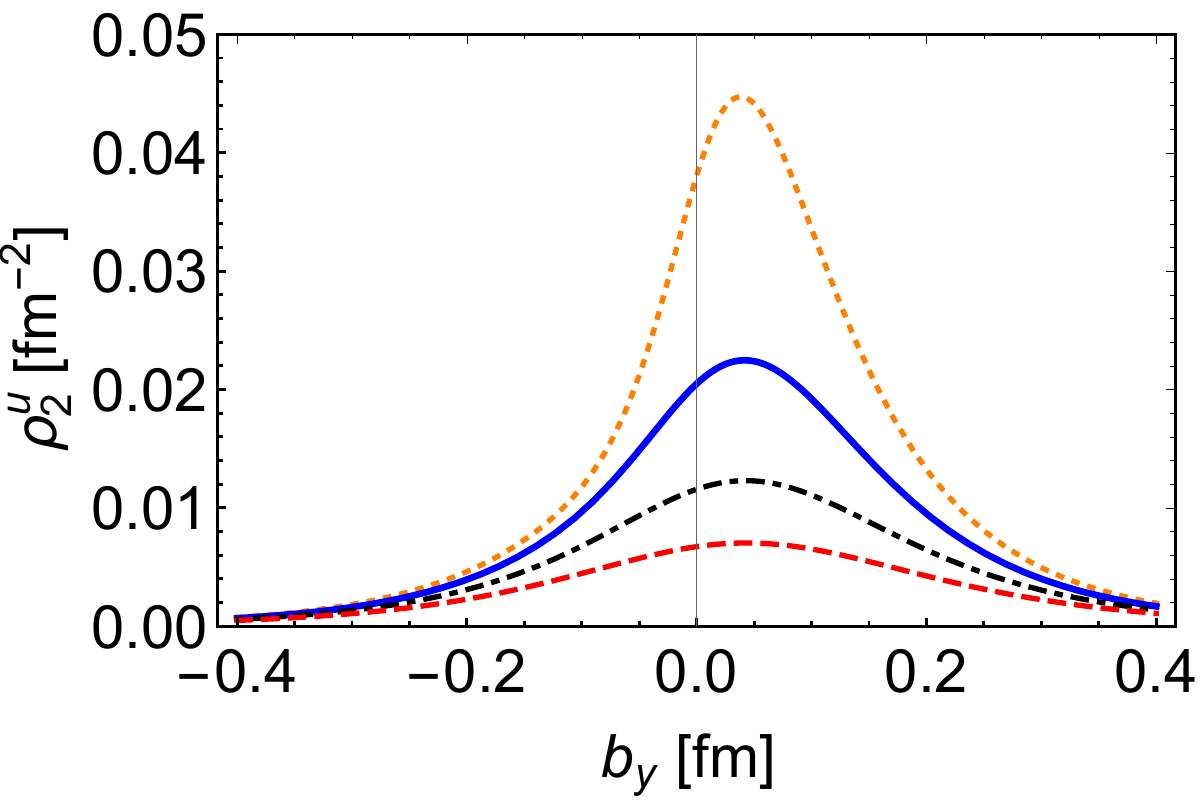}
\caption{The light-front transverse-spin distribution $\rho_1^u\left(\bm{b}_{\bot },\bm{s}_{\perp}\right)$ and $\rho_2^u\left(\bm{b}_{\bot },\bm{s}_{\perp}\right)$, $s_{\bot}=(1,0)$ at constant $b_x$ fm: $b_x=0.1$ fm--orange dotted curve, $b_x=0.15$ fm--blue solid curve, $b_x=0.2$ fm--black dot-dashed curve, $b_x=0.25$ fm--red dashed curve. }\label{rho12}
\end{figure}
\begin{figure}
\centering
\includegraphics[width=0.47\textwidth]{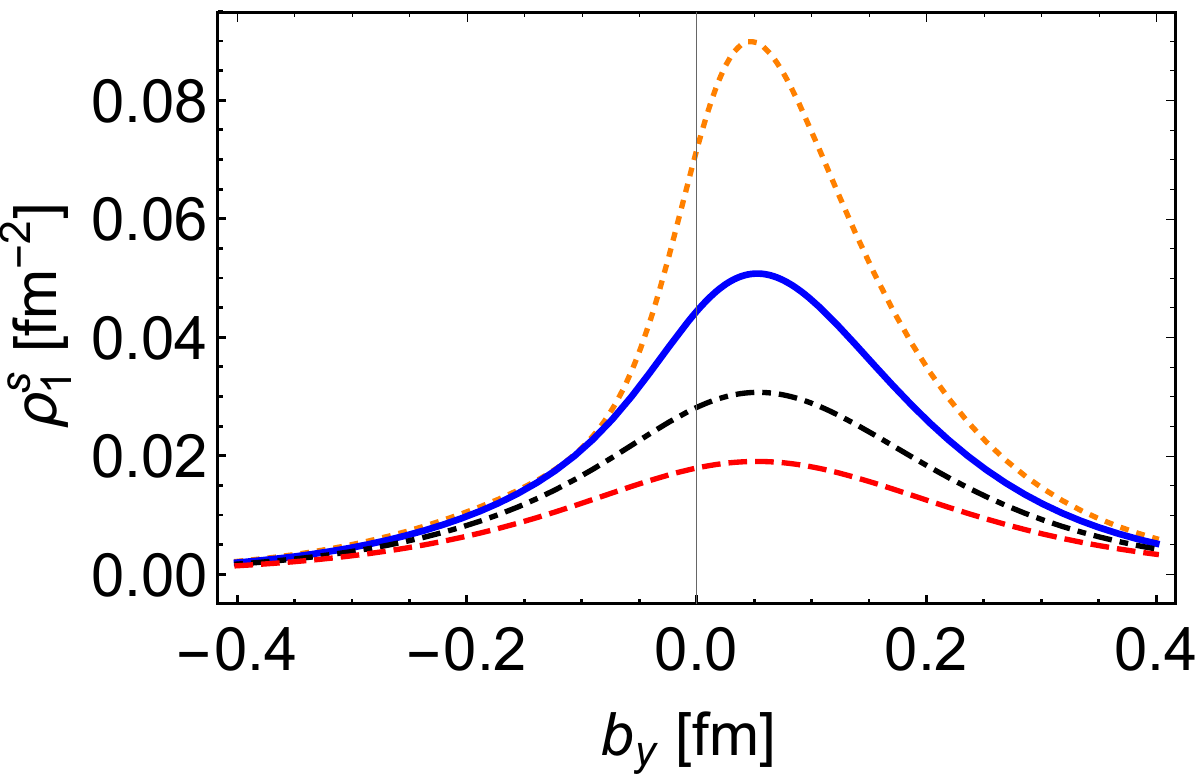}
\qquad
\includegraphics[width=0.47\textwidth]{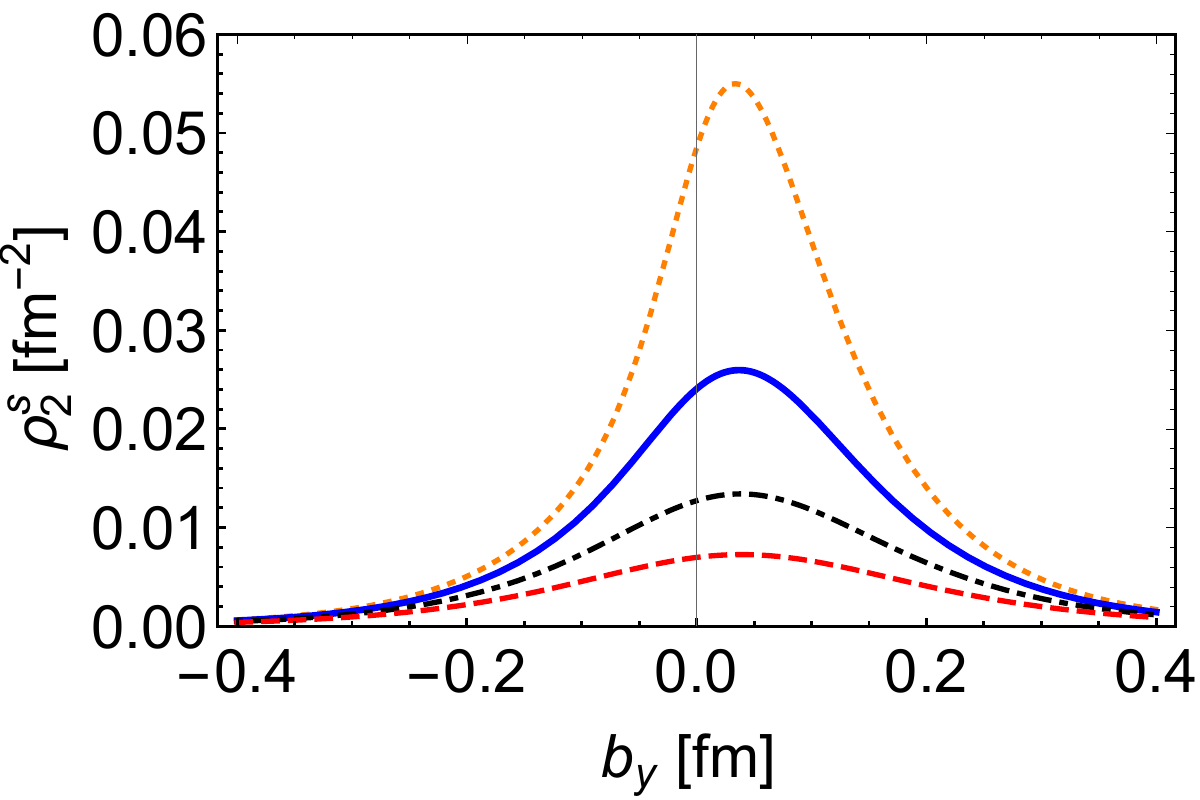}
\caption{The light-front transverse-spin distribution $-\rho_1^s\left(\bm{b}_{\bot },\bm{s}_{\perp}\right)$ and $-\rho_2^s\left(\bm{b}_{\bot },\bm{s}_{\perp}\right)$, $s_{\bot}=(1,0)$ at constant $b_x$ fm: $b_x=0.1$ fm--orange dotted curve, $b_x=0.15$ fm--blue solid curve, $b_x=0.2$ fm--black dot-dashed curve, $b_x=0.25$ fm--red dashed curve. }\label{rho12}
\end{figure}

We present the diagrams illustrating the light-front transverse-spin distribution for a valence quark polarized in the $+x$ direction on the light-front transverse plane. It is observed that the transverse spin density no longer exhibits symmetry around $\bm{b}_{\perp}=(b_x,b_y)$. The peaks have shifted to $(b_x=0, b_y>0)$, indicating a transfer of strength from $b_y<0$ to $b_y>0$.

The average transverse shift is
\begin{align}\label{ats}
\langle b_{\bot}^y\rangle_n^q=\frac{\int d^2\bm{b}_{\bot }b_{\bot }^y \rho_n^q\left(\bm{b}_{\bot },\bm{s}_{\perp}\right)}{\int d^2\bm{b}_{\bot } \rho_n^q\left(\bm{b}_{\bot },\bm{s}_{\perp}\right)},
\end{align}
and the profile of $b_y$ remains symmetric around the line $b_x=0$ for a transverse quark spin $s_{\bot} = (1, 0)$ in the $x$ direction. Table \ref{tb4} compares the average transverse shift of $\langle b_{\bot}^y\rangle_n^q$ in the NJL model and DSE. The results show that the shifts are much smaller than those obtained from the NJL model, primarily due to $F_K\neq 0$, which makes the form factors harder than those from the NJL model. If we set $F_K=0$, we would obtain identical formulas as those from the NJL model. When compared with Ref.~\cite{QCDSF:2007ifr}, our results appear to be small. Additionally, it can be observed from the table that the average transverse shift of $u$ quark is larger than that of $s$ quark, and that the average transverse shifts of $u$ quarks in kaons are slightly larger than those in pions.
\begin{center}
\begin{table}
\centering
\caption{Comparison of the average transverse shift $\langle b_{\bot}^y\rangle_n$ for quarks of pion and kaon in the NJL model and DSE in units of fm.}\label{tb4}
\begin{tabular}{p{1.1cm}p{1.1cm} p{1.1cm} p{1.1cm} p{1.1cm}p{1.1cm}p{1.1cm}}
\hline\hline
 &$\langle b_{\bot}^y\rangle_1^{u,K}$&$\langle b_{\bot}^y\rangle_2^{u,K}$&$\langle b_{\bot}^y\rangle_1^{s,K}$&$\langle b_{\bot}^y\rangle_2^{s,K}$&$\langle b_{\bot}^y\rangle_1^{u,\pi}$&$\langle b_{\bot}^y\rangle_2^{u,\pi}$\\
\hline
NJL &0.116&0.083&0.091&0.063&0.106&0.071\\
DSE &0.056&0.039&0.048&0.025&0.055&0.037\\
\hline\hline
\end{tabular}
\end{table}
\end{center}

\section{Summary and conclusions}\label{excellent}
In this paper, we utilize the contact interaction (CI) to evaluate a range of twist-two, -three and -four kaon generalised transverse momentum dependent parton distribution functions (GTMDs) within the framework of Dyson-Schwinger equations (DSEs), employing proper time regularization with an infrared cutoff to simulate confinement. This approach provides a relatively comprehensive physical depiction of the kaon. Through GTMDs, we derive the kaon twist-two, -three and -four generalized parton distributions (GPDs) and transverse momentum dependent distributions (TMDs). By analyzing the kaon GPDs, we are able to investigate the impact parameter space (IPS) parton distribution functions (PDFs). Additionally, our study also encompasses an examination of the kaon PDFs through analysis of TMDs.

We have conducted a study on the vector and tensor GPDs for both the $u$ quark and $s$ quark. Our findings indicate that the electromagnetic radius of the kaon's $u$-quark is similar to that of the pion's $u$-quark, and it is larger than the electromagnetic radius of the $s$ quark of kaon. Furthermore, it is noted that in this study, a hard CI form factor was used which approximates a non-zero constant value as $Q^2$ approaches infinity. We also study the light-cone energy radius and light-cone charge radius and their ratio. The findings show that the light-cone radius in the NJL model is larger than that in the DSE. Additionally, the light-cone radius of the $u$ quark in a kaon exceeds that of the $s$ quark in a kaon. Moreover, the light-cone radius of the $u$ quark in a kaon is comparable to that of the $u$ quark in a pion. The dressed ratios of the light-cone radius are greater than those of the bare ratios. Specifically, for dressed light-cone radii, the ratios in the NJL model are smaller than those in DSE; however, this does not hold true for bare light-cone radii.

Our calculations indicate that the twist-two TMDs for the kaon, unlike those for the pion, do not yield a zero value for the Boer-Mulders function. Additionally, we have evaluated $h_1^{\perp}(x,\bm{k}_{\perp}^2)$ as well as the twist-three distributions $e$, $f^{\perp}$ and $g^{\perp}$; and the twist-four TMDs $f_3$ and $h_3^{\perp}$. Among these TMDs, $h_1^{\perp}$, $g^{\perp}$ and $h_3^{\perp}$ are derived from chiral odd GTMDs. Since the Wilson line is not considered in our analysis, the formulas of the chiral odd TMDs indicate that without the inclusion of the Wilson line, the ability of chiral odd TMDs to provide substantial information about the internal structure of hadrons is limited.

The Wigner distributions for different quark polarizations in the pion are studied using a CI that includes dynamical spin effects. The three twist-two Wigner distributions from kaon GTMDs are evaluated in impact parameter space as well as in transverse momentum space for different quark polarizations, namely unpolarized, longitudinally polarized, and transversely polarized inside an unpolarized kaon and pion. The diagrams of the three Wigner distributions are plotted in impact parameter space and momentum space.

In both momentum space and impact parameter space, $\rho_{UU}\left( \bm{k}_{\bot },\bm{b}_{\perp}\right)$ of $u$ quark and $s$ quark exhibit circularly symmetric behavior. $\rho_{UL}\left(\bm{k}_{\bot },\bm{b}_{\perp}\right)$ show dipolar distortion patterns in both spaces. In momentum space, $\rho_{UT}\left(\bm{k}_{\bot },\bm{b}_{\perp}\right)$ for $u$ quark and $s$ quark display circularly symmetric behavior, but the axis of symmetry is no longer $(k_x=0,k_y=0)$. For the $u$ quark and $s$ quark of the kaon it shifts to $(k_x<0,k_y=0)$. In impact-parameter space, the $\rho_{UT}\left(\bm{k}_{\bot },\bm{b}_{\perp}\right)$ exhibits dipolar distortion patterns. The symmetry about the $b_x=0$ axis still holds, but unlike the case of pions, not all values in the region $b_y>0$ are negative and positive in the region $b_y<0$. The dividing line between positive and negative values is no longer at $b_y=0$.

Then we studied the mixed transverse Wigner distributions by integrating $b_x$ and $k_y$ of Wigner distributions, all these studies may offer some information of quarks inside the kaon. For $\tilde{\rho}_{UU}\left(k_x, b_y\right)$, the maximum value are at origin of coordinates $\left(k_x=0, b_y=0\right)$, and decreasing with the spread of coordinates. The maximum values of the quark $\tilde{\rho}_{UL}\left(\pm 0.1, k_x, b_y\right)$ are around $\left(|k_x|=0.4, |b_y|=0.1\right)$, in the first and third quadrant, the helicity of quarks are positive and in the second and fourth quadrant are negative. The $\tilde{\rho}_{UT}\left(\pm 0.1,k_x,b_y\right)$ for the $u$ quark is not symmetric at axis $k_x=0$; its maximum values occur at $(k_{x}=\pm 0.1,b_{y}=\mp 0.1)$. It is negative when $b_{y}> 0$, and positive when $b_{y}< 0$. Conversely, the behavior of the $s$ quark is opposite to that of the $u$ quark in this regard. The kaon's behavior in $\tilde\rho _{UT}(\pm 0.1, k_x, b_y )$ is very different from that of pion's behavior. In the isospin limit where $M_u=M_d$, $\tilde\rho _{ UT }(\pm 0.1,k_x,b_y )$ is still axisymmetric about $k_ { x }=0$, but this symmetry no longer exists for $K$ meson.

The spin-OAM correlation of the quark inside the kaon meson is an intriguing area of research. Our results indicate that the absolute value of $C_z$ in the NJL model is greater than that in DSE. In the NJL model, the absolute value of spin-OAM correlation for the $u$ quark in the kaon is similar to that of the $u$ quark in the pion. However, in the DSE, regardless of the GTMD method and GPD method, the spin-OAM correlation for the $u$ quark in the kaon is smaller than that of the $u$ quark in the pion. Additionally, we observed that the absolute value of spin-OAM correlation for the $s$ quark is smaller than that for the $u$ quark in the kaon, which aligns with findings from other models. Furthermore, our results show that  peak distribution for $C_{z,K}^{u}(x,\bm{k}_{\perp}^2)$ occurs at a transverse momentum of approximately $k_{\perp}\simeq0.38$ GeV, while for  $C_{z,K}^{s}(x,\bm{k}_{\perp}^2)$ it occurs at around  $k_{\perp}\simeq0.4$ GeV.

Finally, the information regarding transversity kaons is contained within the tensor GPD. It has been observed that a quark of a kaon in the positive-$x$ direction of the light-front transverse plane causes a distinct distortion of the transverse spin density, resulting in a shift of its peak in the positive-$y$ direction. This distortion decreases as the resolving scale increases. A comparison between the average transverse shifts $\langle b_{\bot}^y\rangle_n$ of $u$ and $s$ quarks in both NJL and DSE reveals that the average transverse shift of $u$ quarks is greater than that of $s$ quarks. In addition, it has been determined that the average transverse shifts of $u$ quarks in kaons are slightly larger than those of $u$ quarks in pions.

This paper is an extension of our previous work on pion GTMDs. In this paper, we compare the results of pion and kaon GTMDs, which complements our earlier findings and provides a comprehensive physical understanding of pseudoscalar mesons. However, this represents only a small step towards obtaining a more realistic physical picture of pseudoscalar mesons. A practical realization of the Wilson line should be considered in order to achieve this goal. Such a practical Wilson line would lead to time-reversal-odd GTMDs, which could provide additional insights that may be valuable for research involving realistic interactions, as compared to the time-reversal-even functions calculated in this study.

Another approach to furthering our understanding is to calculate the realistic light front wave functions of pseudoscalar mesons. These wave functions can be used to predict a wide range of properties associated with pseudoscalar mesons. These are all areas for future research, and efforts are currently underway in these directions.

\appendix
\section{Appendix 1: useful formulas}\label{AppendixT1}
Here we use the gamma-functions ($n\in \mathbb{Z}$, $n\geq 0$)
\begin{subequations}\label{app1}
\begin{align}
\mathcal{C}_0(z)&:=\int_0^{\infty} \mathrm{d}s\, s \int_{\tau_{uv}^2}^{\tau_{ir}^2} \mathrm{d}\tau \, e^{-\tau (s+z)}=z[\Gamma (-1,z\tau_{uv}^2 )-\Gamma (-1,z\tau_{ir}^2 )]\,, \\
\mathcal{C}_n(z)&:=(-)^n\frac{z^n}{n!}\frac{\mathrm{d}^n}{\mathrm{d}\sigma^n}\mathcal{C}_0(z)\,, \\
\bar{\mathcal{C}}_i(z)&:=\frac{1}{z}\mathcal{C}_i(z),
\end{align}
\end{subequations}
where $\tau_{uv,ir}=1/\Lambda_{\text{UV},\text{IR}}$ are, respectively, the infrared and ultraviolet regulators described above, with $\Gamma (\alpha,y )$ being the incomplete gamma-function. They can be illustrated with simple examples:
\begin{subequations}\label{app2}
\begin{align}
\bar{\mathcal{C}}_0(z)&:=\Gamma (-1,z\tau_{ir}^2 )-\Gamma (-1,z\tau_{uv}^2 )\,, \\
\bar{\mathcal{C}}_1(z)&:=\Gamma (0,z\tau_{ir}^2 )-\Gamma (0,z\tau_{uv}^2 )\,, \\
2\bar{\mathcal{C}}_2(z)&:=z\frac{\mathrm{d}^2}{\mathrm{d}\sigma^2}\mathcal{C}_0(z)=\Gamma (-1,z\tau_{ir}^2 )-\Gamma (-1,z\tau_{uv}^2 ).
\end{align}
\end{subequations}
in general
\begin{align}\label{app3}
n! \bar{\mathcal{C}}_n(z)=\Gamma (n-1,z\tau_{ir}^2 )-\Gamma (n-1,z\tau_{uv}^2 ),
\end{align}
such expressions are useful, e.g. in expressing the Bethe-Salpeter kernel in Eq. (\ref{dqm8}).
\begin{subequations}\label{app4}
\begin{align}
\mathcal{K}_{EE}^K &=\int_0^1 d\alpha \bar{\mathcal{C}}_0(\omega(\alpha,Q^2))+\int_0^1 d\alpha[M_uM_s-\alpha\bar{\alpha}Q^2-\omega(\alpha,Q^2)]\bar{\mathcal{C}}_1(\omega(\alpha,Q^2))\,, \\
\mathcal{K}_{EF}^K &=\frac{Q^2}{2M_{us}}\int_0^1 d\alpha (\alpha M_u+\bar{\alpha}M_s)\bar{\mathcal{C}}_1(\omega(\alpha,Q^2))\,, \\
\mathcal{K}_{FE}^K &=\frac{2M_{us}^2}{Q^2}\mathcal{K}_{EF}^K\,, \\
\mathcal{K}_{FF}^K &=-\frac{1}{2}\int_0^1 d\alpha [M_uM_s+\bar{\alpha}M_u^2+\alpha M_s^2] \bar{\mathcal{C}}_1(\omega(\alpha,Q^2))\,,
\end{align}
\end{subequations}
when computing observables, one must employ the canonically normalised amplitude, viz. $\Gamma_K$ rescaled such that
\begin{align}\label{app5}
1=\frac{d}{dQ^2}\Pi_K(Z,Q)|_{Z=Q},
\end{align}
where
\begin{align}\label{app6}
\Pi_K(Z,Q)=6\text{tr}\int \frac{d^4l}{(2\pi)^4}\Gamma_K(-Z)S_u(l+Q)\Gamma_K(Z)S_s(l).
\end{align}
$z$ represent the $\sigma$ functions in the following,
\begin{subequations}\label{aA8}
\begin{align}
\sigma_1^{z,u}&=z+\frac{1-x}{1+u\xi}M_u^2+\frac{x+u\xi}{1+u\xi}M_s^2-\frac{x+u\xi}{1+u\xi}\frac{1-x}{1+u\xi} m_K^2\,, \\
\sigma_2^z&=z+M_u^2-\frac{1}{4}(1+\frac{x}{ \xi })(1-\frac{x}{\xi }) t\,, \\
\sigma_3^z&=z+\bar{\alpha} M_u^2+\alpha M_s^2-\alpha \bar{\alpha} m_K^2-\left(\xi+x-\alpha(1+\xi)\right) \left(\xi-x+\alpha (1-\xi)\right)\frac{t}{4\xi^2}\,, \\
\sigma_4^z&=\sigma_1^{z,0}+\alpha(1-\alpha-x)t\,, \\
\sigma_5&=M_u^2-x(1-x)t\,, \\
\sigma_6&=(x+y)(x+y-1)m_K^2-xyt+(x+y)M_u^2+(1-x-y)M_s^2\,,
\end{align}
\end{subequations}

\section{Appendix 2: Twist-three GTMDs }\label{AppendixT2}
The twist-three kaon GTMDs, which are obtained with the following choices ($r=\bm{k}_{\perp}^2$)
\begin{align}\label{t3gpdo}
\mathcal{H}\rightarrow\{&\mathcal{H}_1= 1,\mathcal{H}_2= i\gamma_5,\mathcal{H}_3=i\gamma_j ,\mathcal{H}_4=i\gamma_j \gamma_5,\mathcal{H}_5=i \gamma_5\sigma_{ij},\mathcal{H}_6=i \gamma_5\sigma_{\mu\nu}n_{\mu}\bar{n}_{\nu}\}.
\end{align}
then we obtain
\begin{subequations}
\begin{align}\label{t3gpd}
W^{[\mathcal{H}_1]}&\rightarrow \frac{m_K}{P\cdot n}E_2\,, \\
W^{[\mathcal{H}_2]}&\rightarrow \frac{m_K}{P\cdot n}\frac{i\varepsilon_{ij}^{\perp}k_i\Delta_j} {m_K^2} \tilde{E}_2\,, \\
W^{[\mathcal{H}_3]}&\rightarrow \frac{m_K}{P\cdot n}\left[\frac{k_j}{m_K}F_2^k+\frac{\Delta_j}{m_K} F_2^{\Delta}\right]\,, \\
W^{[\mathcal{H}_4]}&\rightarrow \frac{m_K}{P\cdot n}\left[\frac{i\varepsilon_{ij}^{\perp}k_i}{m_K}G_2^k+\frac{i\varepsilon_{ij}^{\perp}\Delta_i}{m_K} G_2^{\Delta}\right]\,, \\
W^{[\mathcal{H}_5]}&\rightarrow \frac{m_K}{P\cdot n}\left[i\varepsilon_{ij}^{\perp}H_2\right]\,, \\
W^{[\mathcal{H}_6]}&\rightarrow \frac{m_K}{P\cdot n}\left[\frac{i\varepsilon_{ij}^{\perp}k_i\Delta_j}{m_K^2}\tilde{H_2}\right]\,,
\end{align}
\end{subequations}
\begin{align}\label{t3gpd}
E_2^u(x,r,\xi,t)=& \frac{\check{N}^{EF}N_c}{4\pi^3}  \frac{(M_u-M_s)\theta_{\bar{\xi} 1}}{ m_K(1+\xi)}\frac{\bar{\mathcal{C}}_2(\sigma_1^{r,1})}{\sigma_1^{r,1}}+\frac{\check{N}^{EF} N_c }{4\pi^3} \frac{(M_u-M_s)\theta_{\xi 1} }{m_K(1-\xi)}\frac{\bar{\mathcal{C}}_2(\sigma_1^{r,-1})}{\sigma_1^{r,-1}} \nonumber\\
+&  \frac{\tilde{N}_{EF}N_c}{4\pi^3} \left(\frac{\theta_{\bar{\xi} 1}m_K }{ (1+\xi)}\frac{\bar{\mathcal{C}}_2(\sigma_1^{r,1})}{\sigma_1^{r,1}}+ \frac{\theta_{\xi 1} m_K }{(1-\xi)}\frac{\bar{\mathcal{C}}_2(\sigma_1^{r,-1})}{\sigma_1^{r,-1}}\right) \nonumber\\
+&E_K^2M_u \frac{1}{m_K} \frac{N_c }{2\pi ^3} \frac{\theta_{\bar{\xi} \xi}}{\xi}\frac{1}{\sigma_2^r}\bar{\mathcal{C}}_2(\sigma_2^r)-F_KE_K  \frac{( M_s^2-M_u^2)}{M_sm_K}\frac{N_c }{4\pi ^3} \frac{\theta_{\bar{\xi} \xi}}{\xi}\frac{\bar{\mathcal{C}}_2(\sigma_2^r)}{\sigma_2^r}\nonumber\\
+&F_KE_K  \frac{(M_s+M_u)t}{2M_uM_sm_K}\frac{N_c }{4\pi ^3} \frac{\theta_{\bar{\xi} \xi}}{\xi}\frac{\bar{\mathcal{C}}_2(\sigma_2^r)}{\sigma_2^r}-\hat{N}^{EF}M_u( M_s-M_u)\frac{N_c\theta_{\bar{\xi} \xi}}{2\pi ^3m_K\xi}\frac{\bar{\mathcal{C}}_2(\sigma_2^r)}{\sigma_2^r}\nonumber\\
-&\frac{\hat{N}^{EF}N_c }{4\pi ^3}\frac{t}{m_K} \frac{\theta_{\bar{\xi} \xi}}{\xi}\frac{\bar{\mathcal{C}}_2(\sigma_2^r)}{\sigma_2^r}+\frac{N^{FF}N_c}{8\pi ^3} \frac{xt}{m_K}\frac{\theta_{\bar{\xi} \xi}}{\xi^3}\frac{\bar{\mathcal{C}}_2(\sigma_2^r)}{\sigma_2^r}\nonumber\\
+& \frac{N^{EF}N_c}{16\pi^3} \int_0^1  d\alpha  \frac{\theta_{\alpha \xi}}{m_K\xi}(2M_u(m_K^2-(M_s-M_u)^2)-M_s t) \frac{6\bar{\mathcal{C}}_3(\sigma_3^r)}{[\sigma_3^r]^2},
\end{align}
\begin{align}\label{t3gpd}
\tilde{E}_2^u(x,\bm{k}_{\perp}^2,\xi,t)=0
\end{align}
\begin{align}\label{t3gpd}
F_2^{u,k} (x,r,\xi,t)=&\frac{\bar{N}^{EF}N_c}{2\pi^3} \left(\frac{\theta_{\bar{\xi} 1}}{(1+\xi)} \frac{\bar{\mathcal{C}}_2(\sigma_1^{r,1})}{\sigma_1^{r,1}}+\frac{\theta_{\xi 1}}{(1-\xi)} \frac{\bar{\mathcal{C}}_2(\sigma_1^{r,-1})}{\sigma_1^{r,-1}}\right) \nonumber\\
+&\frac{(E_K^2-2\bar{N}^{EF})N_c}{4\pi^3}  \frac{\theta_{\bar{\xi} \xi}}{\xi }\frac{\bar{\mathcal{C}}_2(\sigma_2^r)}{\sigma_2^r}+  \frac{N^{FF}N_c}{8\pi^3} \frac{xt \theta_{\bar{\xi} \xi}}{\xi^3}\frac{\bar{\mathcal{C}}_2(\sigma_2^r)}{\sigma_2^r}\nonumber\\
+&\frac{N^{EF}N_c}{16\pi^3} \int_0^1  d\alpha \frac{\theta_{\alpha \xi}}{\xi} (2(m_K^2-\left(M_u-M_s\right)^2)-t ) \frac{6\bar{\mathcal{C}}_3(\sigma_3^r)}{[\sigma_3^r]^2},
\end{align}
\begin{align}\label{t3gpd}
& F_2^{u,\Delta} (x,r,\xi,t)\nonumber\\
=&-\frac{N^{EE}N_c }{8\pi^3} \left( \frac{\theta_{\bar{\xi} 1}}{(1+\xi)} \frac{\bar{\mathcal{C}}_2(\sigma_1^{r,1})}{\sigma_1^{r,1}}-\frac{\theta_{\xi 1}}{(1-\xi)} \frac{\bar{\mathcal{C}}_2(\sigma_1^{r,-1})}{\sigma_1^{r,-1}}\right) - \frac{\bar{N}^{EF}N_c }{8\pi^3}\theta_{\bar{\xi} 1} \frac{2x+\xi-1}{(1+\xi)^2} \frac{\bar{\mathcal{C}}_2(\sigma_1^{r,1})}{\sigma_1^{r,1}}\nonumber\\
+& \frac{\bar{N}^{EF}N_c }{8\pi ^3}\theta_{\xi 1}\frac{2x-\xi-1}{(1-\xi)^2} \frac{\bar{\mathcal{C}}_2(\sigma_1^{r,-1})}{\sigma_1^{r,-1}}-(E_K^2-2\bar{N}^{EF})\frac{N_c }{8\pi^3}\theta_{\bar{\xi} \xi} \frac{x}{\xi^2 }\frac{1}{\sigma_2^r}\bar{\mathcal{C}}_2(\sigma_2^r)\nonumber\\
+& \frac{N^{FF}N_c }{16\pi^3}\frac{\theta_{\bar{\xi} \xi}}{\xi^2}(\frac{1}{2}t-\frac{3x^2t}{ 2\xi^2 }-2M_u^2) \frac{\bar{\mathcal{C}}_2(\sigma_2^r)}{\sigma_2^r}\nonumber\\
-&\frac{N^{EF} N_c}{32\pi^3} \int_0^1  d\alpha\frac{\theta_{\alpha \xi}}{\xi^2}  (2(m_K^2-\left(M_u-M_s\right)^2)-t)\frac{(x-\alpha)}{\xi^2}  \frac{6\bar{\mathcal{C}}_3(\sigma_3^r)}{[\sigma_3^r]^2},
\end{align}
\begin{align}\label{t3gpd}
G_2^{u,k}(x,r,\xi,t)=&  \frac{N^{FF}N_c }{4\pi^3}\ (4m_K^2-t)\theta_{\bar{\xi} \xi}\frac{\bar{\mathcal{C}}_2(\sigma_2^r)}{\sigma_2^r}\nonumber\\
-&\frac{N^{EF}N_c}{8\pi^3} \int_0^1  d\alpha  \theta_{\alpha \xi}  (4m_K^2-t)\frac{6\bar{\mathcal{C}}_3(\sigma_3^r)}{[\sigma_3^r]^2},
\end{align}
\begin{align}\label{t3gpd}
G_2^{u,\Delta}(x,r,\xi,t)=&\frac{N^{FF}N_c}{8\pi^3}\theta_{\bar{\xi} 1}  \frac{x t}{ \xi^3}\frac{\bar{\mathcal{C}}_2(\sigma_2^r)}{\sigma_2^r}-\frac{N^{EF}N_c}{16\pi^3} \int_0^1  d\alpha\theta_{\alpha \xi}  \frac{ t(x-\alpha)}{\xi^3}\frac{6\bar{\mathcal{C}}_3(\sigma_3^r)}{[\sigma_3^r]^2},
\end{align}
\begin{align}\label{t3gpd}
H_2^u(x,r,\xi,t)=& \frac{\tilde{N}_{EF}N_c}{2\pi^3}\frac{\theta_{\bar{\xi}1}\left(M_s^2-M_u^2\right)}{m_K (1+\xi)} \frac{\bar{\mathcal{C}}_2(\sigma_1^{r,1})}{\sigma_1^{r,1}}- \frac{\tilde{N}_{EF}N_c}{2\pi^3} \frac{\theta_{\xi 1}\left(M_s^2-M_u^2\right)}{ m_K (1-\xi)} \frac{\bar{\mathcal{C}}_2(\sigma_1^{r,-1})}{\sigma_1^{r,-1}} \nonumber\\
-& \frac{\tilde{N}_{EF}N_c}{2\pi^3}\theta_{\bar{\xi}1}\frac{1-2x-\xi}{(1+\xi)^2} m_K \frac{\bar{\mathcal{C}}_2(\sigma_1^{r,1})}{\sigma_1^{r,1}}+ \frac{\tilde{N}_{EF}N_c}{2\pi^3}\theta_{\xi 1} \frac{1-2x+\xi}{(1-\xi)^2} m_K\frac{\bar{\mathcal{C}}_2(\sigma_1^{r,-1})}{\sigma_1^{r,-1}} \nonumber\\
-& \frac{\hat{N}_{EF}N_c\theta_{\bar{\xi} \xi} xt}{4\pi ^3m_K\xi^2} \frac{\bar{\mathcal{C}}_2(\sigma_2^r)}{\sigma_2^r}-  \frac{N^{FF}N_c}{4\pi^3} \frac{(4m_K^2-t)}{m_K}\frac{\theta_{\bar{\xi} \xi}}{\sigma_2^r}\bar{\mathcal{C}}_2(\sigma_2^r)\nonumber\\
+&  \frac{N^{EF}N_c}{8\pi^3} \int_0^1  d\alpha \frac{\theta_{\alpha \xi} }{m_K} ((M_u+(M_s-M_u)\alpha )(4m_K^2-t))  \frac{6\bar{\mathcal{C}}_3(\sigma_3^r)}{[\sigma_3^r]^2}\nonumber\\
-& \frac{N^{EF}N_c}{8\pi^3} \int_0^1  d\alpha \frac{\theta_{\alpha \xi} (x-\alpha)t(M_s-M_u)}{m_K \xi^2}    \frac{6\bar{\mathcal{C}}_3(\sigma_3^r)}{[\sigma_3^r]^2},
\end{align}
\begin{align}\label{agtmd}
\tilde{H}_2^u(x,r,\xi,t)=& \frac{\tilde{N}_{EF}N_c}{4\pi^3}\left( \frac{\theta_{\bar{\xi}1}m_K}{ (1+\xi)}  \frac{\bar{\mathcal{C}}_2(\sigma_1^{r,1})}{\sigma_1^{r,1}}+\frac{\theta_{\xi1}m_K}{(1-\xi)}\frac{\bar{\mathcal{C}}_2(\sigma_1^{r,-1})}{\sigma_1^{r,-1}}\right)\nonumber\\
-&\hat{N}^{EF} \frac{N_c m_K}{2\pi ^3}  \frac{\theta_{\bar{\xi}\xi}}{\xi}\frac{\bar{\mathcal{C}}_2(\sigma_2^r)}{\sigma_2^r}+ \frac{N^{EF}N_c}{8\pi^3} \int_0^1  d\alpha \frac{\theta_{\alpha \xi}}{\xi} m_K(M_s-M_u)  \frac{6\bar{\mathcal{C}}_3(\sigma_3^r)}{[\sigma_3^r]^2}.
\end{align}
Among the twist-three GTMDs, due to the hermiticity constraint of $\Delta$, $E_2$, $F_2^k$, $G_2^{\Delta}$, and $\tilde{H}_2$ are T-even, $F_2^{\Delta}$, $G_2^{\Delta}$, and $H_2$ are T-odd.

\section{Appendix 3: Twist-four GTMDs}\label{AppendixT3}
The twist-four kaon GTMDs, which are obtained with the following choices ($r=\bm{k}_{\perp}^2$)
\begin{align}\label{t3gpdo}
\mathcal{H}\rightarrow\{&\mathcal{H}_1= i\gamma\cdot \bar{n},\mathcal{H}_2= i\gamma\cdot \bar{n}\gamma_5,\mathcal{H}_3=i\gamma_5\sigma_{j\mu}\bar{n}_{\mu}\}.
\end{align}
then we obtain
\begin{subequations}
\begin{align}\label{t3gpd}
W^{[\mathcal{H}_1]}&\rightarrow \frac{m_K^2}{(P\cdot n)^2}F_3\,, \\
W^{[\mathcal{H}_2]}&\rightarrow \frac{m_K^2}{(P\cdot n)^2}\left[\frac{i\varepsilon_{ij}^{\perp}k_i\Delta_j}{m_K^2}\tilde{G_3}\right]\,, \\
W^{[\mathcal{H}_3]}&\rightarrow \frac{m_K^2}{(P\cdot n)^2}\left[\frac{i\varepsilon_{ij}^{\perp}k_i}{m_K}H_3^k+\frac{i\varepsilon_{ij}^{\perp}\Delta_i}{m_K} H_3^{\Delta}\right]\,,
\end{align}
\end{subequations}
\begin{align}\label{agtmd}
& F_3^u(x,r,\xi,t)\nonumber\\
=&-\frac{N^{EE}}{4m_K^2}\frac{N_c\theta_{\bar{\xi}1}}{4\pi^3}  \frac{(1-\xi)(4m_K^2-t) }{(1+\xi)}  \frac{\bar{\mathcal{C}}_2(\sigma_1^{r,1})}{\sigma_1^{r,1}}-\frac{N^{EE}}{4m_K^2} \frac{N_c\theta_{\xi1}}{4\pi ^3} \frac{(4m_K^2-t) (1+\xi)}{(1-\xi)} \frac{\bar{\mathcal{C}}_2(\sigma_1^{r,-1})}{\sigma_1^{r,-1}}\nonumber\\
-&\frac{\bar{N}^{EF}}{4m_K^2}\frac{N_c\theta_{\bar{\xi}1}}{4\pi^3}  \frac{(2x+\xi-1)(1-\xi)(4m_K^2-t)}{(1+\xi)^2}  \frac{\bar{\mathcal{C}}_2(\sigma_1^{r,1})}{\sigma_1^{r,1}} \nonumber\\
-& \frac{\bar{N}^{EF}}{4m_K^2} \frac{N_c\theta_{\xi1}}{4\pi ^3} \frac{(2x-\xi-1)(1+\xi)(4m_K^2-t)}{(1-\xi)^2} \frac{\bar{\mathcal{C}}_2(\sigma_1^{r,-1})}{\sigma_1^{r,-1}}\nonumber\\
+& \frac{(E_K^2-2\bar{N}^{EF})}{4m_K^2} \frac{N_c}{4\pi^3} x(4m_K^2-t) \frac{\theta_{\bar{\xi}\xi}}{\xi}\frac{\bar{\mathcal{C}}_2(\sigma_2^r)}{\sigma_2^r}+  \frac{(E_K^2-2\bar{N}^{EF})}{4m_K^2} \frac{N_c }{2\pi^3}  \frac{xt\theta_{\bar{\xi}\xi}}{\xi^3} \frac{\bar{\mathcal{C}}_2(\sigma_2^r)}{\sigma_2^r} \nonumber\\
+&\frac{N^{FF}}{4m_K^2}\frac{N_c}{8\pi^3}  \frac{\theta_{\bar{\xi}\xi}}{\xi}(1-\frac{x^2}{\xi^2})t (4m_K^2-t)\frac{\bar{\mathcal{C}}_2(\sigma_2^r)}{\sigma_2^r}-\frac{\bar{N}^{FF}}{2m_K^2} \frac{N_c}{4\pi^3}\frac{\theta_{\bar{\xi}\xi}xt}{\xi^3} \frac{\bar{\mathcal{C}}_2(\sigma_2^r)}{\sigma_2^r} \nonumber\\
+&\frac{N^{EF}}{4m_K^2} \frac{N_c }{16\pi^3}  \int_0^1  d\alpha  \frac{\theta_{\alpha \xi}}{\xi} \frac{6\bar{\mathcal{C}}_3(\sigma_3^r)}{[\sigma_3^r]^2}(4m_K^2-t)((2\alpha-x)(t-2(m_K^2-\left(M_u-M_s\right)^2))-t) \nonumber\\
+& \frac{N^{EF}}{8m_K^2} \frac{N_c}{4\pi^3}  \int_0^1  d\alpha \frac{\theta_{\alpha \xi}}{\xi^3}(2(m_K^2-(M_u-M_s)^2)-t)(x-\alpha)t\frac{6\bar{\mathcal{C}}_3(\sigma_3^r)}{[\sigma_3^r]^2},
\end{align}
\begin{align}\label{agtmd}
\tilde{G}_3^u(x,r,\xi,t)=\frac{N^{FF}N_c }{16\pi^3}\frac{\theta_{\bar{\xi}\xi}}{\xi} \frac{\bar{\mathcal{C}}_2(\sigma_2^r)}{\sigma_2^r}-\frac{N^{EF}N_c}{32\pi^3} \int_0^1  d\alpha \frac{\theta_{\alpha \xi}}{\xi} \frac{6\sigma_3^r}{[\sigma_3^r]^2},
\end{align}
\begin{align}\label{agtmd}
&H_3^{u,\Delta}(x,r,\xi,t)\nonumber\\
=&  \frac{N^{FF}N_c}{16\pi^2} \frac{\theta_{\bar{\xi}\xi}}{ m_K \xi}  \frac{\bar{\mathcal{C}}_2(\sigma_2^r)}{\sigma_2^r}- \frac{N^{EF}N_c}{32\pi^3} \int_0^1  d\alpha \frac{\theta_{\alpha \xi}(M_u+(M_s-M_u)\alpha )}{m_K\xi} \frac{6\bar{\mathcal{C}}_3(\sigma_3^r)}{[\sigma_3^r]^2},
\end{align}
\begin{align}\label{agtmd}
H_3^{u,k}(x,r,\xi,t)=&  \frac{\tilde{N}_{EF}N_c}{8\pi^3} \left( \frac{\theta_{\bar{\xi}1}}{m_K}\frac{\bar{\mathcal{C}}_2(\sigma_1^{r,1})}{\sigma_1^{r,1}} -\frac{\theta_{\xi 1}}{m_K}\frac{\bar{\mathcal{C}}_2(\sigma_1^{r,-1})}{\sigma_1^{r,-1}}\right)\nonumber\\
-&  \frac{\hat{N}^{EF}N_c }{4\pi ^3m_K} \theta_{\bar{\xi}\xi} \frac{\bar{\mathcal{C}}_2(\sigma_2^r)}{\sigma_2^r}- \frac{N^{EF}N_c}{32\pi^3} \int_0^1  d\alpha \theta_{\alpha \xi}\frac{(M_s-M_u)}{m_K}  \frac{6\bar{\mathcal{C}}_3(\sigma_3^r)}{[\sigma_3^r]^2}.
\end{align}
Among the twist-four GTMDs, due to the hermiticity constraint of $\Delta$, $F_3$, $\tilde{G}_3$, and $H_3^{\Delta}$ are T-even, $H_3^k$ is T-odd.

\section{Appendix 4: Twist-three GPDs}\label{AppendixT4}
The twist-three GPDs from GTMDs are defined as
\begin{subequations}
\begin{align}\label{a89}
E_2^u(x,\xi,t)&=\int d^2\bm{k}_{\perp}E_2(x,\bm{k}_{\perp}^2,\xi,t)\,,\\
F_2^u(x,\xi,t)&=\int d^2\bm{k}_{\perp}\left[\frac{\bm{k}_{\perp}\cdot \bm{\Delta}_{\perp}}{\bm{\Delta}_{\perp}^2}F_2^{k}+F_2^{\Delta}\right]\,,\\
G_2^u(x,\xi,t)&=\int d^2\bm{k}_{\perp}\left[\frac{\bm{k}_{\perp}\cdot \bm{\Delta}_{\perp}}{\bm{\Delta}_{\perp}^2}G_2^{k}+G_2^{\Delta}\right]\,,\\
H_2^u(x,\xi,t)&=\int d^2\bm{k}_{\perp}H_2(x,\bm{k}_{\perp}^2,\xi,t),
\end{align}
\end{subequations}

\begin{align}\label{t3gpd}
E_2^u(x,\xi,t)=& \frac{\check{N}^{EF}N_c}{8\pi^2}  \frac{(M_u-M_s)\theta_{\bar{\xi} 1}}{ m_K(1+\xi)}\bar{\mathcal{C}}_1(\sigma_1^{0,1})+\frac{\check{N}^{EF} N_c }{8\pi^2} \frac{(M_u-M_s)\theta_{\xi 1} }{m_K(1-\xi)}\bar{\mathcal{C}}_1(\sigma_1^{0,-1})\nonumber\\
+&  \frac{\tilde{N}_{EF}N_c}{8\pi^2} \left(\frac{\theta_{\bar{\xi} 1}m_K }{ (1+\xi)}\bar{\mathcal{C}}_1(\sigma_1^{0,1})+ \frac{\theta_{\xi 1} m_K }{(1-\xi)}\bar{\mathcal{C}}_1(\sigma_1^{0,-1})\right) \nonumber\\
+&E_K^2M_u \frac{1}{m_K} \frac{N_c }{4\pi ^2} \frac{\theta_{\bar{\xi} \xi}}{\xi}\bar{\mathcal{C}}_1(\sigma_2^0)-F_KE_K  \frac{( M_s^2-M_u^2)}{M_sm_K}\frac{N_c }{8\pi ^2} \frac{\theta_{\bar{\xi} \xi}}{\xi}\bar{\mathcal{C}}_1(\sigma_2^0)\nonumber\\
+&F_KE_K  \frac{(M_s+M_u)t}{2M_uM_sm_K}\frac{N_c }{8\pi ^2} \frac{\theta_{\bar{\xi} \xi}}{\xi}\bar{\mathcal{C}}_1(\sigma_2^0)-\hat{N}^{EF}M_u( M_s-M_u)\frac{N_c\theta_{\bar{\xi} \xi}}{4\pi ^2m_K\xi}\bar{\mathcal{C}}_1(\sigma_2^0)\nonumber\\
-&\frac{\hat{N}^{EF}N_c }{8\pi ^2}\frac{t}{m_K} \frac{\theta_{\bar{\xi} \xi}}{\xi}\bar{\mathcal{C}}_1(\sigma_2^0)+\frac{N^{FF}N_c}{16\pi ^2} \frac{xt}{m_K}\frac{\theta_{\bar{\xi} \xi}}{\xi^3}\bar{\mathcal{C}}_1(\sigma_2^0)\nonumber\\
+& \frac{N^{EF}N_c}{8\pi^2} \int_0^1  d\alpha  \frac{\theta_{\alpha \xi}}{m_K\xi}(2M_u(m_K^2-(M_s-M_u)^2)-M_s t) \frac{\bar{\mathcal{C}}_2(\sigma_3^0)}{\sigma_3^0},
\end{align}
\begin{align}\label{t3gpd}
F_2^u(x,\xi,t)=&-\frac{N^{EE}N_c }{16\pi^2} \left( \frac{\theta_{\bar{\xi} 1}}{(1+\xi)}\bar{\mathcal{C}}_1(\sigma_1^{0,1})-\frac{\theta_{\xi 1}}{(1-\xi)} \bar{\mathcal{C}}_1(\sigma_1^{0,-1})\right) \nonumber\\
-& \frac{\bar{N}^{EF}N_c }{16\pi^2}\theta_{\bar{\xi} 1} \frac{2x+\xi-1}{(1+\xi)^2} \bar{\mathcal{C}}_1(\sigma_1^{0,1})+ \frac{\bar{N}^{EF}N_c }{16\pi ^2}\theta_{\xi 1}\frac{2x-\xi-1}{(1-\xi)^2} \bar{\mathcal{C}}_1(\sigma_1^{0,-1}) \nonumber\\
-&(E_K^2-2\bar{N}^{EF})\frac{N_c }{16\pi^2}\theta_{\bar{\xi} \xi} \frac{x}{\xi^2 }\bar{\mathcal{C}}_1(\sigma_2^0)+ \frac{N^{FF}N_c }{32\pi^2}\frac{\theta_{\bar{\xi} \xi}}{\xi^2}(\frac{1}{2}t-\frac{3x^2t}{ 2\xi^2 }-2M_u^2)\bar{\mathcal{C}}_1(\sigma_2^0)\nonumber\\
-&\frac{ N^{EF}N_c}{16\pi^2} \int_0^1  d\alpha\frac{\theta_{\alpha \xi}}{\xi^2}  (2(m_K^2-\left(M_u-M_s\right)^2)-t)\frac{(x-\alpha)}{\xi^2}  \frac{\bar{\mathcal{C}}_2(\sigma_3^0)}{\sigma_3^0},
\end{align}
\begin{align}\label{t3gpd}
G_2^{u}(x,\xi,t)=&\frac{N^{FF}N_c}{16\pi^2}\theta_{\bar{\xi} 1}  \frac{x t}{ \xi^3}\bar{\mathcal{C}}_1(\sigma_2^0)-\frac{N^{EF}N_c}{8\pi^2} \int_0^1  d\alpha\theta_{\alpha \xi}  \frac{ t(x-\alpha)}{\xi^3}\frac{\bar{\mathcal{C}}_2(\sigma_3^0)}{\sigma_3^0},
\end{align}
\begin{align}\label{t3gpd}
 H_2^u(x,\xi,t)=& \frac{\tilde{N}_{EF}N_c}{4\pi^2}\frac{\left(M_s^2-M_u^2\right)}{m_K} \left(\frac{\theta_{\bar{\xi}1}\bar{\mathcal{C}}_1(\sigma_1^{0,1})}{ (1+\xi)} -\frac{\theta_{\xi 1}\bar{\mathcal{C}}_1(\sigma_1^{0,-1})}{  (1-\xi)} \right)\nonumber\\
-& \frac{\tilde{N}_{EF}N_c}{4\pi^2}\theta_{\bar{\xi}1}\frac{1-2x-\xi}{(1+\xi)^2} m_K \bar{\mathcal{C}}_1(\sigma_1^{0,1})+ \frac{\tilde{N}_{EF}N_c}{4\pi^2}\theta_{\xi 1} \frac{1-2x+\xi}{(1-\xi)^2} m_K\bar{\mathcal{C}}_1(\sigma_1^{0,-1}) \nonumber\\
-& \frac{\hat{N}_{EF}N_c\theta_{\bar{\xi} \xi} xt}{8\pi ^2m_K\xi^2} \bar{\mathcal{C}}_1(\sigma_2^0)-  \frac{N^{FF}N_c}{8\pi^2} \frac{(4m_K^2-t)}{m_K}\theta_{\bar{\xi} \xi}\bar{\mathcal{C}}_1(\sigma_2^0)\nonumber\\
+&  \frac{N^{EF}N_c}{4\pi^2} \int_0^1  d\alpha \frac{\theta_{\alpha \xi} }{m_K} ((M_u+(M_s-M_u)\alpha )(4m_K^2-t))  \frac{\bar{\mathcal{C}}_2(\sigma_3^0)}{\sigma_3^0}\nonumber\\
-& \frac{N^{EF}N_c}{4\pi^2} \int_0^1  d\alpha \frac{\theta_{\alpha \xi} (x-\alpha)t(M_s-M_u)}{m_K \xi^2}    \frac{\bar{\mathcal{C}}_2(\sigma_3^0)}{\sigma_3^0}.
\end{align}
Among the twist-three GPDs, due to the hermiticity constraint of $\Delta$, $E_2$ and $G_2$ are T-even, $F_2$ and $H_2$ are T-odd.

\section{Appendix 5: Twist-four GPDs}\label{AppendixT5}

\begin{subequations}
\begin{align}\label{a89}
F_3(x,\xi,t)&=\int d^2\bm{k}_{\perp}F_3(x,\bm{k}_{\perp}^2,\xi,t)\,,\\
H_3(x,\xi,t)&=\int d^2\bm{k}_{\perp}\left[\frac{\bm{k}_{\perp}\cdot \bm{\Delta}_{\perp}}{\bm{\Delta}_{\perp}^2}H_3^{k}+H_3^{\Delta}\right],
\end{align}
\end{subequations}
\begin{align}\label{agtmd}
& F_3^u(x,\xi,t)\nonumber\\
=&-\frac{N^{EE}}{4m_K^2}\frac{N_c\theta_{\bar{\xi}1}}{8\pi^2}  \frac{(1-\xi)(4m_K^2-t) }{(1+\xi)} \bar{\mathcal{C}}_1(\sigma_1^{9,1}) -\frac{N^{EE}}{4m_K^2} \frac{N_c\theta_{\xi1}}{8\pi ^2} \frac{(4m_K^2-t) (1+\xi)}{(1-\xi)} \bar{\mathcal{C}}_1(\sigma_1^{0,-1})\nonumber\\
-&\frac{\bar{N}^{EF}}{4m_K^2}\frac{N_c\theta_{\bar{\xi}1}}{8\pi^2}  \frac{(2x+\xi-1)(1-\xi)(4m_K^2-t)}{(1+\xi)^2}  \bar{\mathcal{C}}_1(\sigma_1^{0,1}) \nonumber\\
-& \frac{\bar{N}^{EF}}{4m_K^2} \frac{N_c\theta_{\xi1}}{8\pi ^2} \frac{(2x-\xi-1)(1+\xi)(4m_K^2-t)}{(1-\xi)^2} \bar{\mathcal{C}}_1(\sigma_1^{0,-1})\nonumber\\
+& \frac{(E_K^2-2\bar{N}^{EF})}{4m_K^2} \frac{N_c}{8\pi^2} x(4m_K^2-t) \frac{\theta_{\bar{\xi}\xi}}{\xi}\bar{\mathcal{C}}_1(\sigma_2^0)+  \frac{(E_K^2-2\bar{N}^{EF})}{4m_K^2} \frac{N_c }{4\pi^2}  \frac{xt\theta_{\bar{\xi}\xi}}{\xi^3} \bar{\mathcal{C}}_1(\sigma_2^0) \nonumber\\
+&\frac{N^{FF}}{4m_K^2}\frac{N_c}{16\pi^2}  \frac{\theta_{\bar{\xi}\xi}}{\xi}(1-\frac{x^2}{\xi^2})t (4m_K^2-t)\bar{\mathcal{C}}_1(\sigma_2^0) -\frac{\bar{N}^{FF}}{2m_K^2} \frac{N_c}{8\pi^2}\frac{\theta_{\bar{\xi}\xi}xt}{\xi^3} \bar{\mathcal{C}}_1(\sigma_2^0) \nonumber\\
+&\frac{N^{EF}}{4m_K^2} \frac{N_c }{8\pi^2}  \int_0^1  d\alpha  \frac{\theta_{\alpha \xi}}{\xi} \frac{\bar{\mathcal{C}}_2(\sigma_3^0)}{\sigma_3^0}(4m_K^2-t)((2\alpha-x)(t-2(m_K^2-\left(M_u-M_s\right)^2))-t) \nonumber\\
+& \frac{N^{EF}}{8m_K^2} \frac{N_c}{2\pi^2}  \int_0^1  d\alpha \frac{\theta_{\alpha \xi}}{\xi^3}(2(m_K^2-(M_u-M_s)^2)-t)(x-\alpha)t\frac{\bar{\mathcal{C}}_2(\sigma_3^0)}{\sigma_3^0},
\end{align}
\begin{align}\label{agtmd}
H_3^{u}(x,\xi,t)=&  \frac{N^{FF}N_c}{32\pi^2} \frac{\theta_{\bar{\xi}\xi}}{ m_K \xi} \bar{\mathcal{C}}_1(\sigma_2^0)- \frac{N^{EF}N_c}{16\pi^2} \int_0^1  d\alpha \frac{\theta_{\alpha \xi}(M_u+(M_s-M_u)\alpha )}{m_K\xi} \frac{\bar{\mathcal{C}}_3(\sigma_3^0)}{\sigma_3^0}.
\end{align}
Among the twist-four GPDs, due to the hermiticity constraint of $\Delta$, $F_3$ is T-even, $H_3$ is T-odd.

\acknowledgments

Work supported by: the Scientific Research Foundation of Nanjing Institute of Technology (Grant No. YKJ202352), the Natural Science Foundation of Jiangsu Province (Grant No. BK20191472), and the China Postdoctoral Science Foundation (Grant No. 2022M721564).









\bibliographystyle{apsrev4-1}
\bibliography{zhang}
\end{document}